\documentclass[aps,prl,preprint,superscriptaddress,nofootinbib,floatfix]{revtex4-2}

\usepackage{amsmath,amssymb,bm}
\usepackage{graphicx}
\newcommand{\MissingFigure}[3]{%
  \fbox{\parbox[c][#3][c]{#2}{\centering\footnotesize
  \textbf{Figure asset not supplied}\\[3pt]\texttt{\detokenize{#1}}}}}

\usepackage{dcolumn}
\usepackage{slashed}
\usepackage{booktabs,array,multirow}
\usepackage[table]{xcolor}
\usepackage{pgfplots}
\usepgfplotslibrary{fillbetween}
\usepackage{tikz}
\usetikzlibrary{arrows.meta}
\usepackage{enumitem}
\usepackage{microtype}
\usepackage[a4paper,top=25mm,bottom=24mm,left=24mm,right=24mm,
            headheight=28pt,headsep=7mm,footskip=12mm]{geometry}
\usepackage{fancyhdr}
\pgfplotsset{compat=1.18}

\newcommand{\RunningPaperTitle}{Charged Letelier--AdS Black Holes: Thermodynamics and Optics}
\fancypagestyle{plain}{%
  \fancyhf{}%
  \fancyhead[L]{\parbox[b]{0.47\textwidth}{\raggedright\footnotesize\itshape\nouppercase{\leftmark}}}%
  \fancyhead[R]{\parbox[b]{0.47\textwidth}{\raggedleft\footnotesize\itshape\RunningPaperTitle}}%
  \fancyfoot[C]{\small\thepage}%
}

\definecolor{acsblue}{RGB}{17,76,139}
\definecolor{grn}{RGB}{0,140,60}
\usepackage[colorlinks=true,linkcolor=blue,citecolor=red,urlcolor=acsblue,
            pdfstartview=FitH,bookmarksnumbered=true]{hyperref}
\hypersetup{
  pdftitle={Thermodynamic, Optical, and Wave Signatures of Charged Letelier--AdS Black Holes with PFDM and Quintessence},
  pdfauthor={B. Eslam Panah; A. Guvendi; S. Gurtas Dogan; O. Mustafa},
  pdfsubject={Black-hole thermodynamics, heat engines, ray optics, and wave propagation},
  pdfstartview=FitH,
  bookmarksnumbered=true
}

\newcolumntype{C}{>{\centering\arraybackslash}m{4cm}}

\begin{document}

\title[Letelier--AdS Black Holes: Thermodynamics and Wave Optics]{Thermodynamic, Optical, and Wave Signatures of Charged Letelier--AdS Black Holes with PFDM and Quintessence}

\author{B. Eslam Panah}
\affiliation{Department of Theoretical Physics, Faculty of Basic Sciences,
University of Mazandaran, P. O. Box 47416-95447, Babolsar, Iran.}
\affiliation{Center for Theoretical Physics, Khazar University,
41 Mehseti Str., Baku, AZ1096, Azerbaijan.}

\author{S. Gurtas Dogan}
\affiliation{Department of Medical Imaging Techniques,
Hakkari University, 30000, Hakkari, Türkiye.}

\author{A. Guvendi}
\affiliation{Department of Basic Sciences,
Erzurum Technical University, 25050, Erzurum, Türkiye.}

\author{O. Mustafa}
\affiliation{Department of Physics, Eastern Mediterranean University,
99628, G. Magusa, north Cyprus, Mersin 10 - Türkiye.}

\date{\today}

\begin{abstract}
We study a four-dimensional charged Letelier--anti-de Sitter (AdS) black-hole metric containing a cloud of strings, a phenomenological perfect-fluid-dark-matter (PFDM) source, and a Kiselev-type quintessence term. We derive the curvature invariants, horizon temperature, entropy, electric potential, and horizon-based mass function, and analyze local thermal response and the associated extended-phase-space identities. A rectangular benchmark cycle gives $W=0.095000$ and $\eta\simeq15.39\%$ for the explicitly stated geometric-unit parameters; these values are model- and cycle-specific and are not presented as a universal efficiency enhancement. For geometric optics, we derive the photon-sphere condition, critical impact parameter, and finite-radius shadow angle. For a massless, minimally coupled scalar field, we obtain a radial-coordinate Helmholtz equation and identify frequency-dependent barriers, evanescent intervals, and candidate trapping regions; the eikonal barrier-top scale approaches the photon-sphere capture scale. Two qualifications delimit the interpretation. First, with the stated PFDM stress-tensor convention, $\rho_{\rm PFDM}=-\beta/(8\pi r^3)$, so positive $\beta$ describes a negative-density effective source. Second, the mass function follows from the horizon equation, but identifying it with a conserved asymptotic AdS mass requires a separate charge and boundary-normalization analysis. The potential-well analysis likewise does not determine complex quasinormal frequencies. We therefore present the results as a consistent, parameter-specific analysis of the adopted phenomenological geometry, while distinguishing algebraic thermodynamic identities from conclusions requiring asymptotic-charge or spectral calculations.
\end{abstract}

\begingroup
\renewcommand{\thefootnote}{}
\footnotetext{%
\footnotesize
\textit{E-mail addresses:}\\
B. Eslam Panah: eslampanah@umz.ac.ir;
behzad.eslampanah@gmail.com.\\
S. Gurtas Dogan: semragurtasdogan@hakkari.edu.tr (Corresponding Author).\\
A. Guvendi: abdullah.guvendi@erzurum.edu.tr.\\
O. Mustafa: omar.mustafa@emu.edu.tr.
}
\endgroup

\maketitle

\thispagestyle{fancy}

\thispagestyle{fancy}


\section{Introduction}
\label{sec:intro}

Black holes provide a setting in which strong-field gravity, causal structure, and thermodynamics can be studied within a common geometrical framework. The singularity theorems established that gravitational collapse can produce regions from which future-directed causal signals cannot escape, subject to broad assumptions on the matter content and spacetime structure \cite{Penrose1965,Hawking1970}. The resulting black-hole solutions raise questions about the classical description of spacetime, the fate of information, and the relation between gravity and statistical mechanics \cite{Berti2015}. These questions are no longer motivated by theory alone. Gravitational-wave observations of compact-object mergers have made it possible to test strong-field dynamics and infer remnant properties \cite{LIGO2016,LIGO2017,LIGO2020}, while horizon-scale imaging by the Event Horizon Telescope has provided shadow measurements for M87$^\ast$ and Sgr A$^\ast$ \cite{EHT2019,EHT2022}. The interpretation of such observations depends on the geometry of the emitting and propagating regions, and on the assumptions made about the matter surrounding the compact object.

Astrophysical black holes are not generally isolated. Accretion flows, dark-matter distributions, scalar fields, and other environmental components can alter the geometry and affect orbital motion, wave propagation, and the relation between a horizon and distant observables. Studies of environmental effects have therefore considered both the limits of vacuum idealizations and the possibility that matter surrounding a black hole may leave measurable signatures in gravitational-wave or electromagnetic data \cite{Medved2004,Narayan2005,Barausse2014,Macedo2016,Macedo2024,Cardoso2022}. The optical appearance of a black hole is particularly sensitive to the structure of unstable null orbits and to the matter distribution between the photon region and the observer \cite{Cunha2015,Konoplya2021,Rahmatov2025,Hoshimov2025,Bakhodirov2025}. Such effects must be interpreted with care: a change in a metric parameter can modify the horizon, the photon sphere, and the observer's local measurement simultaneously, so a shadow radius alone need not determine the underlying parameters uniquely.

One phenomenological description of surrounding matter is Letelier's cloud of strings, whose averaged stress tensor contributes an anisotropic term proportional to \(r^{-2}\) and modifies the constant part of a static, spherically symmetric metric function \cite{Letelier1979}. String-cloud geometries have been examined in higher-curvature and higher-dimensional gravity, in nonlinear-electrodynamic settings, and in configurations containing dark-energy-like sources \cite{Ghosh2014,Toledo2018,Waseem2023,Rodrigues2022,Muniz2025,Mustafa2022,Atamurotov2022}. Their effects on horizons, circular geodesics, shadows, and thermodynamic quantities have also been investigated \cite{Rincon2018,Li2021,Belhaj2022,He2022,Toledo2019,Toledo2019b,Singh2020,Rodrigues2022b,Sadeghi2024}. In particular, the coexistence of a string cloud and quintessence has been considered in earlier thermodynamic studies, including analyses of the Letelier spacetime and related black-hole families \cite{Dias2019,Zafar2025,AhmedArX2025}. These works provide relevant context for assessing which conclusions follow from the chosen metric and which depend on the detailed matter model.

A second sector considered here is perfect-fluid dark matter (PFDM). The PFDM prescription is a phenomenological effective source used to investigate how a dark-matter-like contribution may affect black-hole geometry; it should not be confused with a unique microscopic model of galactic dark matter \cite{PFDM1,PFDM2,PFDM3}. Several black-hole solutions and their optical or dynamical properties have been studied in this framework \cite{BHPFDM1,BHPFDM2,BHPFDM3,BHPFDM4,ProPFDM1,ProPFDM2,ProPFDM3,ProPFDM4,ProPFDM5,ProPFDM6,ProPFDM7,ProPFDM8,ProPFDM9,ProPFDM10,ProPFDM11,ProPFDM12,ProPFDM13,ProPFDM14,ProPFDM15,ProPFDM16,ProPFDM17,ProPFDM18,ProPFDM19,ProPFDM20,ProPFDM21,ProPFDM22,ProPFDM23,ProPFDM24,ProPFDM25,ProPFDM26,ProPFDM27,ProPFDM28,ProPFDM29,ProPFDM30}. Those applications cover shadows and lensing, wave and quasinormal-mode analyses, thermal behavior, and criticality. The physical interpretation of any PFDM parameter, however, depends on the precise stress-tensor convention. For the convention adopted in this paper, the density associated with the PFDM tensor is \(\rho_{\rm PFDM}=-\beta/(8\pi r^3)\); thus \(\beta>0\) corresponds to negative density. We keep this sign explicit throughout and do not identify the positive-\(\beta\) sector with an ordinary positive-density dark-matter halo.

The third environmental contribution is a Kiselev-type quintessence source, parameterized by a normalization \(\sigma\) and an equation-of-state parameter \(\omega\) in the range \(-1<\omega<-1/3\) \cite{BHPFDM1}. Its radial falloff differs from that of the mass, charge, string cloud, and PFDM terms, and can consequently affect both finite-radius dynamics and the asymptotic form of the metric. When the cosmological constant is negative, the spacetime also has an AdS curvature scale. The resulting metric contains several competing contributions, so horizon structure, thermodynamic branches, null-orbit structure, and finite-frequency wave propagation must be determined from the full metric function rather than inferred from a single term in isolation.

Black-hole thermodynamics provides a second set of diagnostics for this geometry. The Bekenstein--Hawking relations identify the temperature with the surface gravity and the entropy with one quarter of the horizon area in Planck units \cite{Bekenstein1973,Hawking1974,Callen1985}. The thermal response and phase structure of AdS black holes have been studied in connection with the Hawking--Page transition and charged black-hole criticality \cite{Davies1977,Hawking1983,Chamblin1999}. In extended phase space, the cosmological constant is treated as a pressure and the black-hole mass as an enthalpy, giving a generalized first law and a scaling relation of Smarr type \cite{Witten1998,Kastor2009,PLambda2,PLambda3,PLambda4}. This framework has led to a large body of work on \(P\)-\(V\) criticality and related phase behavior in AdS black holes \cite{PV1,PV2,PV3,PV4,PV5,PV6,PV7,PV8,PV9,PV10,PV11,PV12,PV13,PV14,PV15,PV16,PV17,PV18,PV19}. Studies of dark-energy effects on charged AdS criticality and heat-engine performance are especially relevant to the present setting \cite{Li2014,liu2017,Xu2018}. In any such analysis, local stability and global thermodynamic preference must be distinguished: the sign of a heat capacity diagnoses a local response, whereas a phase-preference claim requires a suitable free-energy comparison in a specified ensemble and with consistent boundary conditions.

The same metric also governs geometric optics. In a static, spherically symmetric spacetime, null geodesics determine the photon-sphere condition and the critical impact parameter separating capture from scattering. The theoretical foundations of photon surfaces and strong-field lensing are well established \cite{1,2,3,4,Claudel2001}, and optical-metric methods provide a complementary geometrical treatment of light propagation \cite{8,9,10,11}. Reviews of black-hole shadows clarify how the critical orbit is related to an observer's locally measured angular radius, particularly when the observer is at finite radius rather than at asymptotic infinity \cite{25,Perlick2022,Synge1966}. This distinction is essential in AdS, where the asymptotic region is not flat and an observer-dependent shadow angle is more appropriate than an unqualified asymptotic angular size. Related work on wormholes, curved optical surfaces, and modified-gravity backgrounds further demonstrates the range of geometries in which ray and wave propagation can be analyzed using effective optical descriptions \cite{12,13,14,15,19,26,17,18,20,Spiral2026,Godel2026,CJP2026,EMSchw2026,EslamPanah2026bumblebee}.

Geometric optics does not capture all finite-frequency effects. A minimally coupled scalar field obeys the Klein--Gordon equation, whose radial reduction contains both the angular-momentum barrier and terms built from derivatives of the metric function. In an appropriate radial representation, these terms define an effective potential and a frequency-dependent coefficient that can be used to identify oscillatory and evanescent regions. The standard theory of black-hole perturbations and quasinormal modes provides the context for this analysis \cite{21,22,23,24,Konoplya2011,Horowitz2000,Cardoso2009,Zou2017}. A potential barrier or a bounded oscillatory interval may indicate a candidate trapping region, but it does not by itself establish a quasinormal-mode spectrum: that requires solving the radial boundary-value problem with ingoing behavior at the future horizon and the appropriate outer boundary condition. In the AdS setting considered here, the asymptotic boundary condition is part of the physical problem, not an optional detail.

Against this background, we analyze a four-dimensional, Maxwell-charged
Letelier--AdS metric with the PFDM and Kiselev-type quintessence terms defined
above. We derive the horizon and curvature properties, formulate the
horizon-based thermodynamic identities and a specified rectangular-cycle
benchmark, and calculate photon-sphere, critical-impact-parameter, and
finite-radius shadow observables. We also reduce the massless Klein--Gordon
equation to a radial-coordinate Helmholtz form and compare its leading
eikonal barrier scale with the null-geodesic capture scale. The scope is
deliberately limited: the paper does not claim a conserved asymptotic mass,
a data-derived observational constraint, or a complex quasinormal-mode
spectrum. Those claims would require, respectively, a compatible
boundary-charge construction, an explicit observational likelihood, and a
complex-frequency boundary-value calculation.

The combination of sectors and the use of standard thermodynamic and optical
tools are not, by themselves, claimed as novel. Closely related work includes
an uncharged Schwarzschild--AdS model with a string cloud, PFDM, and
quintessence \cite{AhmedArX2025}, a charged AdS string-cloud/quintessence
analysis using shadow-based diagnostics \cite{Wang2026}, and a model with
Euler--Heisenberg electrodynamics, PFDM, quintessence, and a string cloud
\cite{AhmedSilva2026}. The present paper's specific contribution is the
consistent treatment of the stated Maxwell-charged metric across its
horizon thermodynamics, the explicit rectangular-cycle benchmark, the
finite-radius observer angle, and the scalar Helmholtz/eikonal comparison.
These results are model-specific consistency and observable calculations;
their interpretation remains conditional on the source convention and
the qualifications stated in the relevant sections.

Table~\ref{tab:scope} summarizes the scope of selected related studies. A check mark records that a topic is addressed in the stated scope.

\begin{table*}[tbp]
\centering
\caption{Scope comparison with selected closely related studies. The table
summarizes the published or publicly stated scope of each work and is not a
substitute for a result-by-result comparison.}
\label{tab:scope}
\begingroup
\scriptsize
\setlength{\tabcolsep}{3pt}
\begin{tabular}{lccccc}
\hline
Work & \shortstack{Maxwell\\charge} & PFDM & \shortstack{String\\cloud} & Quintessence & \shortstack{Scalar-wave/QNM\\analysis}\\
\hline
Ahmed--Silva \cite{AhmedArX2025} & -- & \checkmark & \checkmark & \checkmark & --\\
Wang \textit{et al.} \cite{Wang2026} & \checkmark & -- & \checkmark & \checkmark & --\\
Ahmed--Silva \cite{AhmedSilva2026} & \checkmark$^{a}$ & \checkmark & \checkmark & \checkmark & --\\
Present work & \checkmark & \checkmark & \checkmark & \checkmark & \shortstack{Effective potential\\and eikonal limit only}\\
\hline
\end{tabular}
\endgroup
\begin{flushleft}
\footnotesize $^{a}$ The cited work includes nonlinear Euler--Heisenberg
electrodynamics; it is not the same linear Maxwell model considered here.
The table uses ``--'' only to indicate that the corresponding feature is not
part of the stated focus of the cited study; it does not imply that every
possible version of that model excludes it.
\end{flushleft}
\end{table*}

\section{Field Equations and Spacetime}
\label{sec:fe}

\subsection{Field equations and phenomenological sources}

We consider a four-dimensional spacetime with metric signature $(-,+,+,+)$
and use geometrized units, $G=c=1$, so that $\kappa^2=8\pi$. We specify the
Einstein--Maxwell equations and the phenomenological stress tensors directly.
Because covariant matter Lagrangians for the PFDM, string-cloud, and
Kiselev-type sectors are not supplied, we do not claim that these three
effective sources follow from a single variational matter action. This
distinction is important when interpreting the solution as a phenomenological
geometry rather than as a fully specified microscopic matter model.

Let $\mathcal R$ denote the Ricci scalar,
$\mathcal F=F_{\mu\nu}F^{\mu\nu}$, and
$F_{\mu\nu}=\partial_\mu A_\nu-\partial_\nu A_\mu$ the Faraday tensor.
With the stress-tensor convention specified below, the field equations take the form
\begin{align}
 G_\mu{}^\nu+\Lambda\delta_\mu{}^\nu
 &=8\pi T_\mu{}^{\nu\,({\rm tot})},
 \label{FE}\\
 \nabla_\mu F^{\mu\nu}&=0,
 \label{Ftr}
\end{align}
where $G_\mu{}^\nu$ denotes the mixed Einstein tensor.
The Maxwell equation can equivalently be written
$\partial_\mu(\sqrt{-g}F^{\mu\nu})=0$. We define
\begin{equation}
 T_\mu{}^{\nu\,({\rm tot})}
 =T_\mu{}^{\nu\,({\rm Maxwell})}
 +T_\mu{}^{\nu\,({\rm PFDM})}
 +T_\mu{}^{\nu\,({\rm CS})}
 -T_\mu{}^{\nu\,(Q)}.
 \label{TotalT}
\end{equation}
The minus sign in front of $T_\mu{}^{\nu\,(Q)}$ is part of the convention
for the Kiselev-type tensor written in Eq.~(\ref{ComTQ}); it must be kept
consistently in the Einstein equations. All statements about the quintessence
density and its contribution below refer to this explicit convention, not to
a sign convention imported from another Kiselev parametrization. The standard Maxwell stress tensor
is
\begin{equation}
 T_\mu{}^{\nu\,({\rm Maxwell})}
 =\frac{1}{4\pi}\left(
 F_{\mu\lambda}F^{\nu\lambda}
 -\frac14\delta_\mu{}^\nu\mathcal F
 \right).
 \label{TMax}
\end{equation}
Equivalently, the first term may be written
$F_\mu{}^\lambda F^\nu{}_\lambda/(4\pi)$.

For a cloud of strings, we use the Letelier stress tensor \cite{Letelier1979}
\begin{equation}
 T_\mu{}^{\nu\,({\rm CS})}
 =\operatorname{diag}\left(
 -\frac{\alpha}{8\pi r^2},
 -\frac{\alpha}{8\pi r^2},0,0
 \right),
 \label{ComTL}
\end{equation}
where $\alpha$ is the dimensionless string-cloud parameter. For the quintessence sector, we use the Kiselev-type tensor \cite{BHPFDM1}

\begin{widetext}
\begin{equation}
 T_\mu{}^{\nu\,(Q)}
 =\operatorname{diag}\left(
 -\frac{3\omega\sigma}{8\pi r^{3(\omega+1)}},
 -\frac{3\omega\sigma}{8\pi r^{3(\omega+1)}},
 \frac{3\omega\sigma(3\omega+1)}{16\pi r^{3(\omega+1)}},
 \frac{3\omega\sigma(3\omega+1)}{16\pi r^{3(\omega+1)}}
 \right),
 \label{ComTQ}
\end{equation}
\end{widetext}
where $-1<\omega<-1/3$ and $\sigma>0$. Because this tensor enters
Eq.~(\ref{TotalT}) with an overall minus sign, its contribution to the
effective total tensor has energy density
$\rho_Q=-3\omega\sigma/[8\pi r^{3(\omega+1)}]>0$.

The PFDM tensor is taken to be \cite{ProPFDM10}
\begin{equation}
 T_\mu{}^{\nu\,({\rm PFDM})}
 =\operatorname{diag}\left(
 \frac{\beta}{8\pi r^3},
 \frac{\beta}{8\pi r^3},
 -\frac{\beta}{16\pi r^3},
 -\frac{\beta}{16\pi r^3}
 \right),
 \label{ComTPFDM}
\end{equation}
where $\beta$ is an integration parameter controlling the strength of this
sector. With the convention $T_t{}^t=-\rho$, this tensor has
\begin{equation}
 \rho_{\rm PFDM}=-\frac{\beta}{8\pi r^3},\qquad
 p_r=-\rho_{\rm PFDM},\qquad
 p_t=\frac12\rho_{\rm PFDM}.
 \label{PFDMpressures}
\end{equation}
Thus the PFDM density is positive only for $\beta<0$. For $\beta>0$ it is
negative, and $\rho_{\rm PFDM}+p_t=3\rho_{\rm PFDM}/2<0$; consequently the
null and weak energy conditions are violated by this sector. Results for
$\beta>0$ should therefore not be interpreted as those of an ordinary
positive-density matter halo. The string-cloud tensor in
Eq.~(\ref{ComTL}), by contrast, has $\rho_{\rm CS}=\alpha/(8\pi r^2)$.

\subsection{Static, spherically symmetric solution}

We consider the static, spherically symmetric line element
\begin{equation}
 ds^2=-f(r)\,dt^2+\frac{dr^2}{f(r)}
      +r^2(d\theta^2+\sin^2\theta\,d\varphi^2).
 \label{metric}
\end{equation}
For a purely radial electric field, choose
$A_\mu=h(r)\delta_\mu^t$. Equation~(\ref{Ftr}) then gives
\begin{equation}
 \frac{d}{dr}\left(r^2h'(r)\right)=0
 \quad\Longleftrightarrow\quad
 2h'(r)+rh''(r)=0.
 \label{MaxwellODE}
\end{equation}
The solution, after fixing the additive gauge constant to zero, is
\begin{equation}
 h(r)=-\frac{q}{r},
 \label{h(r)}
\end{equation}
where $q$ is the electric-charge parameter. With the convention
$F_{\mu\nu}=\partial_\mu A_\nu-\partial_\nu A_\mu$, the nonzero
radial components are $F_{rt}=h'(r)=q/r^2$ and
$F_{tr}=-h'(r)=-q/r^2$.

For the metric in Eq.~(\ref{metric}), the nonzero diagonal mixed components
of the Einstein tensor are
\begin{align}
 G_t{}^t=G_r{}^r&=\frac{rf'(r)+f(r)-1}{r^2},\\
 G_\theta{}^\theta=G_\varphi{}^\varphi
 &=\frac{f''(r)}{2}+\frac{f'(r)}{r}.
\end{align}
Substitution of the matter tensors and the electric field into
Eq.~(\ref{FE}), followed by multiplication of the temporal and angular
equations by $r^2$, yields

\begin{widetext}
\begin{align}
 {\cal E}_{tt}={\cal E}_{rr}
 &=\Lambda r^2+rf'(r)+f(r)-1+\alpha-\frac{\beta}{r}
   +\frac{q^2}{r^2}
   -\frac{3\omega\sigma}{r^{3\omega+1}}=0,
 \label{eqtt}\\
 {\cal E}_{\theta\theta}={\cal E}_{\varphi\varphi}
 &=\left(\frac{f''(r)}{2}+\Lambda\right)r^2+rf'(r)
   +\frac{\beta}{2r}-\frac{q^2}{r^2}
   +\frac{3\omega\sigma(3\omega+1)}
          {2r^{3\omega+1}}=0.
 \label{eqthethe}
\end{align}
\end{widetext}
Integrating the temporal equation, Eq.~(\ref{eqtt}), gives
\begin{equation}
 f(r)=1-\alpha-\frac{2m_0}{r}-\frac{\Lambda r^2}{3} +\frac{q^2}{r^2} +\frac{\beta}{r}\ln\!\left(\frac{r}{|\beta|}\right) -\frac{\sigma}{r^{3\omega+1}}\label{f(r)}
\end{equation}
Here, $m_0$ is an integration constant identified with the mass parameter.
The logarithmic contribution is defined by its continuous limit as
$\beta\to0$, in which it vanishes; the explicit logarithmic expression
is therefore used only for $\beta\ne0$. Substitution of
Eq.~(\ref{f(r)}) into Eqs.~(\ref{eqtt}) and (\ref{eqthethe}) verifies
that the temporal and angular equations are satisfied.

\subsection{Useful limits}

Several simple limits provide checks on the metric function,
\begin{equation}
 f(r)=1-\alpha-\frac{2m_0}{r}-\frac{\Lambda r^2}{3}
 +\frac{q^2}{r^2}
 +\frac{\beta}{r}\ln\!\left(\frac{r}{|\beta|}\right)
 -\frac{\sigma}{r^{3\omega+1}}.
 \label{flimits}
\end{equation}
The logarithmic term is defined by continuity to vanish at \(\beta=0\).
Setting \(q=0\) gives the uncharged solution; setting \(\beta=0\)
removes the PFDM contribution; setting \(\sigma=0\) removes quintessence.
The simultaneous limit \(\alpha=\beta=\sigma=0\) recovers the
Reissner--Nordström--(A)dS metric,
\begin{equation}
 f(r)=1-\frac{2m_0}{r}-\frac{\Lambda r^2}{3}+\frac{q^2}{r^2}.
 \label{RNAdSlimit}
\end{equation}
Further setting \(q=0\) yields Schwarzschild--(A)dS. These are algebraic
consistency limits; they do not, by themselves, establish the physical
admissibility of every parameter choice in the full model.

\subsection{Curvature invariants and asymptotics}

For the metric (\ref{metric}), the Ricci scalar can be calculated from
\begin{equation}
 \mathcal R=-f''(r)-\frac{4f'(r)}{r}
             -\frac{2[f(r)-1]}{r^2}.
 \label{RicciIdentity}
\end{equation}
Using Eq.~(\ref{f(r)}) gives
\begin{equation}
 \mathcal R=4\Lambda+\frac{2\alpha}{r^2}
 -\frac{\beta}{r^3}
 +\frac{3\omega\sigma(3\omega-1)}
        {r^{3(\omega+1)}}.
 \label{R}
\end{equation}
The charge and mass terms cancel from $\mathcal R$, as expected for this
combination of the Einstein equations. For $\sigma>0$ and
$-1<\omega<-1/3$, the quintessence contribution diverges as $r\to0$;
if $\sigma=0$, the Ricci scalar need not diverge for every parameter
choice. The curvature singularity at the center is more generally
diagnosed using the Kretschmann scalar.

For any metric of the form (\ref{metric}), the Kretschmann scalar has the
compact exact expression
\begin{equation}
 K\equiv R_{\mu\nu\rho\sigma}R^{\mu\nu\rho\sigma}
 = [f''(r)]^2+4\left[\frac{f'(r)}{r}\right]^2
  +4\left[\frac{1-f(r)}{r^2}\right]^2
 \label{Kcompact}
\end{equation}
This compact expression is more transparent than a fully expanded formula
and avoids possible ambiguities in auxiliary abbreviations. For explicit evaluation, the derivatives of
Eq.~(\ref{f(r)}) are

\begin{widetext}
\begin{align}
 f'(r)&=\frac{2m_0}{r^2}-\frac{2\Lambda r}{3}
 -\frac{2q^2}{r^3}
 +\frac{\beta}{r^2}\left[1-\ln\!\left(\frac{r}{|\beta|}\right)\right]
 +(3\omega+1)\sigma r^{-3\omega-2},
 \label{fprime}\\
 f''(r)&=-\frac{4m_0}{r^3}-\frac{2\Lambda}{3}
 +\frac{6q^2}{r^4}
 +\frac{\beta}{r^3}\left[2\ln\!\left(\frac{r}{|\beta|}\right)-3\right]
 -(3\omega+1)(3\omega+2)\sigma r^{-3\omega-3}.
 \label{fsecond}
\end{align}
\end{widetext}
Terms containing $\beta\ln(r/|\beta|)$ are omitted by taking the
continuous $\beta\to0$ limit when $\beta=0$. For generic nonzero $m_0$,
$q$, $\alpha$, $\beta$, or $\sigma$, the curvature invariants signal a
central curvature singularity; exceptional parameter cancellations
should be assessed from Eqs.~(\ref{R}) and (\ref{Kcompact}), rather than
asserting that the Ricci scalar alone diverges in every limit.

As $r\to\infty$, the curvature invariants satisfy
\begin{equation}
 \mathcal R\longrightarrow4\Lambda,\qquad
 K\longrightarrow\frac{8\Lambda^2}{3},
 \label{asymptoticcurvature}
\end{equation}
for the stated range of $\omega$ and finite parameters. When $\Lambda<0$
the leading asymptotic curvature is AdS-like, and when $\Lambda>0$ it is
dS-like. If $\Lambda=0$, the invariants tend to zero, but for
$\sigma\ne0$ the metric generally contains the growing term
$-\sigma r^{-3\omega-1}$; therefore vanishing asymptotic curvature
invariants alone do not establish standard asymptotic flatness.

\subsection{Horizons and parameter dependence}

The horizons are located at the positive real roots of
\begin{equation}
 f(r_h)=0.
 \label{horizoneq}
\end{equation}
An extremal horizon corresponds to a double root and therefore satisfies both
\begin{equation}
 f(r_e)=0,\qquad f'(r_e)=0.
 \label{extremality}
\end{equation}
Consequently, any critical value of $\sigma$, $\omega$, $\alpha$, $\beta$,
$q$, or $m_0$ depends on the fixed values of the other parameters and on
the radial domain considered. It cannot be specified universally from
Eq.~(\ref{f(r)}) alone. For each parameter scan, the number and locations
of horizons should be established by solving Eq.~(\ref{horizoneq}) and
checking the double-root conditions (\ref{extremality}); plots by
themselves should not be used to claim a critical value without reporting
the parameters held fixed.

\begin{figure*}[tbp]
 \centering
 \IfFileExists{1a.eps}{\includegraphics[width=0.30\textwidth]{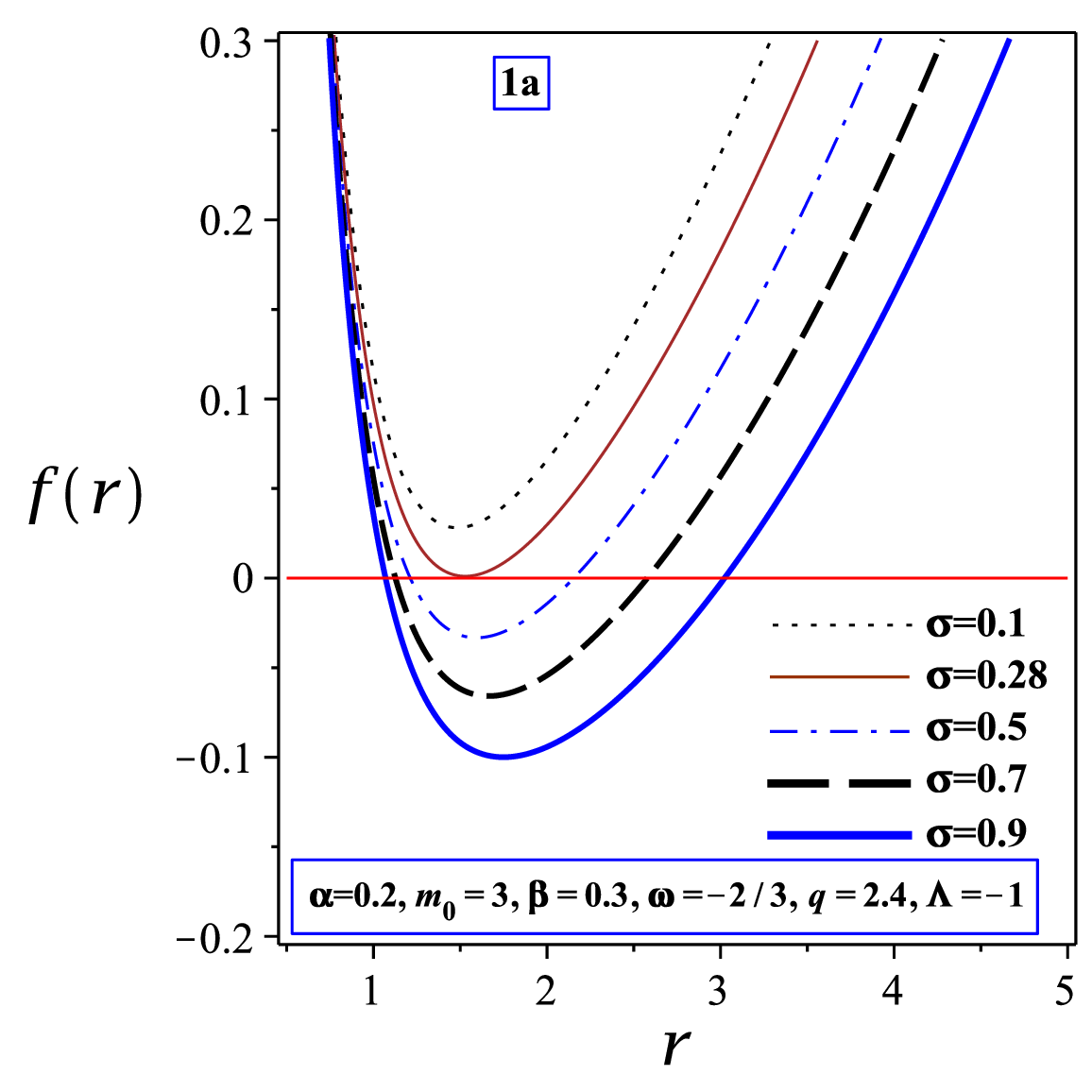}}{\MissingFigure{1a.eps}{0.28\textwidth}{0.12\textheight}}
 \IfFileExists{1b.eps}{\includegraphics[width=0.30\textwidth]{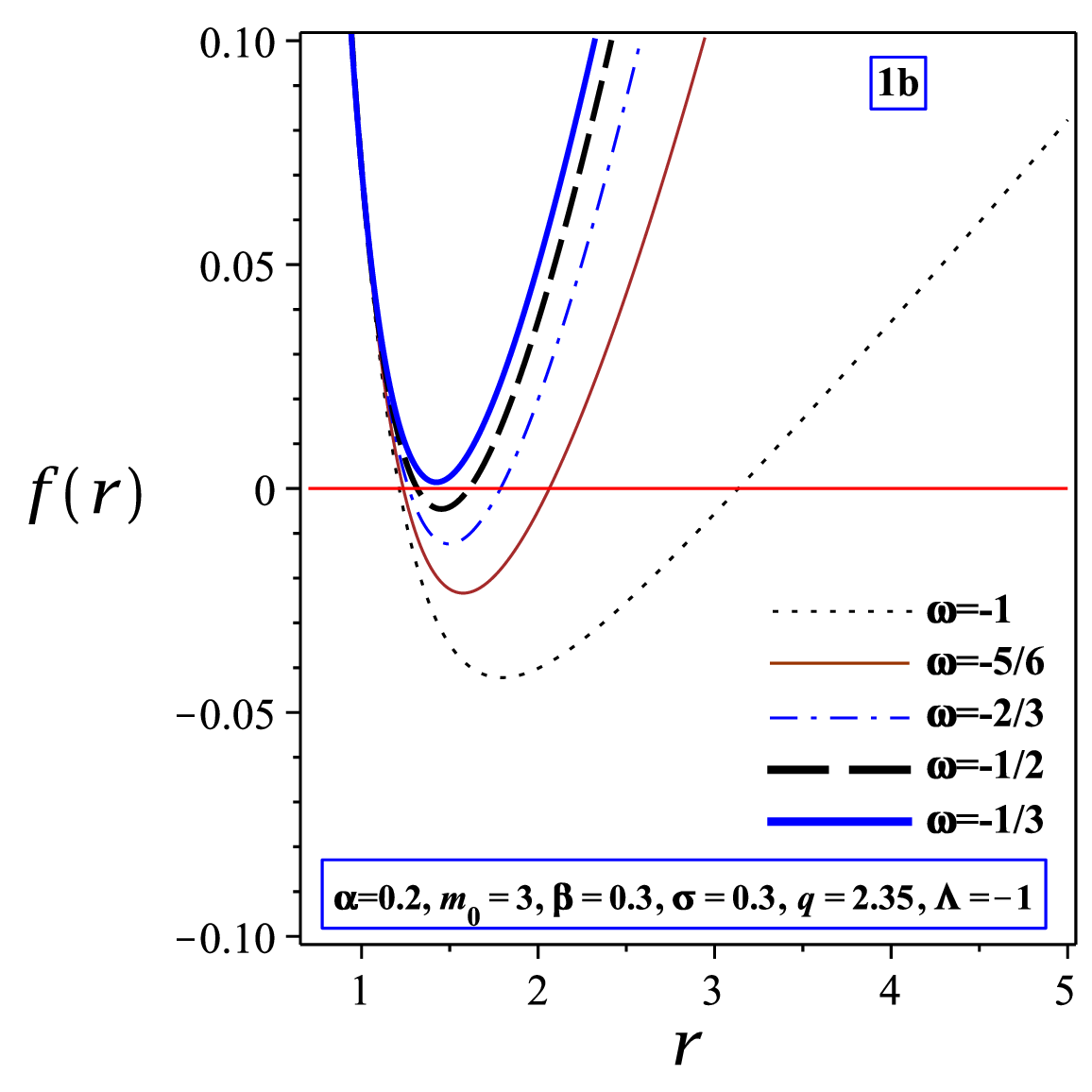}}{\MissingFigure{1b.eps}{0.28\textwidth}{0.12\textheight}}
 \IfFileExists{1c.eps}{\includegraphics[width=0.30\textwidth]{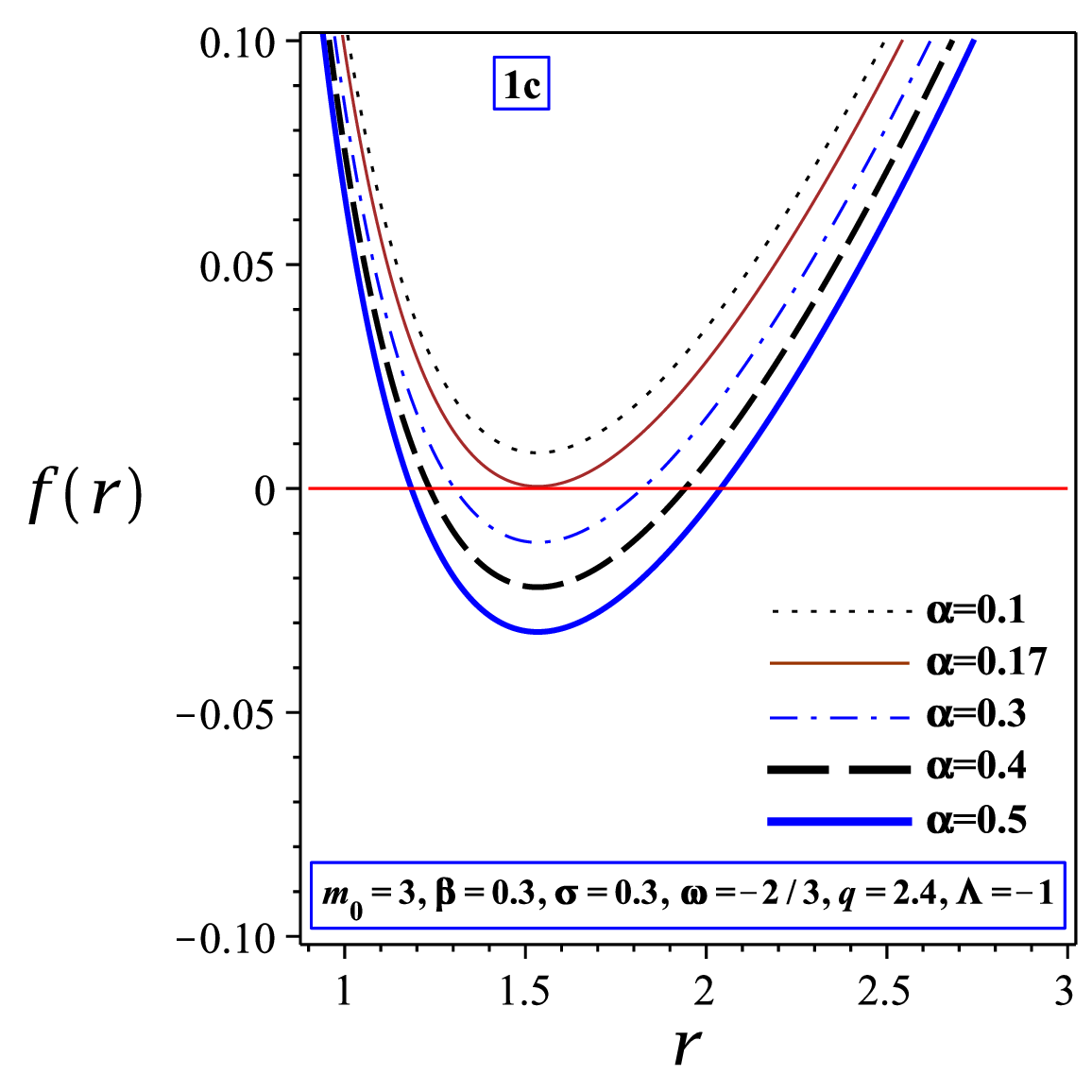}}{\MissingFigure{1c.eps}{0.28\textwidth}{0.12\textheight}}\\
 \IfFileExists{1d.eps}{\includegraphics[width=0.30\textwidth]{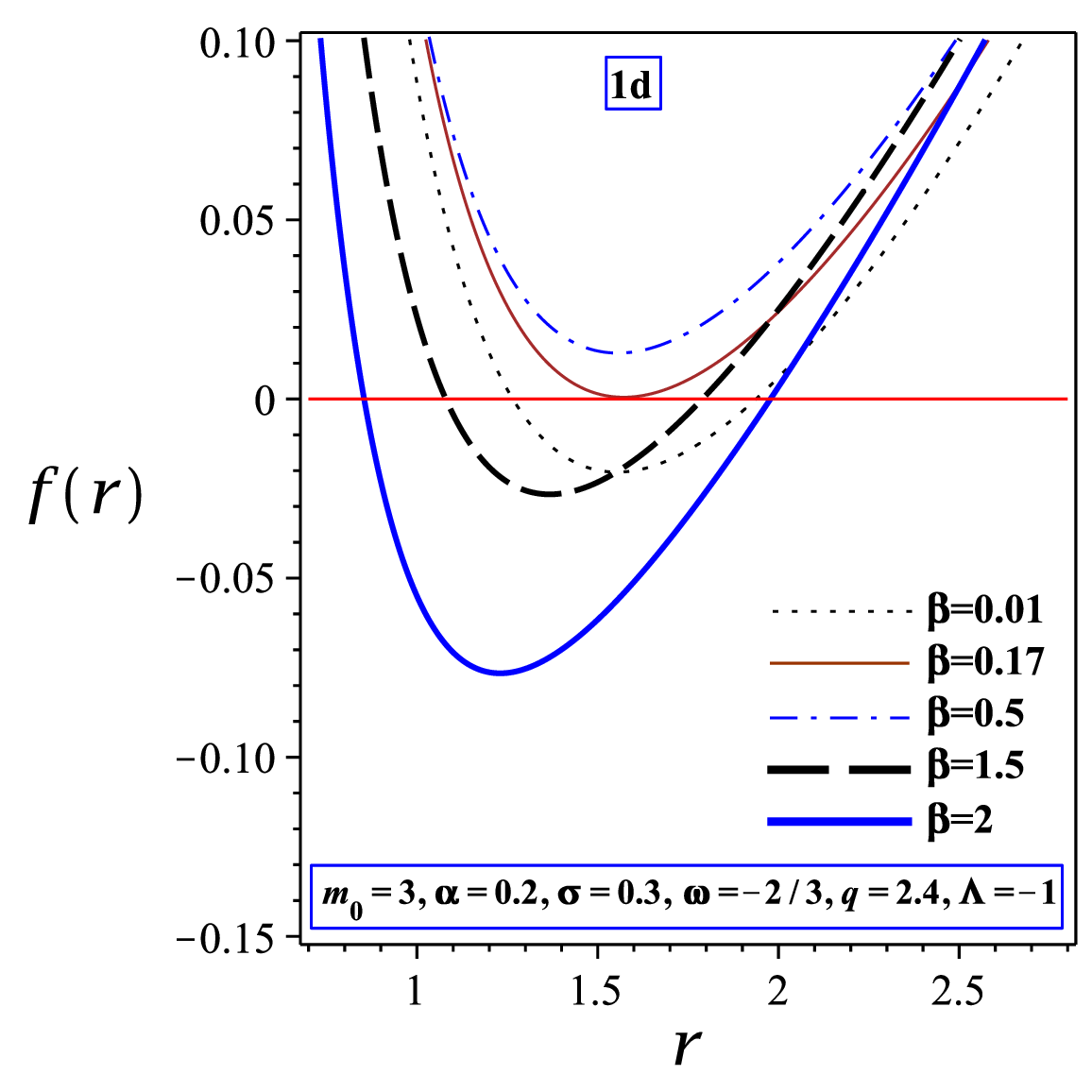}}{\MissingFigure{1d.eps}{0.28\textwidth}{0.12\textheight}}
 \IfFileExists{1e.eps}{\includegraphics[width=0.30\textwidth]{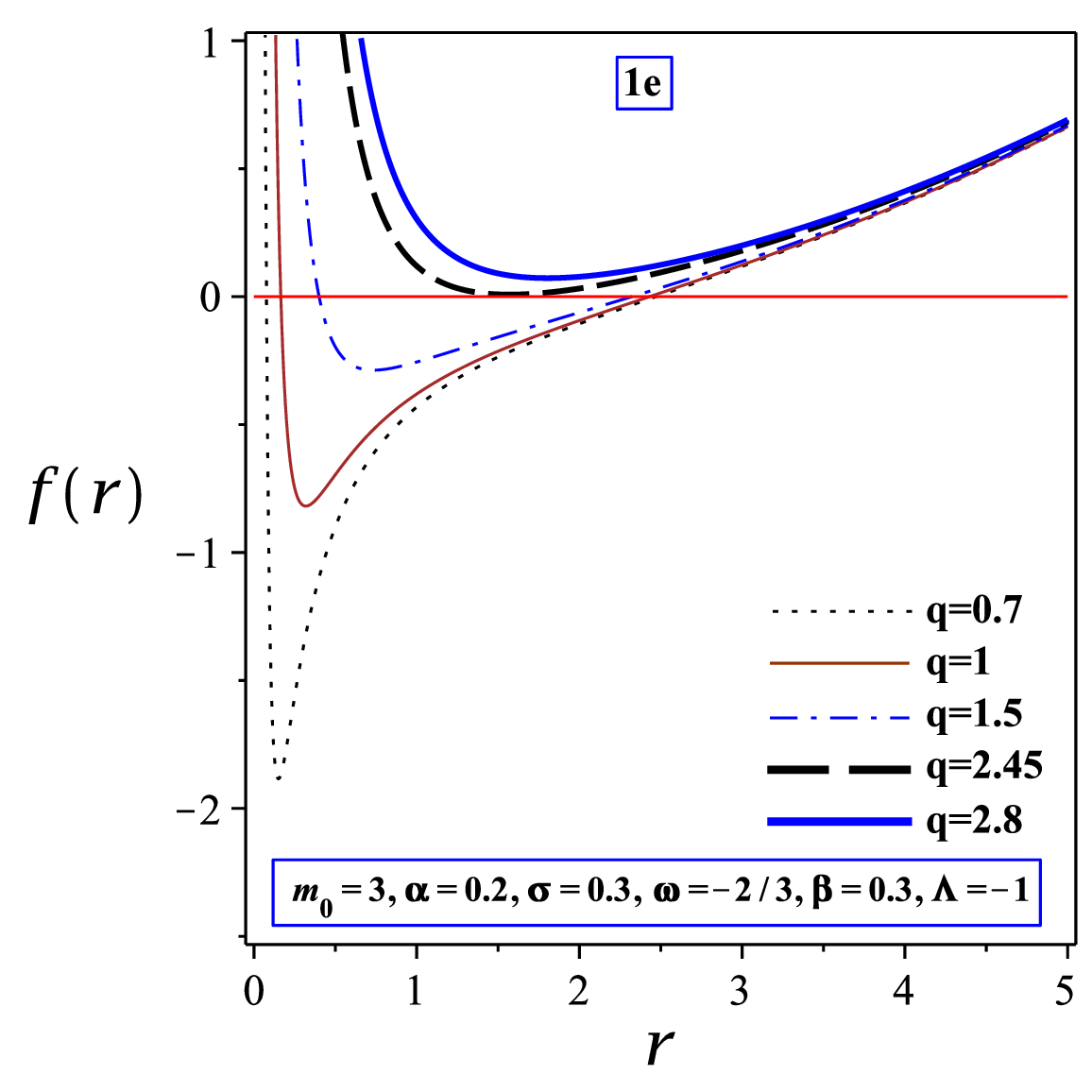}}{\MissingFigure{1e.eps}{0.28\textwidth}{0.12\textheight}}
 \IfFileExists{1f.eps}{\includegraphics[width=0.30\textwidth]{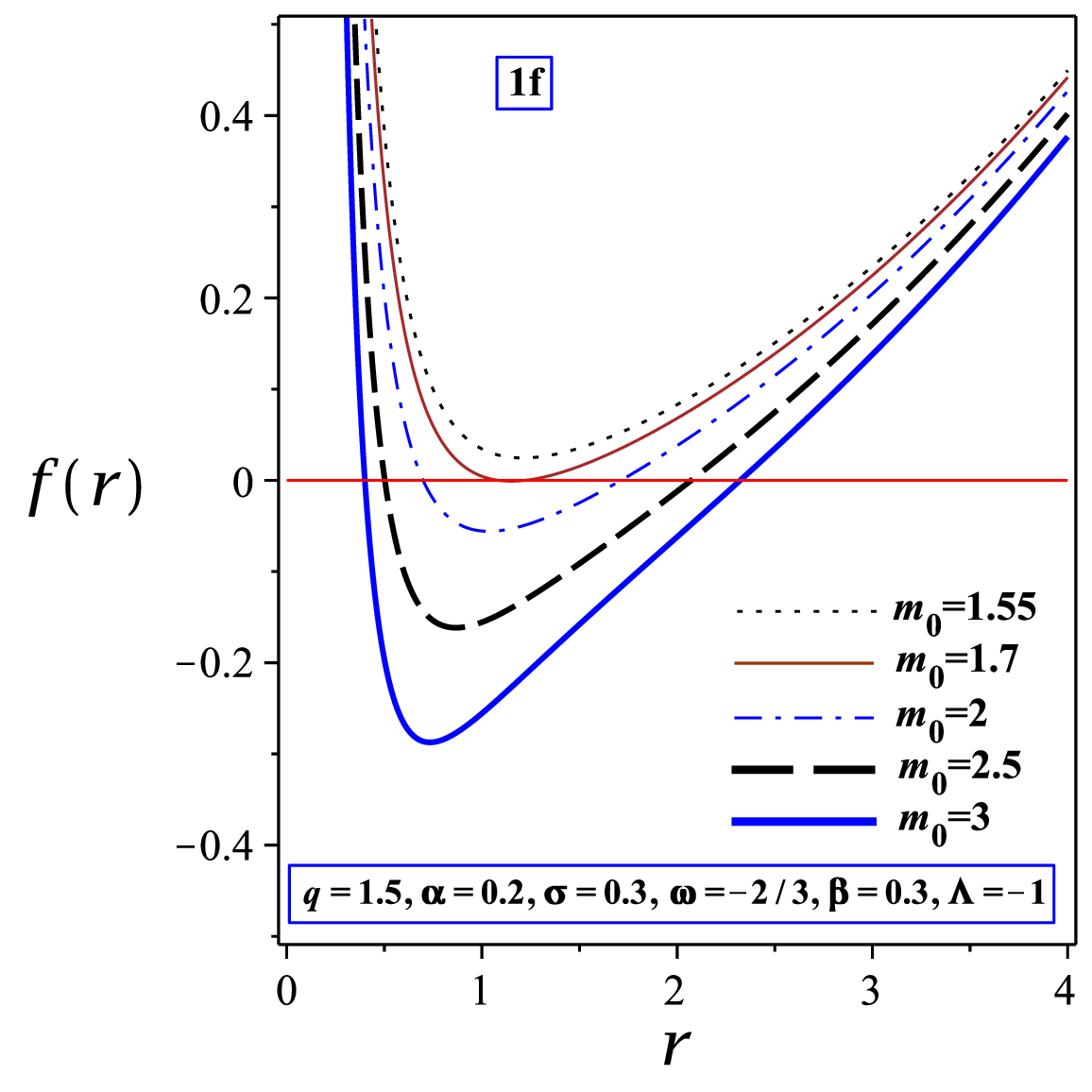}}{\MissingFigure{1f.eps}{0.28\textwidth}{0.12\textheight}}
 \caption{Metric function $f(r)$ from Eq.~(\ref{f(r)}) versus $r$ for variations of (a) $\sigma$, (b) $\omega$, (c) $\alpha$, (d) $\beta$, (e) $q$, and (f) $m_0$. The nonvaried parameters are held fixed within each scan.}
\label{fig1}
\end{figure*}

Figure~\ref{fig1} shows representative scans of $f(r)$ as the model parameters are varied. A positive root of $f(r)$ marks a horizon, while the number and locations of such roots depend on the remaining fixed parameters. The curves therefore describe the parameter sets used in the numerical scans rather than universal trends across the full parameter space.

\section{Thermodynamics}
\label{sec:thermo}

We now translate the geometry into horizon thermodynamics. The state is parametrized by the outer horizon radius $r_+$, defined as the relevant largest positive root of $f(r_+)=0$ in the static region. Unless explicitly varied in the extended phase space, $(\alpha,\beta,\sigma,\omega,\Lambda)$ are held fixed. Every branch statement below is conditional on the existence of that horizon and on the chosen parameter set; zeros and extrema seen in a scan are not universal properties of the entire solution family.

For a static metric normalized so that $g_{tt}=-f(r)$ and $g_{rr}=f(r)^{-1}$,
the horizon-generating Killing field is $\chi=\partial_t$, with the time normalization fixed by the metric ansatz. Assuming a
non-extremal horizon at $r=r_+$ and the stated normalization of $t$, the
surface gravity is $\kappa=|f'(r_+)|/2$. On the outer, positive-temperature
branch, this gives
\begin{equation}
T=\frac{\kappa}{2\pi}
=\frac{|f'(r_+)|}{4\pi}
=\frac{f'(r_+)}{4\pi},\label{Temp1}
\end{equation}
Here the last equality applies to the outer branch with $f'(r_+)>0$.
Using $f(r_+)=0$ to eliminate the integration parameter $m_0$, the Hawking
temperature can be written as
\begin{equation}
T=\frac{1-\alpha }{4\pi r_{+}}+\frac{\beta }{4\pi r_{+}^{2}}-\frac{q^{2}}{%
4\pi r_{+}^{3}}-\frac{\Lambda r_{+}}{4\pi }+\frac{3\omega \sigma }{4\pi
r_{+}^{3\omega +2}},  \label{Temp2}
\end{equation}
The temperature depends on the parameters appearing in $f(r)$ through the
horizon radius and the explicit terms in Eq.~(\ref{Temp2}). The formula is
valid on a non-extremal horizon; extremality is characterized by $T=0$.

For $\omega=-2/3$ and fixed nonzero charge $q$, the leading formal
small-$r_+$ behavior is
\begin{equation}
T(r_+)\sim -\frac{q^{2}}{4\pi r_{+}^{3}},
\end{equation}%
Thus the formal small-radius limit is negative when $q\ne0$, because the
charge term dominates the other terms in Eq.~(\ref{Temp2}). This asymptotic
statement does not establish that arbitrarily small radii describe admissible
outer horizons: the horizon equation and the existence of the relevant branch
must also be checked. If $q=0$, the leading-order conclusion changes.

\begin{figure*}[tbp]
\centering
\IfFileExists{2a.eps}{\includegraphics[width=0.30\textwidth]{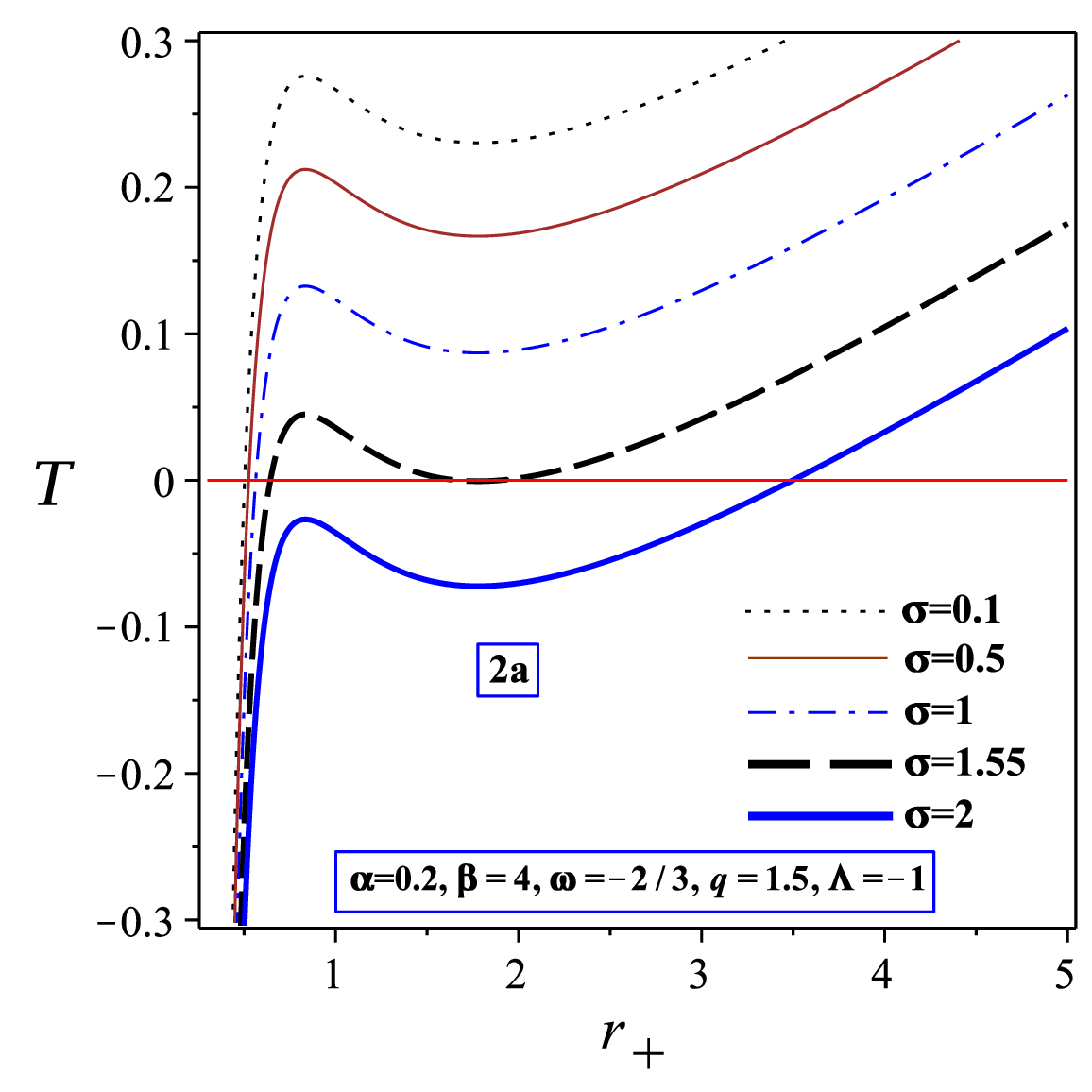}}{\MissingFigure{2a.eps}{0.28\textwidth}{0.12\textheight}} \IfFileExists{2b.eps}{\includegraphics[width=0.30\textwidth]{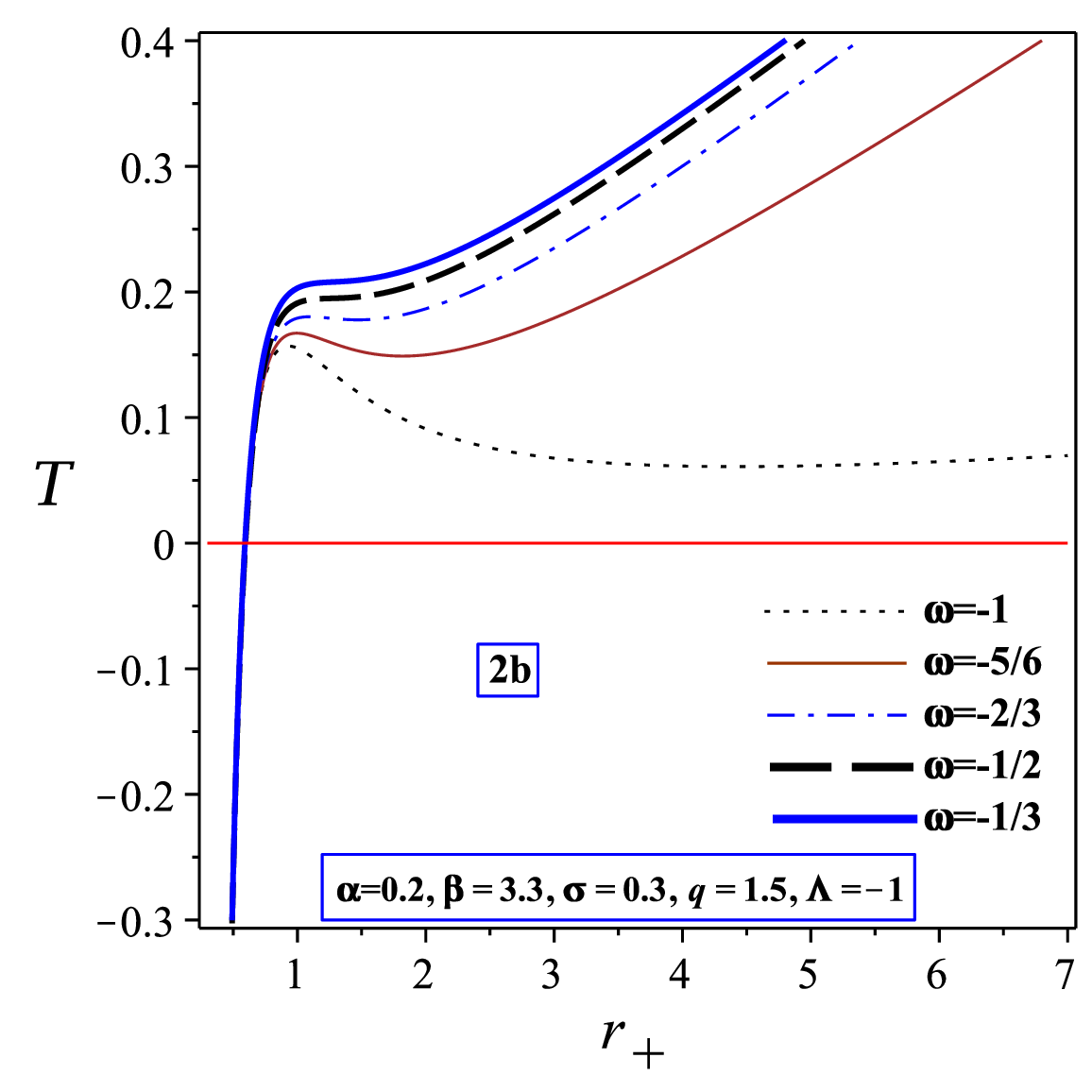}}{\MissingFigure{2b.eps}{0.28\textwidth}{0.12\textheight}} %
\IfFileExists{2c.eps}{\includegraphics[width=0.30\textwidth]{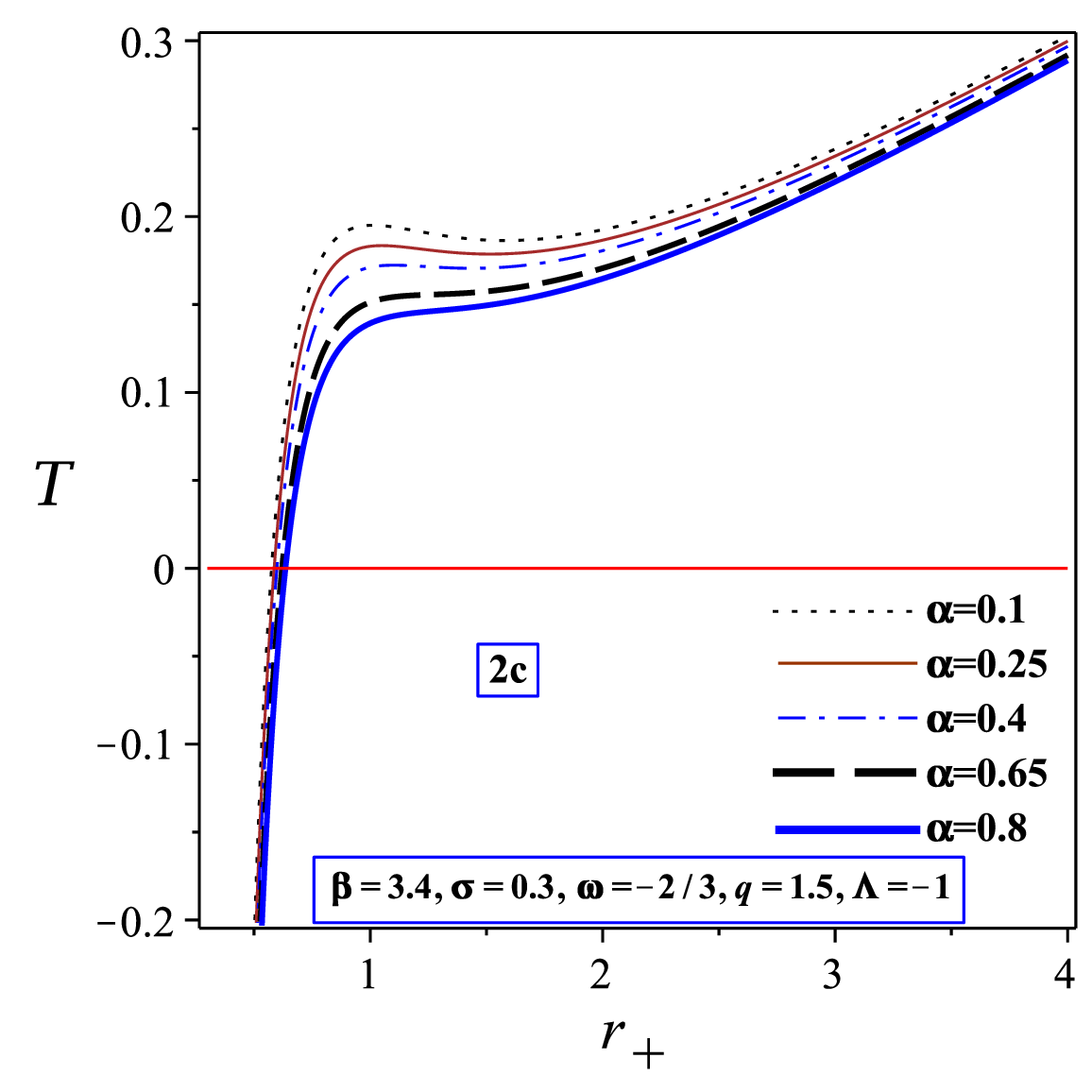}}{\MissingFigure{2c.eps}{0.28\textwidth}{0.12\textheight}} \par
\IfFileExists{2d.eps}{\includegraphics[width=0.30\textwidth]{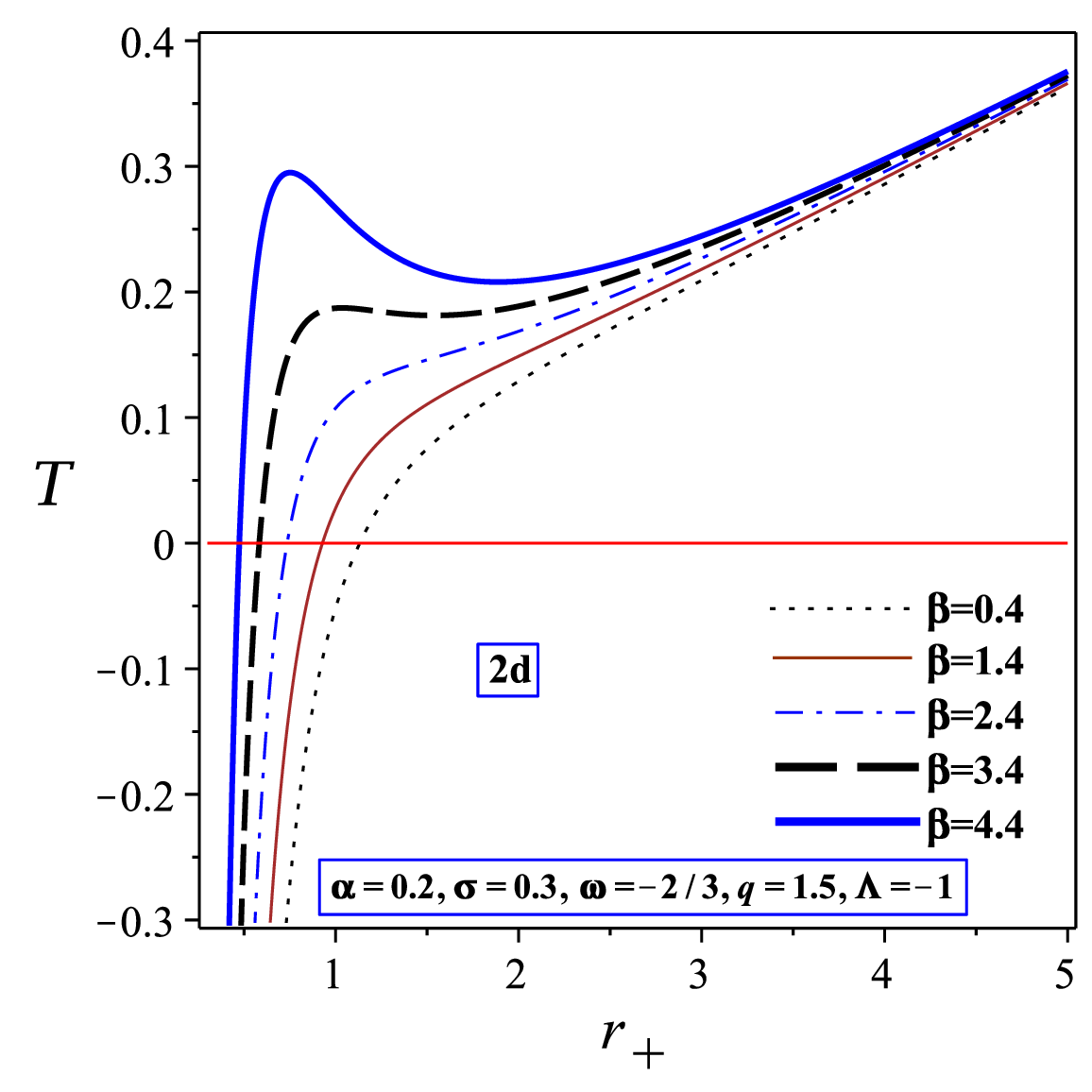}}{\MissingFigure{2d.eps}{0.28\textwidth}{0.12\textheight}} \IfFileExists{2e.eps}{\includegraphics[width=0.30\textwidth]{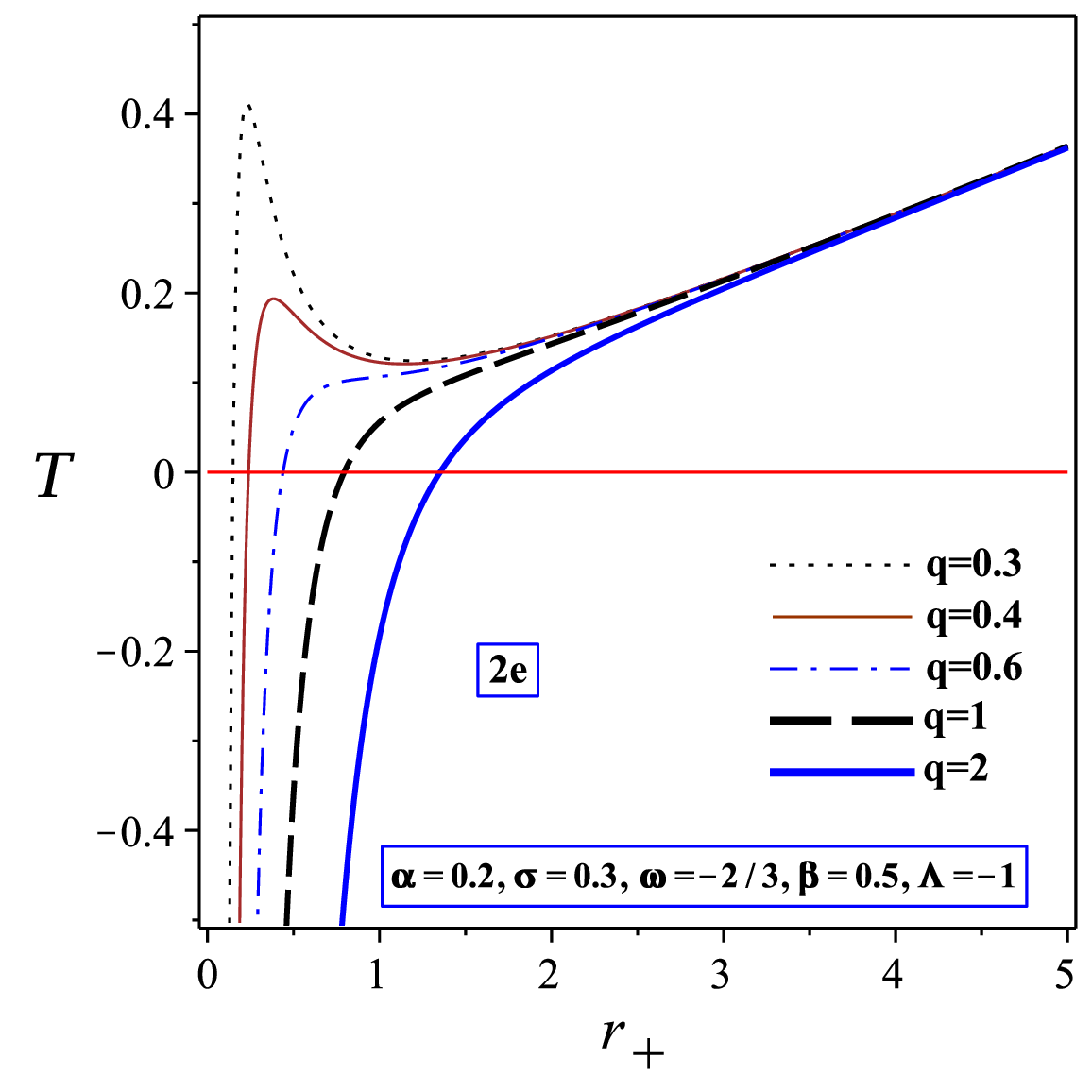}}{\MissingFigure{2e.eps}{0.28\textwidth}{0.12\textheight}}
\caption{Hawking temperature $T$ from Eq.~(\ref{Temp2}) versus $r_{+}$ for variations of (a) $\sigma$, (b) $\omega$, (c) $\alpha$, (d) $\beta$, and (e) $q$.}
\label{fig2}
\end{figure*}

For $\omega=-2/3$ and $\Lambda<0$, the leading large-radius term is
\begin{equation}
T(r_+)\sim -\frac{\Lambda r_+}{4\pi},\qquad r_+\to\infty,
\end{equation}
and is positive. Figure~\ref{fig2} shows the finite-radius behavior for
the parameter choices used in the numerical scan. The root and extremum
counts quoted below describe those plotted choices, rather than universal
statements independent of the other parameters.

For the displayed values, changing the quintessence normalization $\sigma$ changes the number and location of temperature zeros. The positive-temperature branches and any critical
value $\sigma_{\rm crit}$ should be understood as properties of this
particular scan. Increasing $\sigma$ shifts the roots in the direction shown
in the figure and changes the interval on which $T>0$.

The plotted curves show a change in the number of temperature extrema as the state parameter $\omega$ varies.
A critical value at which two extrema merge can be identified numerically by
solving $T'(r_+)=T''(r_+)=0$ for the chosen fixed parameters. Temperature
extrema correspond to zeros of $T'(r_+)$ and hence, away from $T=0$, to poles
of the heat capacity in Eq.~(\ref{C}); they are not by themselves proof of a
global phase transition.

The plotted scan shows that varying the string-cloud parameter $\alpha$ changes the temperature extrema
and shifts the zero of $T$. Any quoted $\alpha_{\rm crit}$ is conditional on
the remaining fixed parameters and the radius interval displayed.

The sign and magnitude of the PFDM parameter $\beta$ affect both the logarithmic mass term and
the temperature term proportional to $\beta/r_+^2$. For the values shown in
the figure, increasing $\beta$ shifts the temperature zero as plotted; the
direction and number of extrema should not be generalized beyond the scanned
range.

The electric charge enters the temperature through $-q^2/(4\pi r_+^3)$, so the
temperature depends on $q^2$ rather than on the sign of $q$. The root and
extremum structure changes with $|q|$ for the parameter choices shown.

In summary, $T>0$ is a necessary condition for a non-extremal outer horizon
to lie on a positive-temperature branch, but it is not by itself a complete
criterion for physical admissibility. The horizon condition and the global
structure of the metric must also be respected. The parameter trends in
Fig.~\ref{fig2} are specific to the values held fixed in each scan.

The entropy of these black holes can be obtained by using the area law, in
the following form
\begin{equation}
S=\frac{A}{4}=\pi r_{+}^{2},  \label{entropy}
\end{equation}%
where $A=4\pi r_+^2$ is the horizon area. We use $G=\hbar=c=k_B=1$.

With the standard Maxwell normalization used for this solution, the electric
charge obtained from the flux integral is
\begin{equation}
Q=\frac{1}{4\pi}\int_{S^2_\infty}{}^\star F=q,
\label{Q}
\end{equation}%
Here the angular integration uses the induced area element on a sphere of
constant $t$ and $r$; the resulting charge is $Q=q$ in the normalization of
the metric and gauge potential adopted here.

Choosing the gauge potential $A_t=-q/r$ and taking the reference potential
at infinity to vanish, the horizon-to-infinity electric potential is
\begin{equation}
\Phi =A_{\mu }\chi ^{\mu }\left\vert _{r\rightarrow \infty }\right. -A_{\mu
}\chi ^{\mu }\left\vert _{r\rightarrow r_{+}}\right. =\frac{q}{r_{+}}.
\label{U}
\end{equation}
where $A_{\mu }=(\frac{-q}{r},0,0,0)$.

Solving the horizon condition $f(r_+)=0$ gives the mass parameter in terms of
the horizon radius:%
\begin{widetext}
\begin{equation}
\begin{aligned}
M=m_0={}&\frac{(1-\alpha)r_+}{2}
+\frac{\beta}{2}\ln\!\left(\frac{r_+}{|\beta|}\right)
+\frac{q^2}{2r_+}-\frac{\Lambda r_+^3}{6}
-\frac{\sigma}{2r_+^{3\omega}}.
\end{aligned}
\label{MM}
\end{equation}
\end{widetext}%
At $\beta=0$, the logarithmic contribution is understood by continuity and vanishes.
We denote this horizon-derived quantity by $M=m_0$ in the fixed-$\Lambda$
thermodynamic discussion; in extended phase space we write the same
horizon-derived function as $\mathcal M$ to emphasize its enthalpy role. It should not be identified automatically with an
Ashtekar--Magnon--Das conserved mass: the PFDM logarithm and, for
$-1<\omega<-1/3$, the nonstandard asymptotic behavior induced by the
quintessence term require a separate asymptotic-charge analysis and
normalization prescription.

For $\omega=-2/3$ and fixed nonzero $q$, the leading formal small-radius
behavior of the mass parameter is charge-dominated:
\begin{equation}
M(r_+)\sim \frac{q^{2}}{2r_{+}},
\end{equation}%
For fixed $q\ne0$, the charge term dominates this formal small-radius limit
and makes $M$ positive sufficiently close to $r_+=0$. This does not by itself
establish the existence or physical admissibility of a black-hole solution at
every such radius.

For $\omega=-2/3$, the large-radius mass-parameter behavior is
\begin{equation}
M(r_+)\sim -\frac{\Lambda r_{+}^{3}}{%
6},
\end{equation}%
For $\Lambda<0$, the cubic term dominates the large-radius behavior when
$\omega=-2/3$, and the mass parameter is therefore asymptotically positive.
This conclusion concerns the stated limit and does not rule out a finite
intermediate range with negative $M=m_0$.

\begin{figure*}[tbp]
\centering
\IfFileExists{3a.eps}{\includegraphics[width=0.30\textwidth]{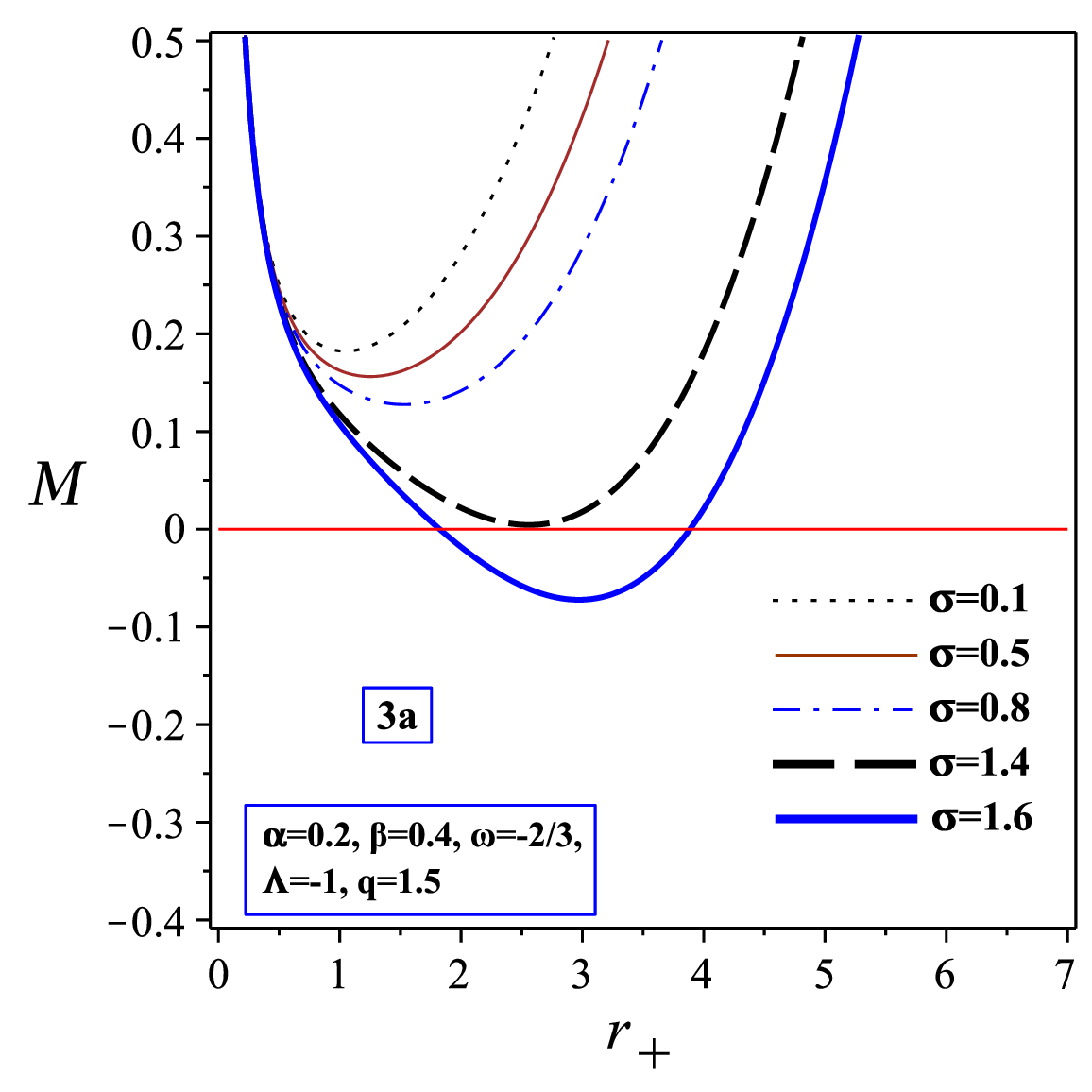}}{\MissingFigure{3a.eps}{0.28\textwidth}{0.12\textheight}} \IfFileExists{3b.eps}{\includegraphics[width=0.30\textwidth]{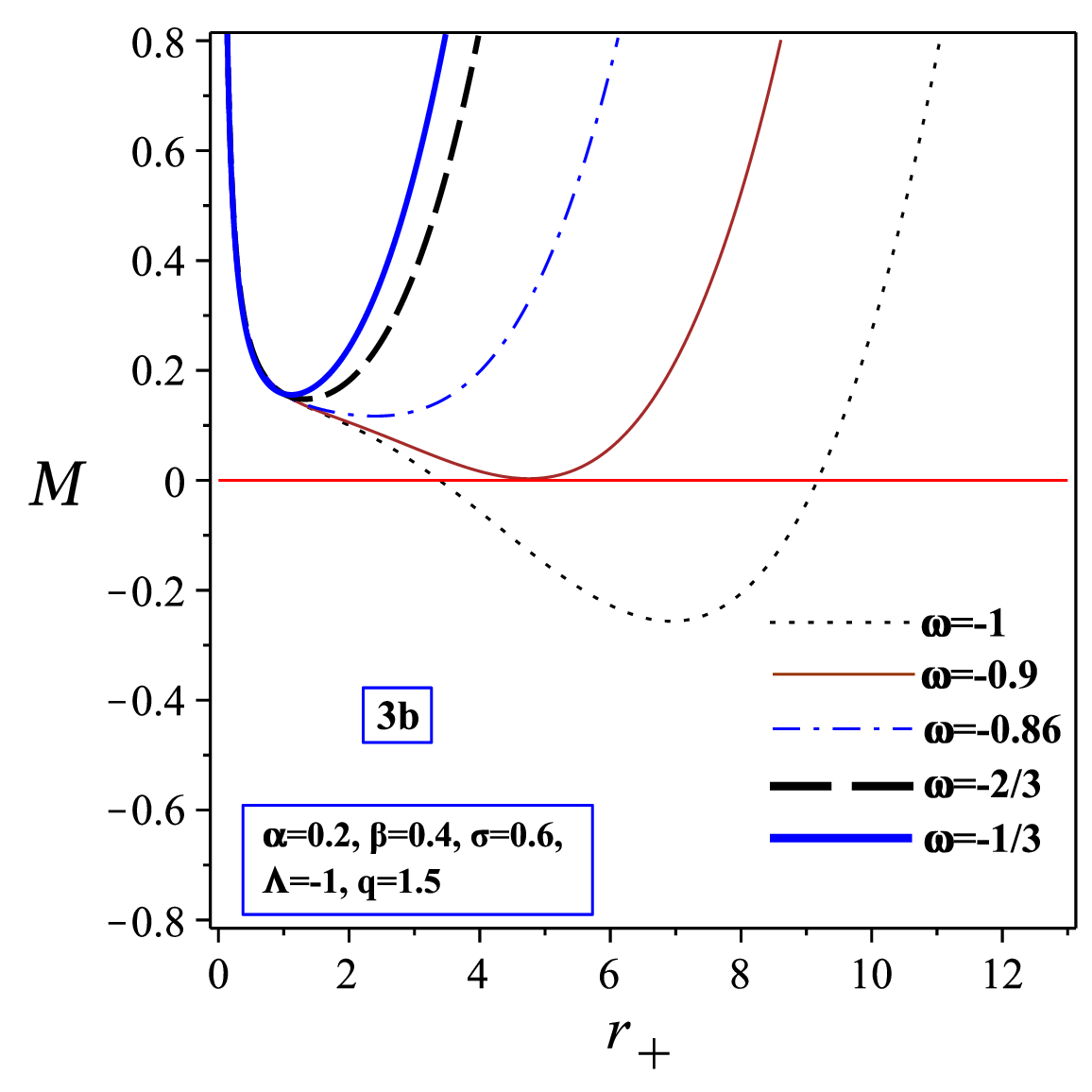}}{\MissingFigure{3b.eps}{0.28\textwidth}{0.12\textheight}} %
\IfFileExists{3c.eps}{\includegraphics[width=0.30\textwidth]{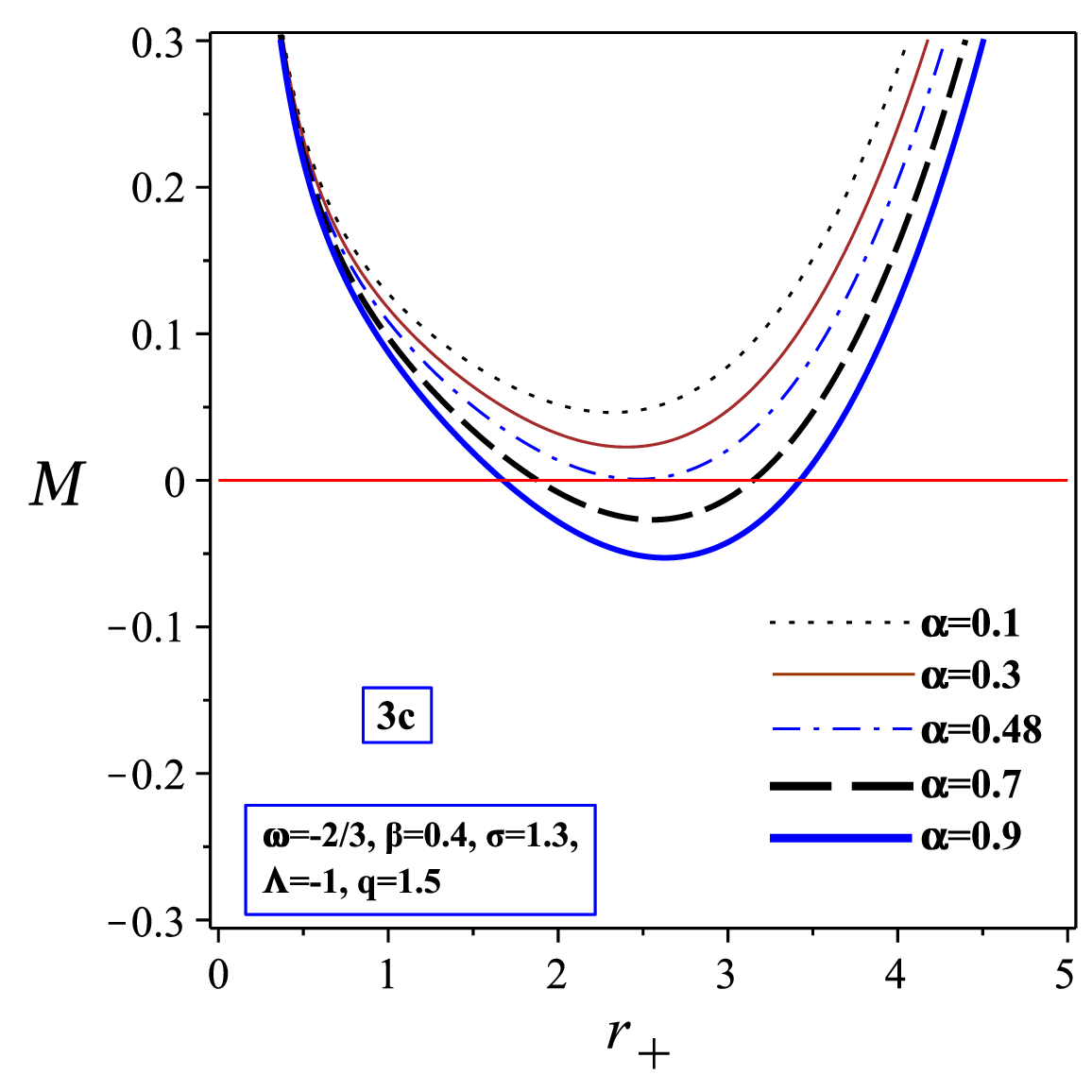}}{\MissingFigure{3c.eps}{0.28\textwidth}{0.12\textheight}} \par
\IfFileExists{3d.eps}{\includegraphics[width=0.30\textwidth]{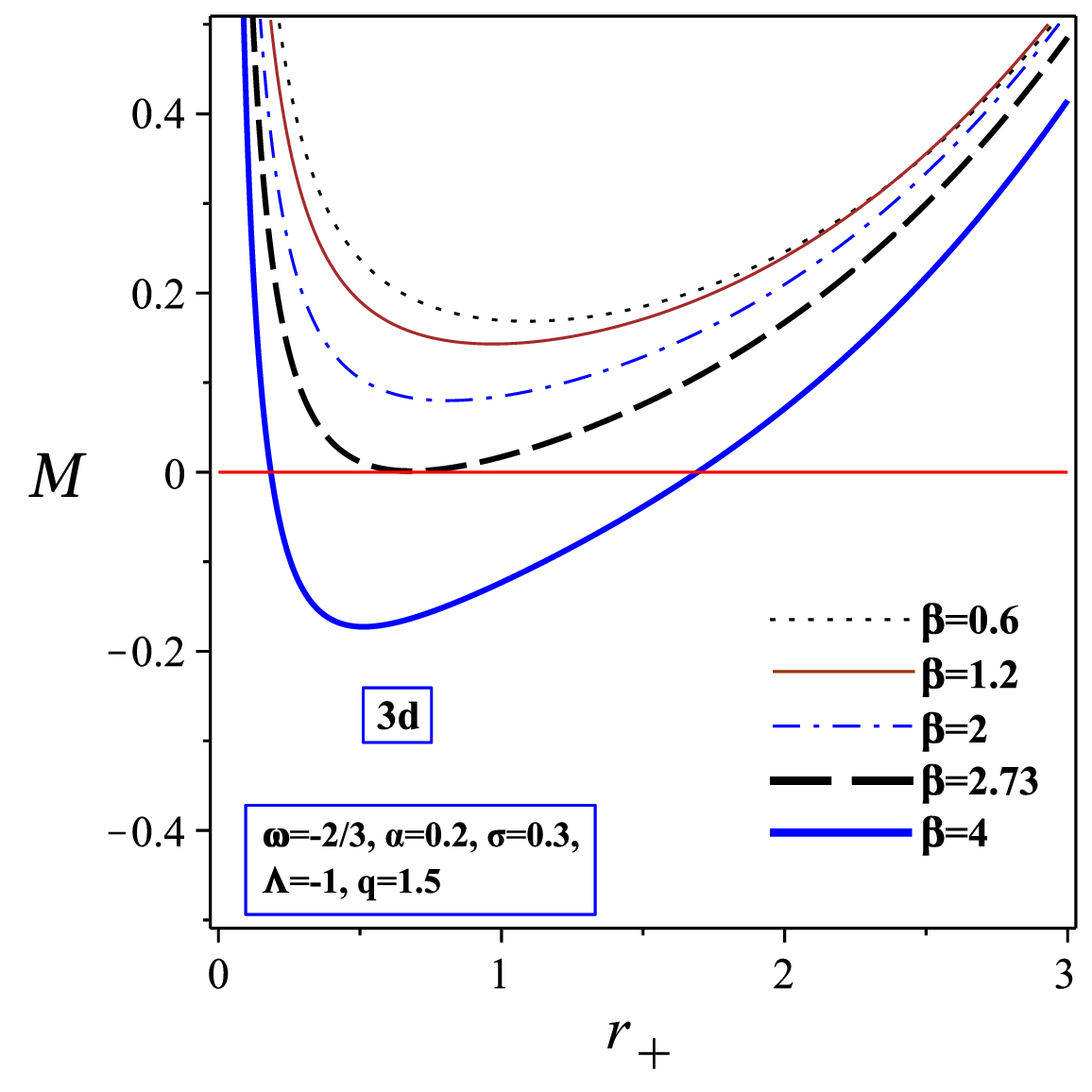}}{\MissingFigure{3d.eps}{0.28\textwidth}{0.12\textheight}} \IfFileExists{3e.eps}{\includegraphics[width=0.30\textwidth]{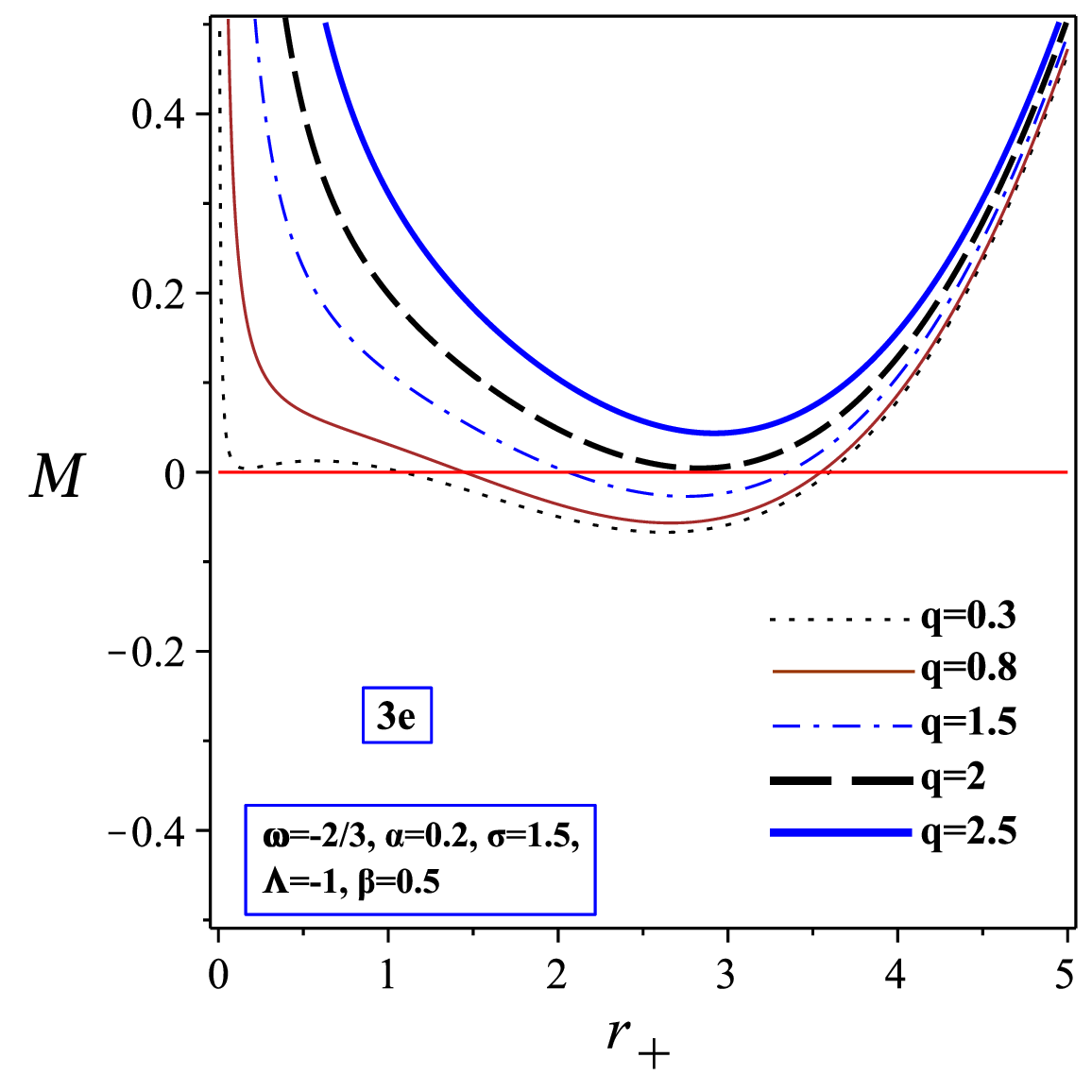}}{\MissingFigure{3e.eps}{0.28\textwidth}{0.12\textheight}}
\caption{Horizon-derived mass parameter $M=m_0$ versus $r_{+}$ for variations of (a) $\sigma$, (b) $\omega$, (c) $\alpha$, (d) $\beta$, and (e) $q$.}
\label{fig3}
\end{figure*}

Figure~\ref{fig3} plots the horizon-derived mass parameter $M=m_0$ against
$r_+$ for the parameter choices shown. The sign of this quantity is a useful
diagnostic within the adopted parametrization, but a negative value does not
by itself prove that a geometry is unphysical; that conclusion requires an
analysis of the horizon structure, curvature, and admissible asymptotics.
Likewise, critical values inferred from these plots are conditional on the
other parameters held fixed.

The displayed curves show that increasing $\sigma$ can lower $M$ over an
intermediate range of horizon radii and can produce a region with $M<0$ for
the selected parameters.

Changing the state parameter $\omega$ changes the radius dependence of the quintessence contribution, $-\sigma/(2r_+^{3\omega})$. The sign and extent of any
intermediate negative-mass-parameter interval are therefore dependent on the
chosen scan.

The string-cloud contribution $(1-\alpha)r_+/2$ makes the mass parameter decrease with
increasing $\alpha$ at fixed $r_+$; the resulting sign changes depend on the
remaining terms.

The PFDM logarithmic term is $\beta\ln(r_+/|\beta|)/2$. Its effect is not
globally monotonic in $r_+$ or in the sign of $\beta$, so any threshold
identified in the figure applies only to the displayed parameter range.
At $\beta=0$, this term is understood by its continuous limit and vanishes.

The electric charge contributes $q^2/(2r_+)$, which is nonnegative and can raise the
mass parameter, especially at smaller radii. The charge dependence is even
in $q$; only $|q|$ affects this mass function.

The figure therefore shows how the model parameters alter the sign and
shape of the horizon-derived function $M(r_+)$. Its sign is not, by itself,
a test of geometric regularity or global thermodynamic preference. No
parameter-independent inequality involving $\sigma$, $\omega$, $\alpha$,
$\beta$, and $q$ guarantees positivity for every radius; the remaining
parameters, horizon branch, and admissible domain must be specified. This
distinction is important when the same function is reinterpreted as enthalpy
in extended phase space.

Replacing Eqs.~(\ref{entropy}) and (\ref{Q}) in Eq.~(\ref{MM}) gives
$M=M(S,Q)$:

\begin{widetext}
\begin{equation}
\begin{split}
M(S,Q)=\frac{1-\alpha}{2}\sqrt{\frac{S}{\pi}}
+\frac{\beta}{2}\ln\!\left(\frac{\sqrt{S}}{|\beta|\sqrt{\pi}}\right)
+\frac{Q^2}{2\sqrt{S/\pi}}-\frac{\Lambda S^{3/2}}{6\pi^{3/2}}
-\frac{\sigma}{2(S/\pi)^{3\omega/2}}.
\end{split}
\label{MS}
\end{equation}
\end{widetext}
Differentiating the mass function with respect to $S$ and $Q$ at fixed
$(\alpha,\beta,\sigma,\omega,\Lambda)$ reproduces $T$ and $\Phi$,
respectively. Because $M$ is obtained from the horizon equation, this is an
internal differential-consistency identity, not an independent measurement
of a conserved asymptotic mass. Accordingly, the first law within this
fixed-parameter thermodynamic model is
\begin{equation}
dM=TdS+\Phi dQ.
\end{equation}
This is the fixed-$(\alpha,\beta,\sigma,\omega,\Lambda)$ first law. If any of these model parameters are varied, their conjugate work terms must be included, as in the extended first law below.

\section{Thermal Stability}

\subsection{Heat capacity}

In the canonical ensemble, with $Q$ and the external model parameters fixed, $C_Q>0$ is the local thermal-stability criterion against small fluctuations. It is a local criterion: it does not establish global preference over another black-hole branch or a reference background.

The heat capacity is defined as
\begin{equation}
C_{q}=\frac{T}{\left( \frac{\partial T}{\partial S}\right) _{q}}=T\left(
\frac{\frac{\partial S}{\partial r_{+}}}{\frac{\partial T}{\partial r_{+}}}%
\right) _{q},  \label{defC}
\end{equation}%
by applying Eqs. (\ref{Temp2}), and (\ref{entropy}) within (\ref{defC}), we
can get the heat capacity as
\begin{equation}
C_{q}=\frac{2\pi r_{+}^{2}\left( \Lambda r_{+}^{2}+\alpha -1-\frac{\beta }{%
r_{+}}+\frac{q^{2}}{r_{+}^{2}}-\frac{3\omega \sigma }{r_{+}^{3\omega +1}}%
\right) }{\Lambda r_{+}^{2}+1-\alpha +\frac{2\beta }{r_{+}}-\frac{3q^{2}}{%
r_{+}^{2}}+\frac{3\left( 3\omega +2\right) \omega \sigma }{r_{+}^{3\omega +1}%
}}.  \label{C}
\end{equation}

\begin{figure*}[tbp]
\centering
\IfFileExists{4a.eps}{\includegraphics[width=0.30\textwidth]{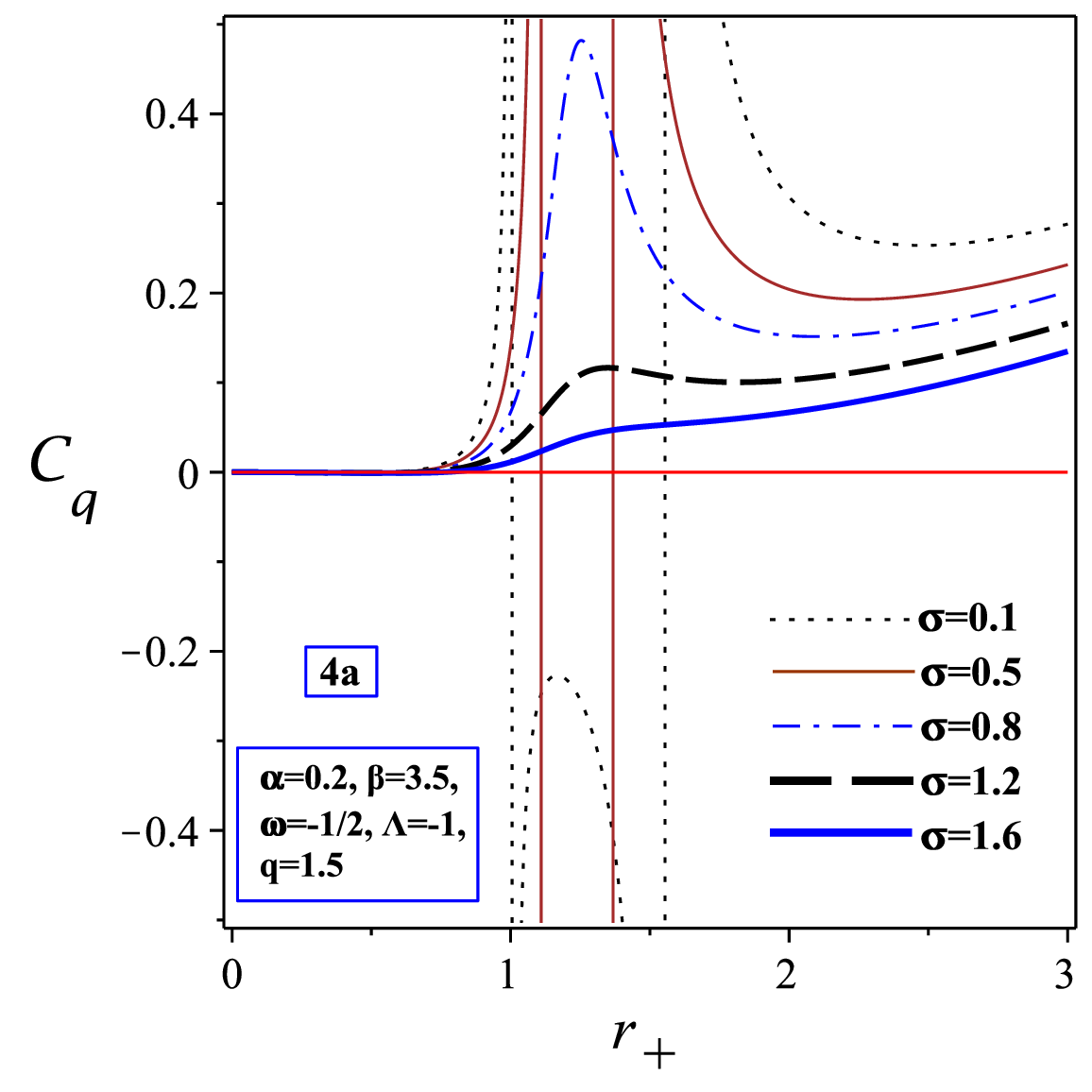}}{\MissingFigure{4a.eps}{0.28\textwidth}{0.12\textheight}} \IfFileExists{4b.eps}{\includegraphics[width=0.30\textwidth]{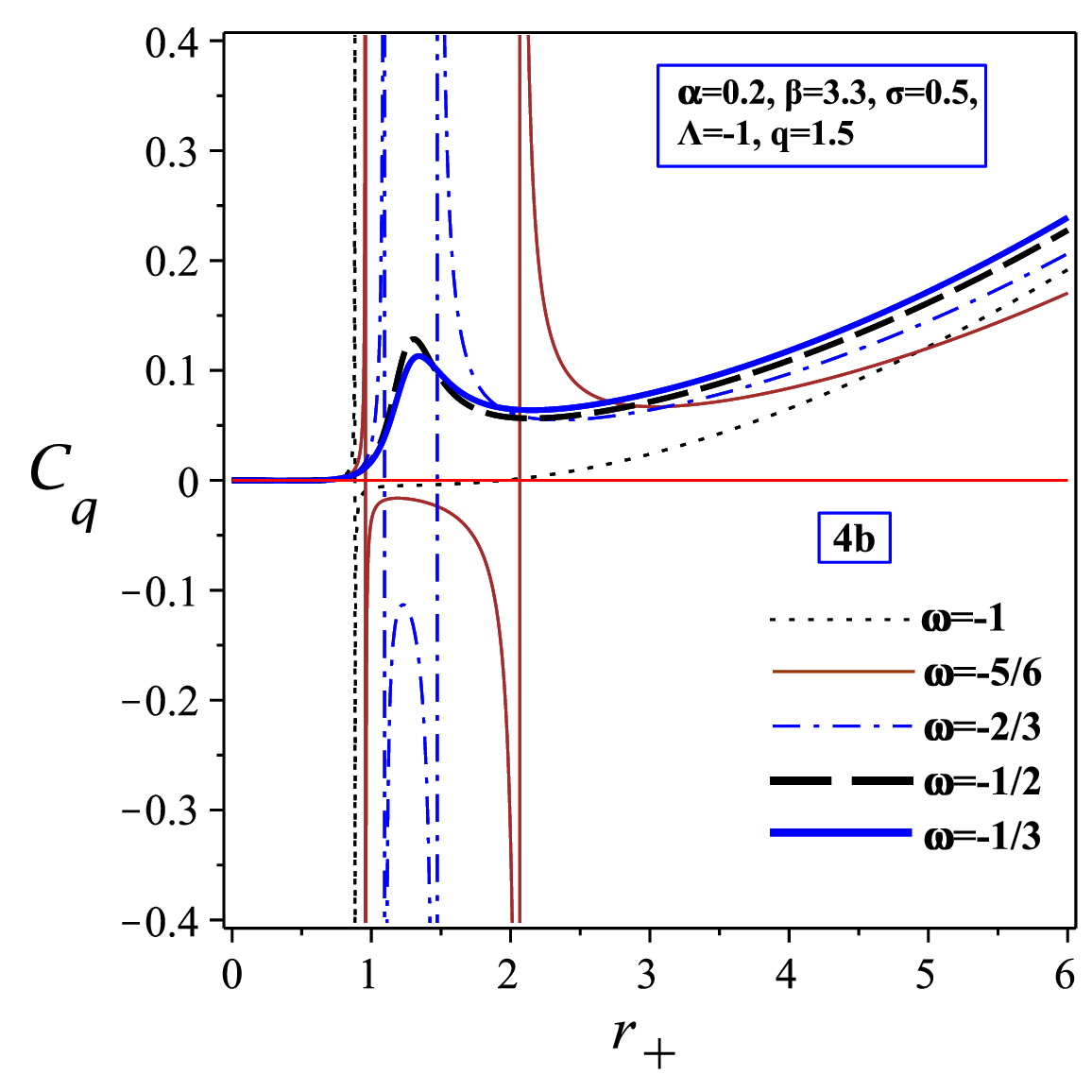}}{\MissingFigure{4b.eps}{0.28\textwidth}{0.12\textheight}} %
\IfFileExists{4c.eps}{\includegraphics[width=0.30\textwidth]{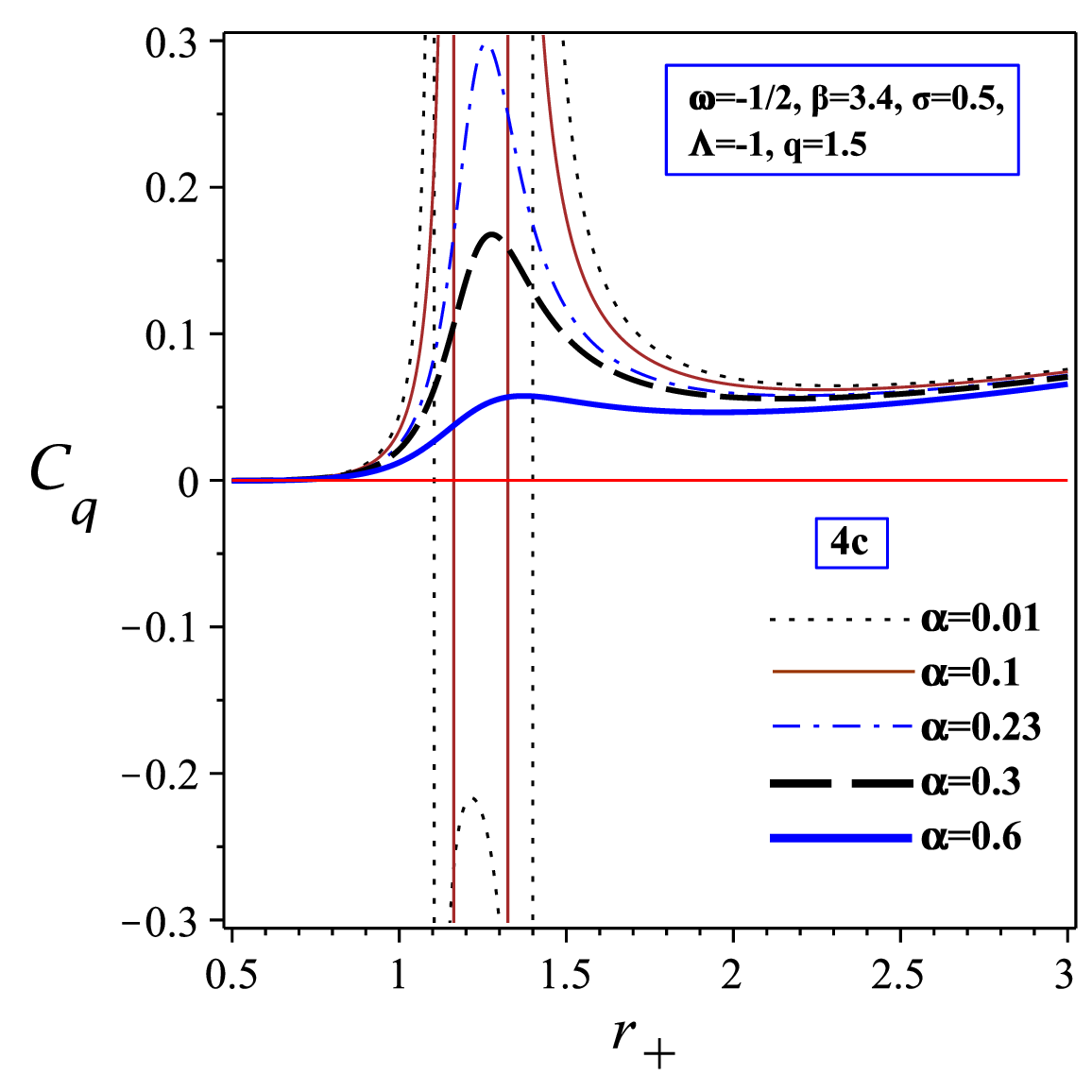}}{\MissingFigure{4c.eps}{0.28\textwidth}{0.12\textheight}} \par
\IfFileExists{4d.eps}{\includegraphics[width=0.30\textwidth]{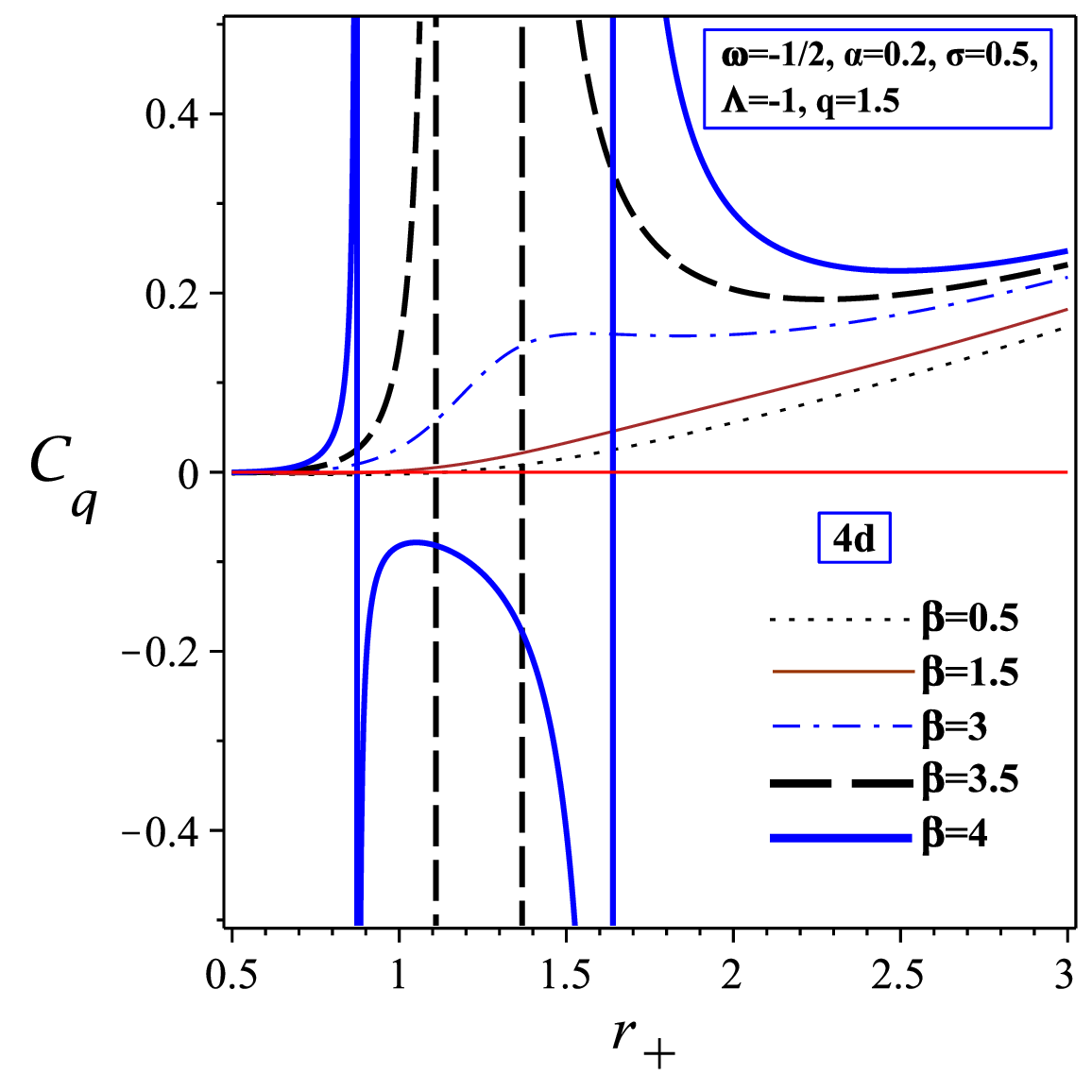}}{\MissingFigure{4d.eps}{0.28\textwidth}{0.12\textheight}} \IfFileExists{4e.eps}{\includegraphics[width=0.30\textwidth]{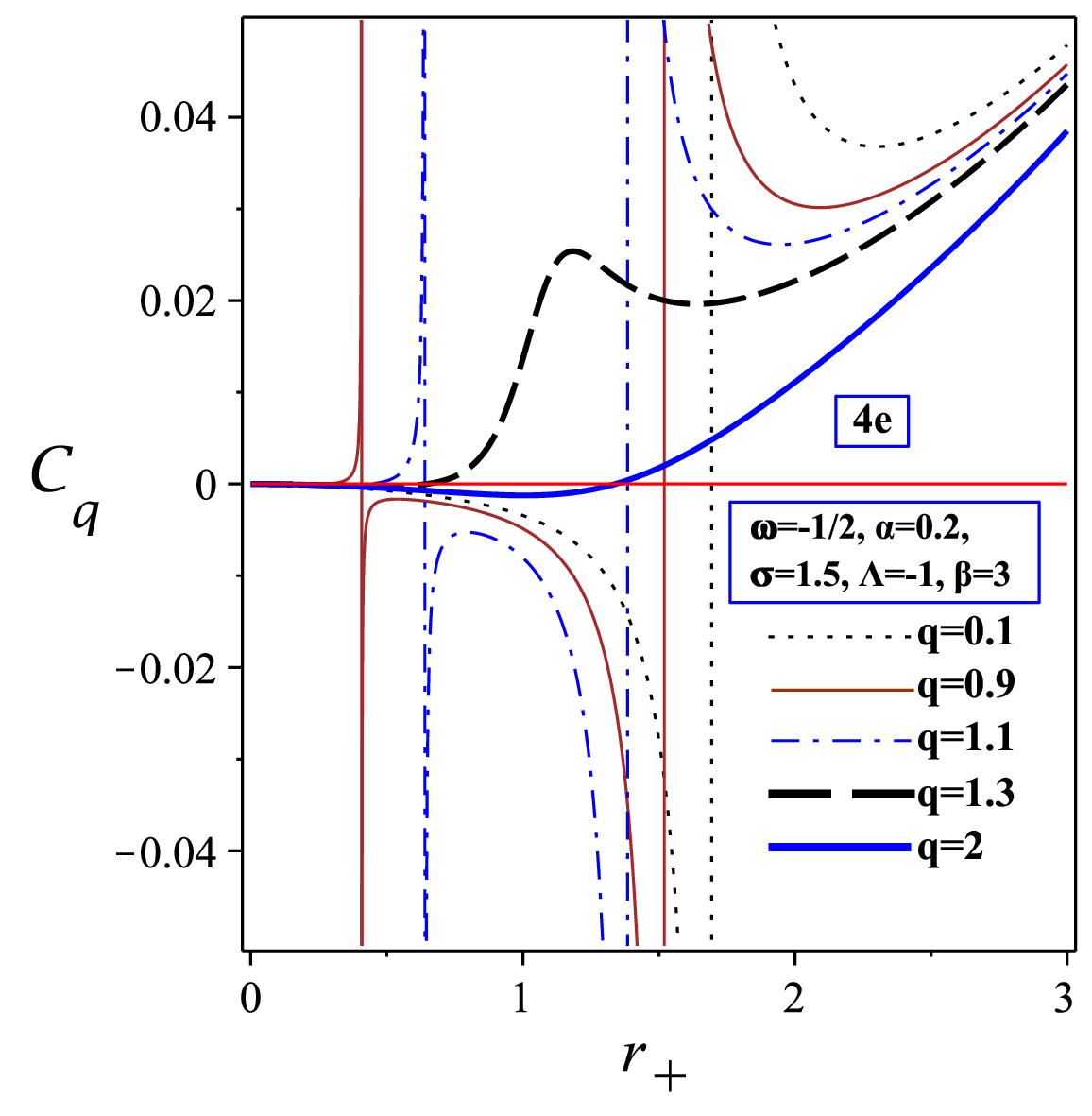}}{\MissingFigure{4e.eps}{0.28\textwidth}{0.12\textheight}}
\caption{Heat capacity $C_Q$ from Eq.~(\ref{C}) versus $r_{+}$ for variations of (a) $\sigma$, (b) $\omega$, (c) $\alpha$, (d) $\beta$, and (e) $q$.}
\label{fig4}
\end{figure*}

The heat-capacity curves in Fig.~\ref{fig4} examine local thermal
stability for the parameter values chosen in each scan. For fixed charge and
fixed external parameters,
\begin{equation}
C_Q=\frac{T}{(\partial T/\partial S)_Q}
=\frac{2\pi r_+T(r_+)}{T'(r_+)},
\end{equation}
so a pole can occur when $T'(r_+)=0$ while $T(r_+)\ne0$. A zero of $C_Q$
usually coincides with $T=0$ (an extremal boundary), provided $T'$ is finite.
The sign of $C_Q$ must be evaluated on each branch; a pole alone does not
establish a global phase transition.

For the displayed $\sigma$ scan, the curves show two poles over part of the parameter range, with a negative-heat-capacity interval between them. As $\sigma$
changes, the pole locations move and the interval can narrow or disappear.
These observations are specific to the fixed parameters used for the figure.

The plotted curves show changes in the number of poles and zeros as the state parameter $\omega$ varies. The intervals of positive heat capacity are those on which the
explicit expression in Eq.~(\ref{C}) is positive; they should be identified
from the plotted roots and poles for each chosen parameter set.

Varying the string-cloud parameter $\alpha$ changes the denominator of Eq.~(\ref{C}) and hence the
locations of any poles. The figure shows the corresponding changes in the
positive- and negative-heat-capacity branches. A threshold
$\alpha_{\rm crit}$, if quoted, is conditional on the other fixed parameters.

The PFDM parameter $\beta$ affects both $T$ and $T'$. For the selected values, the
figure exhibits the pole structure shown in the panel. Because $\beta$ may
have either sign, the trend should not be extrapolated outside the plotted
range.

The electric-charge dependence enters through $q^2$. The displayed curves indicate
how increasing $|q|$ changes the locations or number of heat-capacity poles.
Local stability on any branch is determined by $C_Q>0$, not by the sign of
the charge itself.

Thus, the heat-capacity plots provide a parameter-specific map of local
thermal stability, complementing the temperature branches identified in
Sec.~\ref{sec:thermo}. A first-order small/large-black-hole transition cannot
be inferred from a pole alone: it requires competing branches at the same
control parameters and a free-energy comparison in a specified ensemble with
consistent boundary conditions. The potential in the next paragraph is
therefore interpreted only after its ensemble is stated.

\subsection{Thermodynamic potentials and ensemble dependence}

For fixed temperature and electric potential, the grand-canonical potential
is $G=M-TS-\Phi Q$. For a canonical ensemble with fixed electric charge, the
appropriate potential is instead the Helmholtz free energy
$F=M-TS$. Global preference can be assessed only by comparing admissible
phases at the same ensemble control parameters and with identical boundary
conditions; the sign of either potential by itself is not a stability
criterion. The expression for $G$ below is included as an algebraic
diagnostic along the displayed horizon family. Because that family holds
$Q$ fixed while $r_+$ (and hence $\Phi=Q/r_+$) changes, the plotted curves
must not be interpreted as a grand-canonical coexistence diagram.
\begin{equation}
F=M-TS,\qquad G=M-TS-\Phi Q.  \label{G}
\end{equation}

\begin{figure*}[tbp]
\centering
\IfFileExists{5a.eps}{\includegraphics[width=0.30\textwidth]{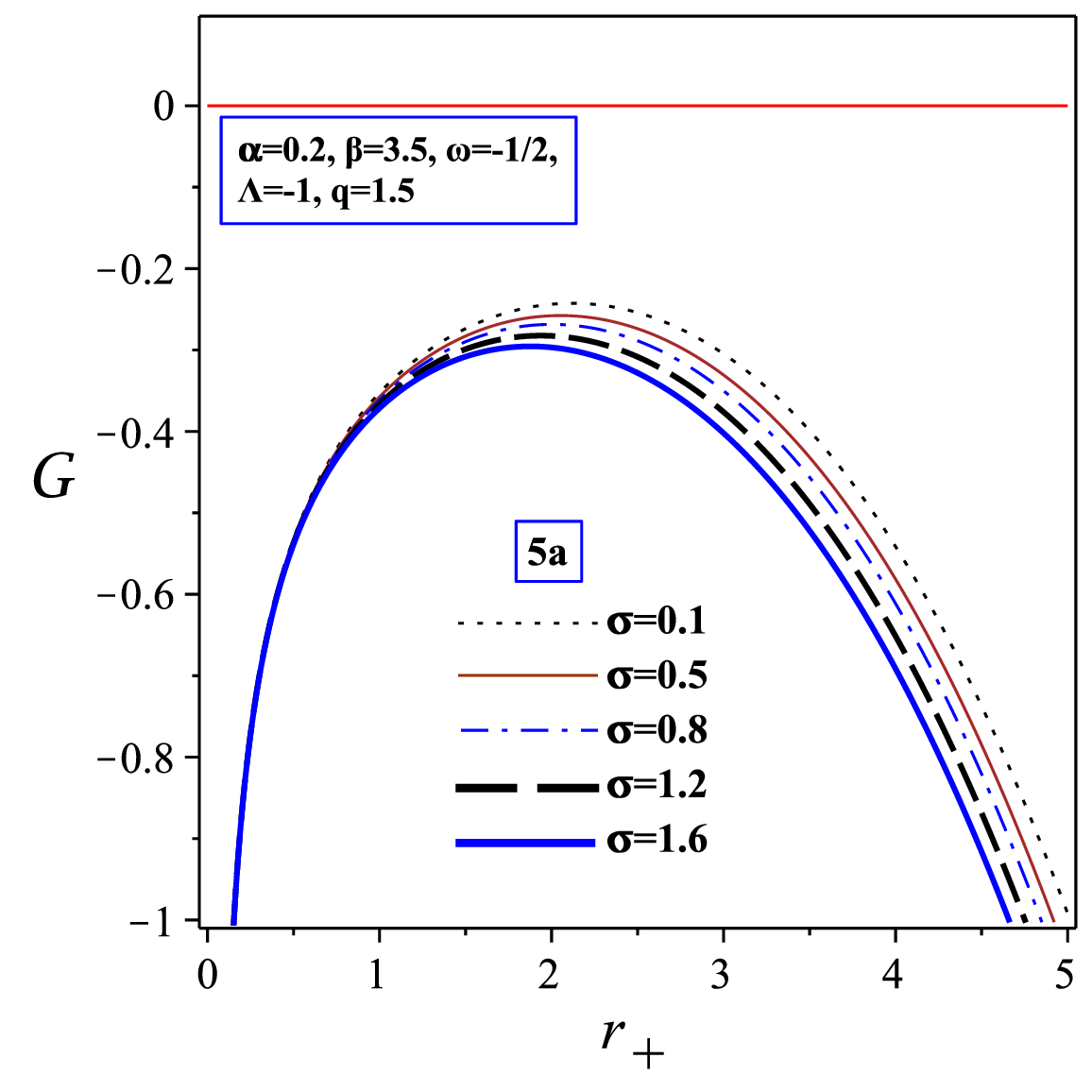}}{\MissingFigure{5a.eps}{0.28\textwidth}{0.12\textheight}}
\IfFileExists{5b.eps}{\includegraphics[width=0.30\textwidth]{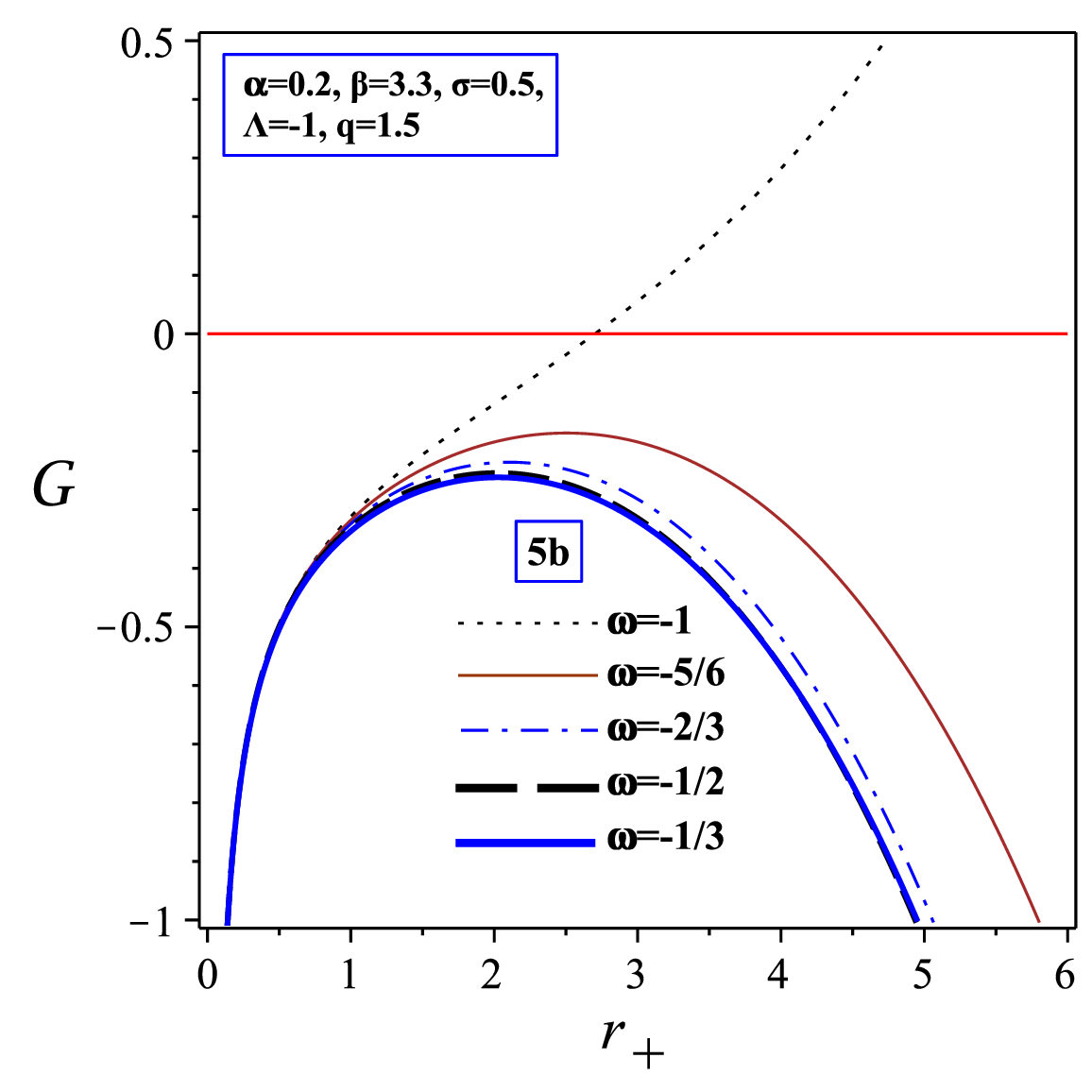}}{\MissingFigure{5b.eps}{0.28\textwidth}{0.12\textheight}}
\IfFileExists{5c.eps}{\includegraphics[width=0.30\textwidth]{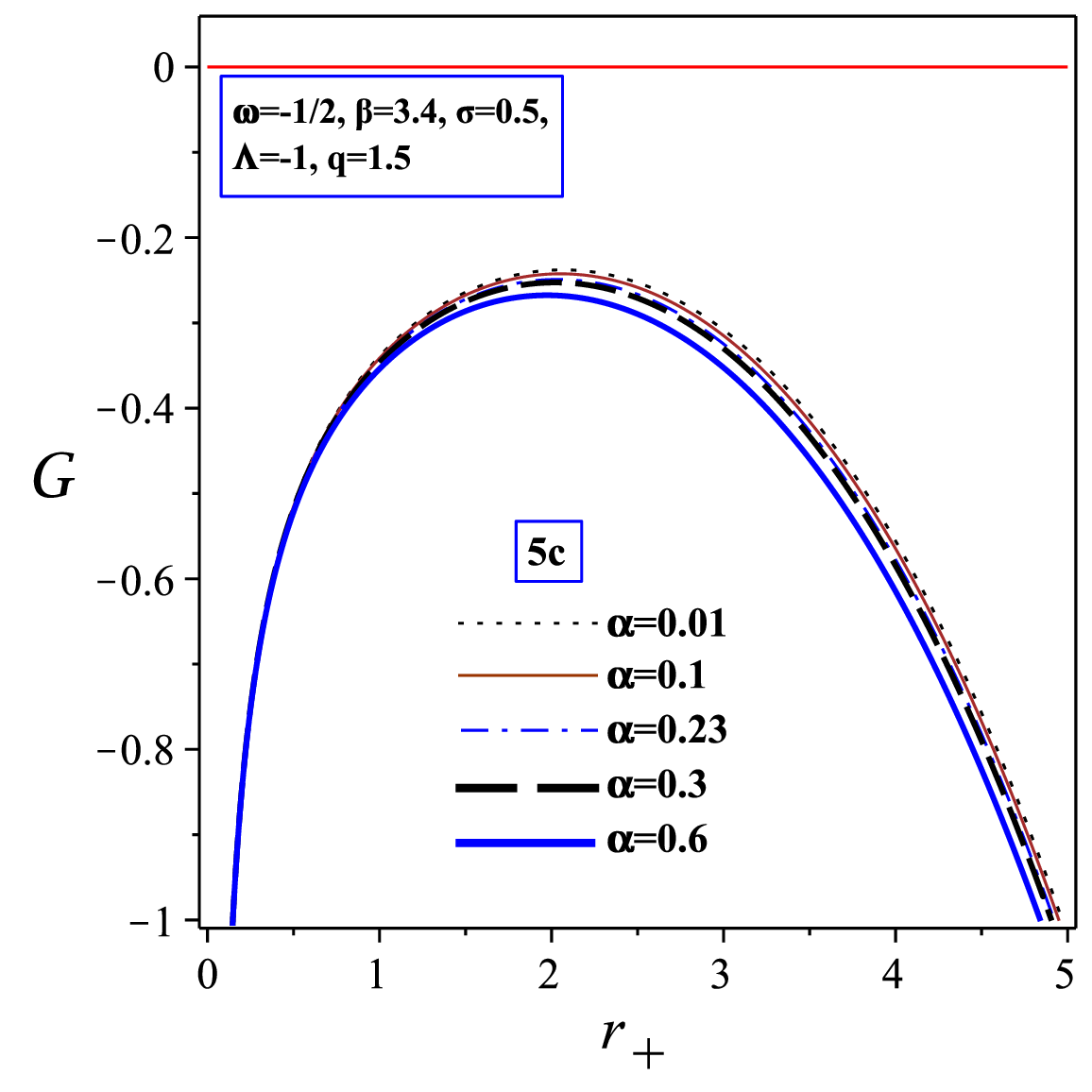}}{\MissingFigure{5c.eps}{0.28\textwidth}{0.12\textheight}}\par\medskip
\IfFileExists{5d.eps}{\includegraphics[width=0.30\textwidth]{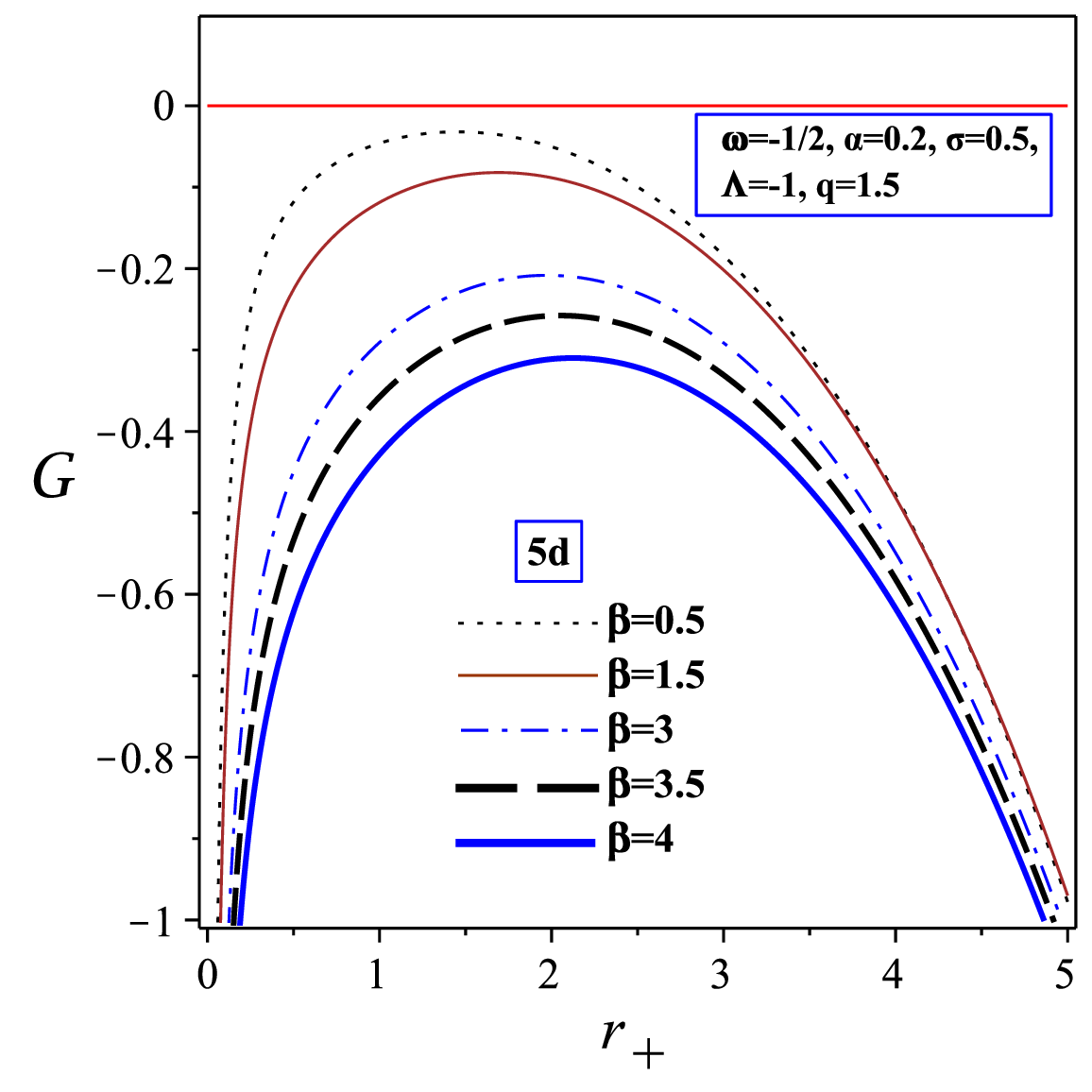}}{\MissingFigure{5d.eps}{0.28\textwidth}{0.12\textheight}}
\IfFileExists{5e.eps}{\includegraphics[width=0.30\textwidth]{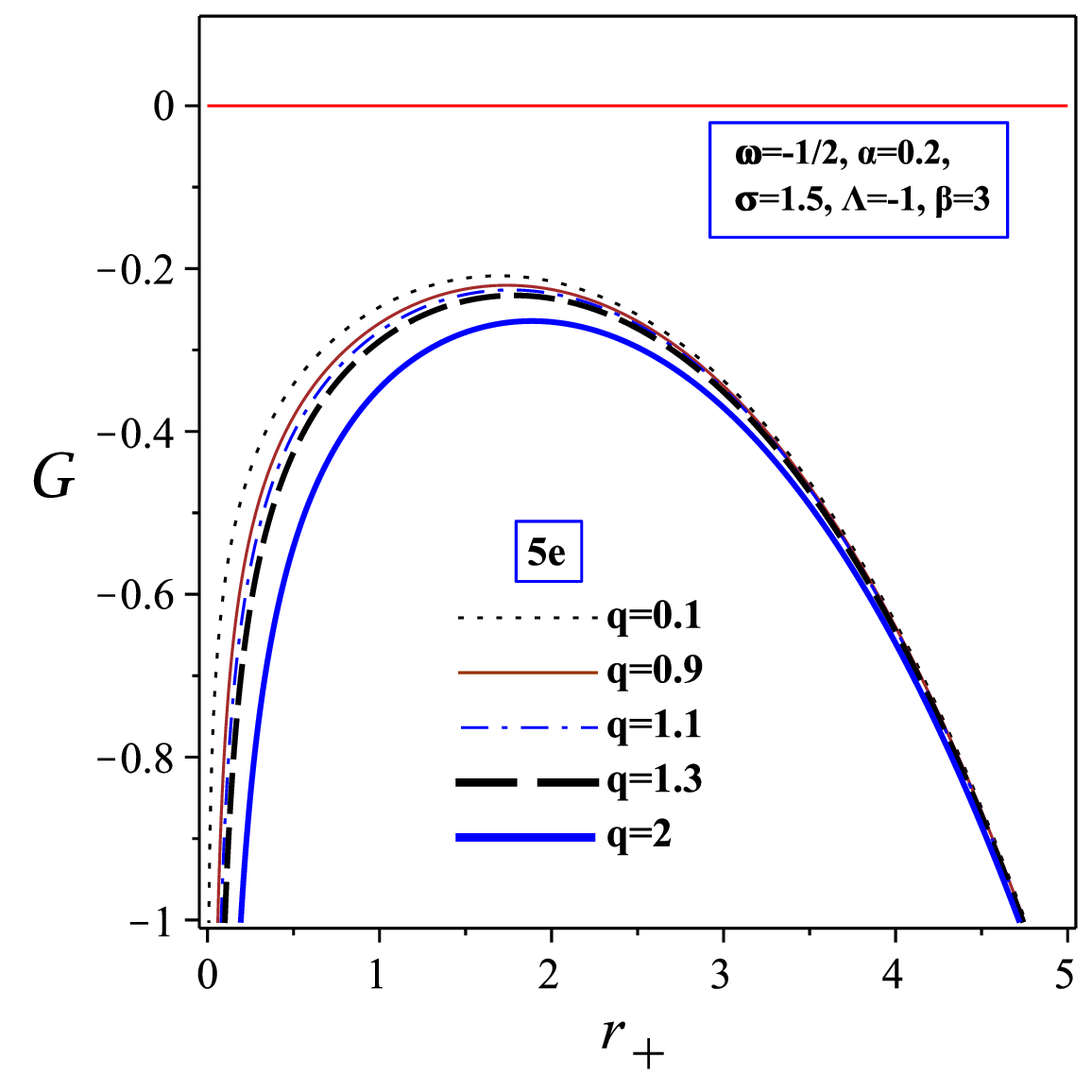}}{\MissingFigure{5e.eps}{0.28\textwidth}{0.12\textheight}}
\caption{Gibbs potential $G$ from Eq.~(\ref{GG}) versus $r_{+}$ for variations of (a) $\sigma$, (b) $\omega$, (c) $\alpha$, (d) $\beta$, and (e) $q$.}
\label{fig5}
\end{figure*}

Substitution of Eqs.~(\ref{Temp2}), (\ref{entropy}), (\ref{Q}),
(\ref{U}), and (\ref{MM}) gives
\begin{equation}
\begin{split}
G=\frac{\Lambda r_{+}^{3}}{12}+\frac{r_{+}}{4}\left( 1-\alpha \right) -\frac{%
Q^{2}}{4r_{+}}+\frac{\beta }{4}\left( 2\ln \left( \frac{r_{+}}{\left\vert
\beta \right\vert }\right) -1\right) -\frac{\left( 3\omega +2\right) \sigma
}{4r_{+}^{3\omega }}.  \label{GG}
\end{split}
\end{equation}
Figure~\ref{fig5} displays $G$ for the parameter choices used in the
numerical scan. These curves describe the chosen thermodynamic potential, but
a claim of global stability requires a reference background or competing
black-hole branch evaluated with the same ensemble and boundary conditions.
For a canonical fixed-charge analysis, the corresponding potential is the
Helmholtz free energy $F=M-TS$, rather than $G$.

\section{Extended Phase Space}
\label{sec:extended}

The preceding first law holds at fixed cosmological constant and fixed matter-sector parameters. In extended phase space, we promote $P=-\Lambda/(8\pi)$ to a thermodynamic variable \cite{PLambda2,PLambda3,PLambda4}. In the standard framework the mass is interpreted as enthalpy. Here we apply that formal thermodynamic construction to the horizon-derived function $M=m_0$; identifying it with a conserved asymptotic mass still requires a separate charge and boundary-normalization analysis because of the PFDM and quintessence asymptotics. The state parameter $\omega$ is held fixed in each thermodynamic family. Since $[\sigma]=L^{3\omega+1}$ depends on $\omega$, varying $\omega$ while holding a dimensionful $\sigma$ fixed would mix parameter families with different dimensional assignments; comparisons between different $\omega$ values below are therefore model comparisons, not thermodynamic variations $d\omega$.

Substituting $\Lambda=-8\pi P$ into Eq.~(\ref{MS}) gives
\begin{widetext}
\begin{align}
\mathcal{M}=M(S,Q,P,\beta,\omega,\sigma,\alpha)&=\frac{(1-\alpha)}{2}\sqrt{\frac{S}{\pi}}
+\frac{\beta}{2}\ln\!\left(\frac{\sqrt{S}}{|\beta|\sqrt{\pi}}\right)
+\frac{Q^2}{2\sqrt{S/\pi}}\nonumber\\
&+\frac{4PS^{3/2}}{3\sqrt{\pi}}
-\frac{\sigma}{2(S/\pi)^{3\omega/2}}.
\label{MSQPb}
\end{align}
\end{widetext}

From Eq.~(\ref{MSQPb}), we obtain the thermodynamic conjugates associated
with $S$, $Q$, $P$, $\beta$, $\sigma$, and $\alpha$, with $\omega$ held
fixed. For the derivative with respect to $\beta$, we assume $\beta\ne0$:

\begin{widetext}
\begin{align}
T &=
\left(\frac{\partial \mathcal{M}}{\partial S}\right)_{Q,P,\beta,\omega,\sigma,\alpha}
=2P\sqrt{\frac{S}{\pi}}
+\frac{\beta}{4S}
+\frac{1-\alpha}{4\sqrt{\pi S}}
-\frac{\sqrt{\pi}\,Q^2}{4S^{3/2}}
+\frac{3\omega\sigma}
{4S\left(S/\pi\right)^{3\omega/2}},
\label{TSQPb}\\[4pt]
V &=
\left(\frac{\partial \mathcal{M}}{\partial P}\right)_{S,Q,\beta,\omega,\sigma,\alpha}
=\frac{4S^{3/2}}{3\sqrt{\pi}},
\label{VSQPb}\\[4pt]
\Phi &=
\left(\frac{\partial \mathcal{M}}{\partial Q}\right)_{S,P,\beta,\omega,\sigma,\alpha}
=\frac{Q}{\sqrt{S/\pi}},
\label{PhiSQPb}\\[4pt]
B &=
\left(\frac{\partial \mathcal{M}}{\partial \beta}\right)_{S,Q,P,\omega,\sigma,\alpha}
=\frac{1}{2}\left[
\ln\!\left(\frac{\sqrt{S}}{|\beta|\sqrt{\pi}}\right)-1
\right],
\label{BSQPb}\\[4pt]
\Sigma &=
\left(\frac{\partial \mathcal{M}}{\partial \sigma}\right)_{S,Q,P,\beta,\omega,\alpha}
=-\frac{1}{2\left(S/\pi\right)^{3\omega/2}},
\label{GSQPb}\\[4pt]
A &=
\left(\frac{\partial \mathcal{M}}{\partial \alpha}\right)_{S,Q,P,\beta,\omega,\sigma}
=-\frac{1}{2}\sqrt{\frac{S}{\pi}}.
\label{ASQPb}
\end{align}
\end{widetext}
For the thermodynamic function in Eq.~(\ref{MSQPb}), differentiation with
respect to all listed variables gives the extended first law:
\begin{equation}
d\mathcal{M}=TdS+\Phi dQ+VdP+Bd\beta+\Sigma d\sigma+Ad\alpha .
\end{equation}

\subsection{Smarr Relation}

In Eq.~(\ref{MSQPb}), the mass function depends on $S$, $Q$, $P$, $\beta$,
$\sigma$, and $\alpha$, with $\omega$ treated as a fixed dimensionless
label of the source model. Deriving its Smarr relation requires the scaling
dimension of each varied variable and parameter. We use natural units,
$c=G=1$.

For the black hole solution under consideration in four dimensional
spacetime, the length dimensions of the relevant quantities are $\left[ M%
\right] =\left[ Q\right] =\left[ \beta \right] =L$, $\left[ S\right] =L^{2}$%
, $\left[ P\right] =L^{-2}$, $\left[ \sigma \right] =L^{3\omega +1}$, and $%
\left[ \alpha \right] =\left[ \omega \right] =L^{0}$. Here, $L$ denotes a
characteristic length scale. In particular, both $\alpha $ and $\omega $ are
dimensionless.

Under a uniform rescaling of the length scale, $L\longrightarrow \lambda L$,
the thermodynamic variables and model parameters transform according to

\begin{equation}
\left\{
\begin{array}{c}
Q\rightarrow \lambda Q,~~~\&~~~\beta \rightarrow \lambda \beta
,~~~\&~~~M\rightarrow \lambda M \\
\\
S\rightarrow \lambda ^{2}S,\ \ \ \&\ \ \ P\rightarrow \lambda
^{-2}P,~~~\sigma \rightarrow \lambda ^{3\omega +1}\sigma \\
\\
\alpha \rightarrow \alpha ,\ \ \ \&\ \ \ \omega \rightarrow \omega%
\end{array}%
\right. .
\end{equation}

Consequently, the mass function obeys the weighted homogeneity relation

\begin{widetext}
\begin{equation}
\mathcal{M}\!\left(
\lambda^{2}S,\lambda^{-2}P,\lambda Q,\lambda\beta,
\omega,\lambda^{3\omega+1}\sigma,\alpha
\right)
=\lambda\,\mathcal{M}(S,Q,P,\beta,\omega,\sigma,\alpha).
\label{lambdaM}
\end{equation}
\end{widetext}
Thus, $\mathcal{M}$ is a weighted homogeneous function of degree one.

Applying the weighted Euler theorem to Eq.~(\ref{lambdaM}) yields

\begin{widetext}
\begin{align}
\mathcal{M}={}&
2S\left(\frac{\partial\mathcal{M}}{\partial S}\right)_{Q,P,\beta,\omega,\sigma,\alpha}
-2P\left(\frac{\partial\mathcal{M}}{\partial P}\right)_{S,Q,\beta,\omega,\sigma,\alpha}
+Q\left(\frac{\partial\mathcal{M}}{\partial Q}\right)_{S,P,\beta,\omega,\sigma,\alpha}
\nonumber\\
&+\beta\left(\frac{\partial\mathcal{M}}{\partial\beta}\right)_{S,Q,P,\omega,\sigma,\alpha}
+(3\omega+1)\sigma
\left(\frac{\partial\mathcal{M}}{\partial\sigma}\right)_{S,Q,P,\beta,\omega,\alpha}.
\label{EulerMass}
\end{align}
\end{widetext}
No explicit term proportional to $\alpha$ appears in Eq.~(\ref{EulerMass}),
because it has zero scaling weight. The parameter $\omega$ also has zero
scaling weight but is held fixed, rather than assigned a thermodynamic
conjugate: differentiating with respect to an exponent that changes the
dimension of $\sigma$ would produce a logarithm of a dimensionful quantity
unless an additional reference scale and its conjugate were introduced.

Replacing the thermodynamic conjugate quantities (Eqs.~(\ref{TSQPb})-(\ref%
{ASQPb})) into Eq. (\ref{EulerMass}), one obtains the modified Smarr
relation
\begin{equation}
\mathcal{M}=2ST-2PV+Q\Phi +\beta B+(3\omega +1)\sigma \Sigma .
\label{ModifiedSmarr}
\end{equation}

The modified Smarr relation separates the electric, pressure, PFDM, and quintessence-normalization terms. Its algebraic reduction in simpler limits is immediate, but interpreting it as a relation for a conserved mass requires suitable asymptotics and a compatible mass normalization. The $\beta$-derivative is defined only for $\beta\ne0$; although $\mathcal M$ is continuous as $\beta\to0$, this derivative is singular there.

\section{Black-hole heat engine}
\label{sec:heatengine}

We now quantify a rectangular-cycle benchmark in the formal extended
thermodynamic description. Related studies have examined heat-engine
efficiency and dark-energy effects in AdS black-hole thermodynamics
\cite{liu2017,Xu2018}. The cosmological constant is represented by the
pressure $P=-\Lambda/(8\pi)$, and the horizon-derived function
$\mathcal M$ in Eq.~(\ref{MSQPb}) is used as enthalpy for this calculation.
Because its identification with a conserved asymptotic mass has not been
established for the present PFDM/quintessence asymptotics, the cycle should be
read as a reproducible result within the stated horizon-thermodynamic model;
its interpretation as a physical AdS black-hole engine is conditional on a
separate asymptotic-charge analysis. Throughout a cycle we hold
$Q,\alpha,\beta,\sigma$ and $\omega$ fixed. We consider a clockwise
rectangular cycle with $P_{\rm H}>P_{\rm C}$ and
$V_{\rm R}>V_{\rm L}$, whose corners are
\begin{equation}
\begin{aligned}
A&=(P_{\rm H},V_{\rm L}),\quad
B=(P_{\rm H},V_{\rm R}),\\
C&=(P_{\rm C},V_{\rm R}),\quad
D=(P_{\rm C},V_{\rm L}).
\end{aligned}
\label{eq:enginecorners}
\end{equation}
For this static spherical solution, the entropy and thermodynamic
volume are $S=\pi r_+^2$ and $V=4\pi r_+^3/3$. Hence $S$ depends only on
$V$, and $C_V=T(\partial S/\partial T)_V=0$ when the matter parameters are
held fixed. The vertical branches are therefore isochores and exchange no
heat. The net mechanical work is the area of the rectangle,
\begin{equation}
W=(P_{\rm H}-P_{\rm C})(V_{\rm R}-V_{\rm L})>0.
\label{eq:enginework}
\end{equation}

An explicit expression follows by writing the enthalpy as

\begin{widetext}
\begin{align}
\mathcal{M}(P,r_+)&=\frac{4\pi P r_+^3}{3}+H_0(r_+),\nonumber\\
H_0(r)&=\frac{1-\alpha}{2}r
+\frac{\beta}{2}\ln\!\left(\frac{r}{|\beta|}\right)
+\frac{q^2}{2r}-\frac{\sigma}{2r^{3\omega}}.
\label{eq:engineenthalpy}
\end{align}
\end{widetext}
Define $\Delta V=4\pi(r_{\rm R}^3-r_{\rm L}^3)/3$ and
$\Delta H_0=H_0(r_{\rm R})-H_0(r_{\rm L})$, where
$r_{\rm L}=(3V_{\rm L}/4\pi)^{1/3}$ and
$r_{\rm R}=(3V_{\rm R}/4\pi)^{1/3}$. The heat absorbed on the upper
isobar and the magnitude of the heat rejected on the lower isobar are

\begin{widetext}
\begin{align}
Q_{\rm H}
&=\mathcal{M}(P_{\rm H},V_{\rm R})
-\mathcal{M}(P_{\rm H},V_{\rm L})
=\Delta H_0+P_{\rm H}\Delta V,\nonumber\\
Q_{\rm C}
&=\mathcal{M}(P_{\rm C},V_{\rm R})
-\mathcal{M}(P_{\rm C},V_{\rm L})
=\Delta H_0+P_{\rm C}\Delta V.
\label{eq:engineheat}
\end{align}
\end{widetext}
These expressions obey $Q_{\rm H}-Q_{\rm C}=W$, providing a direct
first-law consistency check. A cycle is a heat engine only if $Q_{\rm H}>0$, $Q_{\rm C}\geq0$ (with
$Q_{\rm C}$ defined here as the magnitude of rejected heat), and the Hawking
temperature remains positive along both isobars. These conditions are
independent checks: a positive rectangle area alone does not guarantee a
physically admissible engine. When they hold, the efficiency is
\begin{equation}
\eta=\frac{W}{Q_{\rm H}}=1-\frac{Q_{\rm C}}{Q_{\rm H}}
=\frac{(P_{\rm H}-P_{\rm C})\Delta V}
{\Delta H_0+P_{\rm H}\Delta V}.
\label{eq:engineefficiency}
\end{equation}

\paragraph{Benchmark cycle.}
To make the result quantitative without implying a universal
parameter trend, we give one explicitly specified benchmark. We use the
reference matter parameters adopted in the ray-optics analysis,
$\alpha=0.1$, $\beta=0.2$, $q=0.4$, $\sigma=0.02$, and
$\omega=-2/3$, and choose $r_{\rm L}=2$, $r_{\rm R}=3$,
$P_{\rm C}=0.03/(8\pi)$, and $P_{\rm H}=0.06/(8\pi)$ in the geometric
units of the manuscript. The enthalpy expression gives
$\Delta V=79.5870$ and $\Delta H_0=0.427213$; consequently,
\begin{align}
Q_{\rm H}&=0.617213,\qquad
Q_{\rm C}=0.522213,\nonumber\\
W&=0.095000,\qquad
\eta=0.153918\simeq15.39\%.
\label{eq:enginebenchmark}
\end{align}
The identity $Q_{\rm H}-Q_{\rm C}=W$ holds to the precision shown.
Because the reference choice uses $\beta=0.2$, its PFDM density is negative
under the stress-tensor convention in Sec.~\ref{sec:fe}; this benchmark is
therefore a reproducible calculation within the phenomenological model, not
a claim about an ordinary positive-density dark-matter halo. For the four
corners $(P,r_+)$, the Hawking temperatures are approximately
$T(P_{\rm C},r_{\rm L})=0.03979$, $T(P_{\rm C},r_{\rm R})=0.02915$,
$T(P_{\rm H},r_{\rm L})=0.04456$, and
$T(P_{\rm H},r_{\rm R})=0.03631$. A one-parameter-at-a-time scan with the
same cycle corners gives the representative efficiencies in
Table~\ref{tab:engine-scan}; the minimum temperature is evaluated over both
isobars for $2\leq r_+\leq3$. All listed cases remain at positive temperature
throughout these branches.

\begin{table*}[tbp]
\centering
\caption{Efficiency of the same rectangular heat-engine cycle under
one-parameter changes from the reference set. Unless specified in the first
column, $\alpha=0.1$, $\beta=0.2$, $q=0.4$, $\sigma=0.02$, and
$\omega=-2/3$. The cycle corners are $r_{\rm L}=2$, $r_{\rm R}=3$,
$P_{\rm C}=0.03/(8\pi)$, and $P_{\rm H}=0.06/(8\pi)$. The reported
$T_{\min}$ is the minimum Hawking temperature along the two isobars.}
\label{tab:engine-scan}
\begin{ruledtabular}
\begin{tabular}{lccc}
Case & $Q_{\rm H}$ & $T_{\min}$ & $\eta$ (\%)\\
\hline
Reference & 0.617213 & 0.029149 & 15.3918\\
$\alpha=0.2$ & 0.567213 & 0.026496 & 16.7486\\
$\beta=-0.2$ & 0.536120 & 0.025612 & 17.7199\\
$\beta=0.4$ & 0.657760 & 0.030917 & 14.4430\\
$q=0$ & 0.630547 & 0.029621 & 15.0663\\
$q=0.6$ & 0.600547 & 0.028559 & 15.8189\\
$\sigma=0$ & 0.667213 & 0.032332 & 14.2383\\
$\sigma=0.06$ & 0.517213 & 0.022783 & 18.3677\\
$\omega=-0.8$ & 0.580327 & 0.026404 & 16.3701\\
$\omega=-0.5$ & 0.643535 & 0.030954 & 14.7622\\
\end{tabular}
\end{ruledtabular}
\end{table*}

For this cycle alone, increasing $\alpha$ from $0.1$ to $0.2$ raises
$\eta$ from $15.39\%$ to $16.75\%$; changing $\beta$ from $0.2$ to $0.4$
lowers it to $14.44\%$, while $\beta=-0.2$ gives $17.72\%$. Increasing
$q$ from $0.4$ to $0.6$ raises the efficiency modestly, and increasing
$\sigma$ from $0.02$ to $0.06$ raises it to $18.37\%$ for the fixed corners.
Changing $\omega$ from $-2/3$ to $-0.8$ raises the efficiency to $16.37\%$,
whereas $\omega=-0.5$ lowers it to $14.76\%$. These are controlled,
cycle-specific comparisons, not universal monotonicity laws: the enthalpy
change $\Delta H_0$ controls how each parameter enters the efficiency, and a
different cycle can produce different trends. Since the isobars are not
isothermal, a comparison with a Carnot engine must additionally specify the
hot- and cold-reservoir temperatures; no universal Carnot comparison is
asserted here.

\section{Ray optics}
\label{sec:rayoptics}
In the geometric-optics limit light propagates along null geodesics of the
background, so the photon sphere, the critical impact parameter and the
shadow are fixed entirely by the metric function $f(r)$. These quantities are
the most direct link between the matter content of the solution and what an
observer can actually measure, which is why null geodesics have become a
standard probe of black hole spacetimes since the early work on lensing and
photon surfaces \cite{1,2,3,4,Claudel2001} and, more recently, of the
horizon-scale images reported by the Event Horizon Telescope
\cite{EHT2019,EHT2022,25,Perlick2022}. An equivalent description is provided
by the optical metric, in which light rays are geodesics of an effective
Riemannian (or Randers--Finsler) manifold and the deflection angle follows
from the Gauss--Bonnet theorem \cite{8,9,10,11}. The same viewpoint has been
used to treat curved optical surfaces and optical wormholes as media with a
position-dependent refractive index, where the curvature of the optical
manifold decides which rays are trapped and which escape
\cite{12,13,14,15,19,26}, and to follow how topological defects,
Lorentz-violating terms or modified gravity reshape the photon region
\cite{17,18,20}. For AdS black holes carrying several matter sectors, a
combined thermodynamic and optical treatment has proved useful because both
sets of observables are controlled by the same function $f(r)$
\cite{EslamPanah2026bumblebee}.

In the metric function in Eq.~(\ref{f(r)}), the Letelier string cloud changes
the constant term to $1-\alpha$, the PFDM sector contributes a logarithmic
term, the Maxwell field contributes $q^{2}/r^{2}$, and the Kiselev-type
quintessence sector contributes $-\sigma r^{-(3\omega+1)}$. We determine how
these terms affect the photon sphere, the critical impact parameter, and the
shadow measured by a static observer at a specified finite radius.

\subsection{Null geodesics and effective potential}

Because the spacetime (\ref{metric}) is static and spherically symmetric, we
may restrict the motion to the equatorial plane $\theta=\pi/2$ without loss
of generality. Null geodesics follow from
\begin{equation}
2\mathcal{L}=-f(r)\,\dot{t}^{2}+\frac{\dot{r}^{2}}{f(r)}+r^{2}\dot{\varphi}^{2}=0,
\label{eq3.2}
\end{equation}
where the dot denotes the derivative with respect to an affine parameter
$\lambda$. We write $\mathcal{L}$ for the Lagrangian to avoid confusion with
the angular momentum $L$. The Killing vectors $\partial_{t}$ and
$\partial_{\varphi}$ give the conserved energy and angular momentum
\begin{equation}
E=f(r)\,\dot{t},\qquad L=r^{2}\dot{\varphi},
\label{eq3.3}
\end{equation}
and Eq.~(\ref{eq3.2}) reduces to a one-dimensional problem,
\begin{equation}
\dot{r}^{2}+V_{\rm eff}(r)=E^{2},\qquad
V_{\rm eff}(r)=\frac{L^{2}f(r)}{r^{2}} .
\label{eq3.7}
\end{equation}
It is convenient to write out $V_{\rm eff}/L^{2}$ explicitly,
\begin{equation}
\frac{V_{\rm eff}}{L^{2}}=\frac{1-\alpha}{r^{2}}-\frac{2m_{0}}{r^{3}}
+\frac{q^{2}}{r^{4}}+\frac{\beta}{r^{3}}\ln\frac{r}{|\beta|}
-\frac{\sigma}{r^{3(\omega+1)}}-\frac{\Lambda}{3}.
\label{Veff2}
\end{equation}
The cosmological constant appears in $V_{\rm eff}/L^2$ only through the
additive constant $-\Lambda/3$. Consequently, it drops out of the extremum
condition and does not shift the photon-sphere radius for this metric ansatz.
It does, however, affect $f(r_{\rm ph})$, the critical impact parameter, and
the shadow angle measured by a finite-radius observer. In terms of $u=1/r$ and the impact parameter $b=L/E$,
the orbit equation reads
\begin{equation}
\left(\frac{du}{d\varphi}\right)^{2}=\frac{1}{b^{2}}-u^{2}f(1/u),
\qquad
\frac{d^{2}u}{d\varphi^{2}}=-\frac{1}{2}\frac{d}{du}\left[u^{2}f(1/u)\right],
\label{orbit}
\end{equation}
and the second-order form is independent of $\Lambda$ explicitly. The
cosmological constant still enters the initial data corresponding to a chosen
impact parameter and therefore must be retained when setting up a ray at a
specified initial radius.

\subsection{Photon sphere, critical impact parameter and shadow}

Circular null orbits are extrema of $f(r)/r^{2}$ and therefore satisfy
$rf'(r)=2f(r)$. An extremum is an unstable photon sphere when it is a local
maximum of $f/r^{2}$, equivalently when
$\left.d^{2}(f/r^{2})/dr^{2}\right|_{r_{\rm ph}}<0$. For general parameter
values, the algebraic condition becomes
\begin{equation}
\begin{aligned}
2(1-\alpha)r_{\rm ph}^{2}-6m_{0}r_{\rm ph}+4q^{2}-\beta r_{\rm ph}\left[1-3\ln\!\left(\frac{r_{\rm ph}}{|\beta|}\right)\right]
-3\sigma(\omega+1)r_{\rm ph}^{1-3\omega}=0.
\end{aligned}
\label{rph}
\end{equation}
For $\beta=\sigma=0$, the outer solution is
\[
 r_{\rm ph}=\frac{3m_{0}+\sqrt{9m_{0}^{2}-8(1-\alpha)q^{2}}}{2(1-\alpha)},
\]
provided the discriminant is nonnegative and the root lies in the static
exterior. This is the charged Letelier result; it reduces to $3m_{0}$ when
$\alpha=q=0$. For $q\ne0$, the inner algebraic root, when present, is not the
outer photon sphere relevant to the usual shadow boundary. The critical impact parameter separating captured from
scattered rays is
\begin{equation}
b_{c}=\frac{r_{\rm ph}}{\sqrt{f(r_{\rm ph})}} .
\label{bc}
\end{equation}
Because the spacetime is asymptotically AdS rather than asymptotically flat,
we define the observable shadow angle locally for a static observer at a
finite radius $r_{O}$ in the static region, where $f(r_{O})>0$. The angular
radius is \cite{Synge1966,Perlick2022}
\begin{equation}
\sin\theta_{\rm sh}=\frac{r_{\rm ph}}{r_{O}}\sqrt{\frac{f(r_{O})}{f(r_{\rm ph})}}
=\frac{b_{c}}{r_{O}}\sqrt{f(r_{O})} .
\label{shadow}
\end{equation}
As a consistency check, in the continuous limit
$\alpha=q=\beta=\sigma=0$, Eq.~(\ref{bc}) gives the
Schwarzschild--AdS result
$b_{c}=3\sqrt{3}\,m_{0}/\sqrt{1-9\Lambda m_{0}^{2}}$, provided the
photon sphere lies in the static region and $f(3m_{0})>0$.

\begin{figure*}[tbp]
\centering
\IfFileExists{fig_veff.eps}{\includegraphics[width=0.94\textwidth]{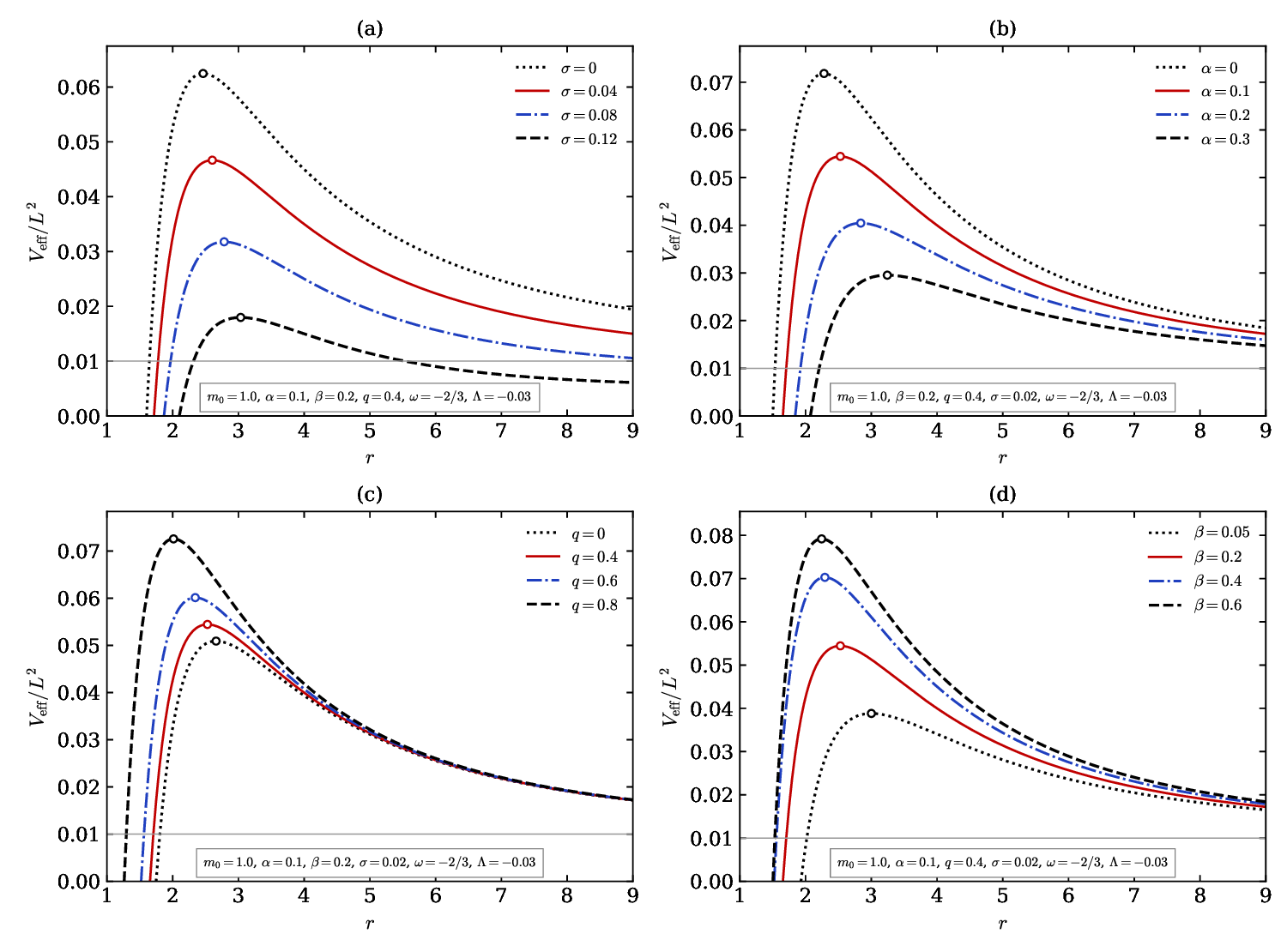}}{\MissingFigure{fig_veff.eps}{0.92\textwidth}{0.20\textheight}}
\caption{Effective potential $V_{\rm eff}/L^{2}$ of Eq.~(\ref{Veff2})
outside the event horizon for different values of (a) $\sigma$, (b) $\alpha$,
(c) $q$ and (d) $\beta$. Open circles mark the photon sphere $r_{\rm ph}$. The
thin gray line is the asymptotic value $-\Lambda/3$. The fixed parameters are
given in each panel.}
\label{figray1}
\end{figure*}

\begin{figure*}[tbp]
\centering
\IfFileExists{fig_orbits.eps}{\includegraphics[width=0.96\textwidth]{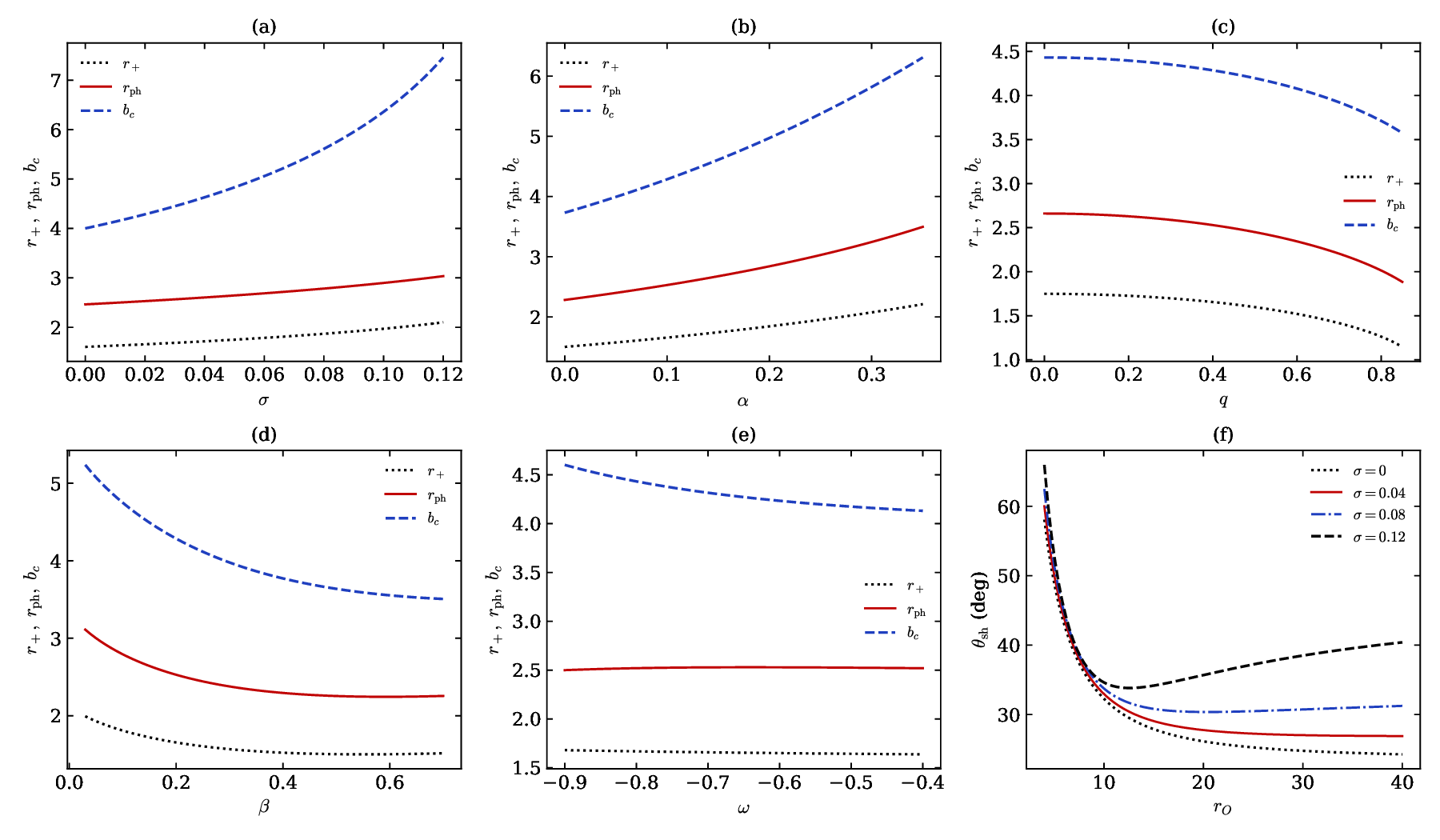}}{\MissingFigure{fig_orbits.eps}{0.92\textwidth}{0.20\textheight}}
\caption{Event horizon $r_{+}$, photon-sphere radius $r_{\rm ph}$ and
critical impact parameter $b_{c}$ as functions of (a) $\sigma$, (b)
$\alpha$, (c) $q$, (d) $\beta$ and (e) $\omega$, with the remaining
parameters fixed at their reference values. Panel (f) gives the angular
radius of the shadow, Eq.~(\ref{shadow}), versus the observer position
$r_{O}$ for several values of $\sigma$.}
\label{figray2}
\end{figure*}

Equation~(\ref{rph}) has to be solved numerically for general parameters.
The results are shown in Figs.~\ref{figray1} and \ref{figray2} and
collected in Table~\ref{tabray}. Unless stated otherwise we take $m_{0}=1$,
$\alpha=0.1$, $\beta=0.2$, $q=0.4$, $\sigma=0.02$, $\omega=-2/3$ and
$\Lambda=-0.03$. We do not use the parameter set of Fig.~\ref{fig1} here,
for the following reason. For $m_{0}=3$, $\omega=-2/3$ and
$\sigma\gtrsim0.3$, the quintessence term $-\sigma r$ in $f(r)$ dominates the
mass term. In that regime $f/r^{2}$ increases monotonically everywhere outside
the event horizon, so no circular null orbit exists and the black hole casts
no shadow in the usual sense. The same mechanism appears in
Fig.~\ref{figray1}(a): increasing $\sigma$ lowers and flattens the potential
barrier. For the reference values above, the photon sphere survives only for
$\sigma\lesssim0.17$; approaching this value, $f(r_{\rm ph})\to0$ and $b_{c}$
grows without bound.

Figure~\ref{figray1} shows that, for the parameter ranges displayed, the
string-cloud and quintessence variations lower the potential barrier and push
its maximum outward (panels (b) and (a), respectively). The cloud does this by reducing the
effective ``$1$'' in $f(r)$ to $1-\alpha$. Quintessence does it by adding the
negative contribution $-\sigma/r^{3(\omega+1)}$, which decays more slowly
than every other term in Eq.~(\ref{Veff2}). A lower barrier means that
photons with larger impact parameters are captured, so $b_{c}$ and the shadow
grow. The electric charge and the PFDM parameter work the other way. The
$q^{2}/r^{4}$ term is repulsive and raises the barrier (panel c), in the same
way as for the Reissner--Nordstr\"{o}m solution. Over the range shown, larger
$\beta$ also shrinks the horizon and the photon sphere and raises the peak
(panel d). Beyond $r\simeq5$ the curves in panel (c) are indistinguishable,
because the charge term dies off as $r^{-4}$. The PFDM term, which falls off
only as $r^{-3}\ln r$, still separates the curves in panel (d) at large $r$.

The direction of the $\beta$ dependence is tied to the PFDM energy density
discussed in Sec.~\ref{sec:fe}. With the conventions adopted there,
$\rho_{\rm PFDM}=-\beta/(8\pi r^{3})$, so positive energy density requires
$\beta<0$. The logarithmic term contributes
$+\tfrac{\beta}{2}\ln(r/|\beta|)$ to the horizon-derived mass function.
For the radii in the reference family, $r>|\beta|$ and the logarithm is
positive: positive $\beta$ therefore raises this mass contribution, whereas
negative $\beta$ lowers it. At fixed $m_0$, the corresponding metric term
also depends on $\beta\ln(r/|\beta|)$, which is not monotonic in $\beta$
over the entire positive-$\beta$ range. This explains why the tabulated
$\beta=0.05$ case does not lie on a simple monotonic continuation of the
$\beta=0.2$ to $0.4$ trend. The sign-dependent behavior should not be read
as a universal monotonicity statement for arbitrary parameter choices. With the other parameters at their reference values, $\beta=-0.2$ gives $r_{+}=2.6784$, $r_{\rm ph}=4.3838$, $b_{c}=6.7988$ and $\theta_{\rm sh}=45.18^{\circ}$, compared with $1.6556$, $2.5292$, $4.2857$ and $26.86^{\circ}$ for $\beta=0.2$ (Table~\ref{tabray}). A PFDM halo with positive energy density therefore enlarges the horizon, the photon sphere and the shadow, as a halo of ordinary matter would.

The dependence on each parameter is made quantitative in
Fig.~\ref{figray2}. For the parameter variations in this table, increasing $\alpha$ or
$\sigma$ increases the horizon radius $r_+$, photon-sphere radius
$r_{\rm ph}$, and critical impact parameter $b_c$, while increasing $|q|$
decreases them over the displayed range. The dependence on positive $\beta$
is not monotonic across all tabulated values: relative to the reference
$\beta=0.2$ case, $\beta=0.05$ enlarges these radii, whereas $\beta=0.4$ and
$0.6$ reduce them. These are numerical trends for the stated reference set,
not parameter-independent theorems. The critical impact parameter responds much more strongly than
$r_{\rm ph}$. As $\sigma$ goes from $0$ to $0.12$, $r_{\rm ph}$ grows by about
$24\%$ while $b_{c}$ nearly doubles. The reason is that $\sigma$ also lowers
$f(r_{\rm ph})$ in the denominator of Eq.~(\ref{bc}). The equation-of-state
parameter $\omega$ has almost no effect on $r_{\rm ph}$ (panel e), but it
does change $b_{c}$: a more negative $\omega$ makes the quintessence tail
flatter and gives a larger critical impact parameter. Panel (f) shows a
feature that is absent for asymptotically flat black holes. Close to the
hole, $\theta_{\rm sh}$ decreases with $r_{O}$ as expected. Further out,
however, $f(r_{O})$ is controlled by the competition between the AdS term
$-\Lambda r_{O}^{2}/3$ and the quintessence term $-\sigma r_{O}$. When
$\sigma$ is large enough, $\theta_{\rm sh}$ goes through a minimum and then
increases again. Any comparison with observations therefore has to specify
where the observer sits.

\begin{table}[tbph]
\centering
\caption{Event horizon, photon sphere, critical impact parameter and shadow
angular radius (for $r_{O}=20$). In each row one parameter differs from the
reference set $m_{0}=1$, $\alpha=0.1$, $\beta=0.2$, $q=0.4$, $\sigma=0.02$,
$\omega=-2/3$, $\Lambda=-0.03$.}
\label{tabray}
\begin{tabular}{lcccc}
\hline\hline
 & $r_{+}$ & $r_{\rm ph}$ & $b_{c}$ & $\theta_{\rm sh}\,(^{\circ})$\\
\hline
reference            & 1.6556 & 2.5292 & 4.2857 & 26.86\\
$\sigma=0$           & 1.6030 & 2.4631 & 4.0013 & 26.13\\
$\sigma=0.06$        & 1.7855 & 2.6868 & 5.0583 & 28.88\\
$\sigma=0.10$        & 1.9699 & 2.8969 & 6.3603 & 32.45\\
$\alpha=0$           & 1.5015 & 2.2815 & 3.7310 & 23.44\\
$\alpha=0.2$         & 1.8429 & 2.8400 & 4.9716 & 31.21\\
$\alpha=0.3$         & 2.0741 & 3.2439 & 5.8188 & 36.84\\
$q=0$                & 1.7489 & 2.6604 & 4.4318 & 27.86\\
$q=0.6$              & 1.5207 & 2.3437 & 4.0782 & 25.47\\
$q=0.8$              & 1.2626 & 2.0133 & 3.7121 & 23.04\\
$\beta=0.05$         & 1.9300 & 3.0015 & 5.0756 & 32.23\\
$\beta=0.4$          & 1.5235 & 2.2945 & 3.7720 & 23.52\\
$\beta=0.6$          & 1.5039 & 2.2448 & 3.5539 & 22.16\\
$\beta=-0.05$ & 2.3392 & 3.7233 & 6.0347 & 39.19\\
$\beta=-0.2$ & 2.6784 &4.3838 & 6.7988 & 45.18\\
$\beta=-0.4$ & 2.9582 & 5.0022 & 7.4179 & 50.45\\
$\omega=-0.8$        & 1.6686 & 2.5201 & 4.4327 & 24.57\\
$\omega=-0.5$        & 1.6435 & 2.5253 & 4.1746 & 27.08\\
\hline\hline
\end{tabular}
\end{table}

\subsection{Light trajectories}

\begin{figure*}[tbp]
\centering
\IfFileExists{fig_rays.eps}{\includegraphics[width=0.88\textwidth]{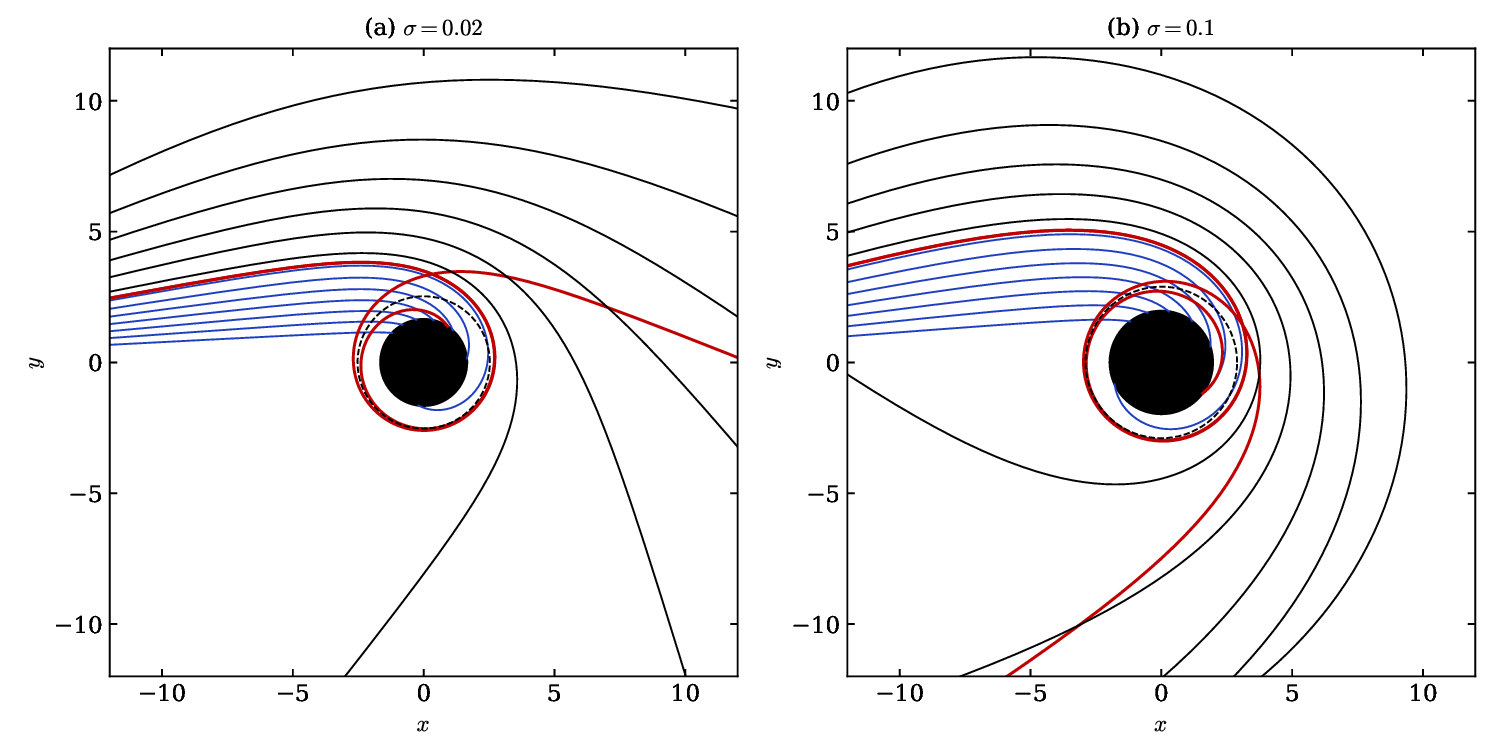}}{\MissingFigure{fig_rays.eps}{0.86\textwidth}{0.19\textheight}}
\caption{Null geodesics in the equatorial plane for (a) $\sigma=0.02$ and (b)
$\sigma=0.1$, with the other parameters at their reference values. Blue:
captured rays ($b<b_{c}$); black: scattered rays ($b>b_{c}$); red:
near-critical rays with $b=b_{c}(1\mp10^{-3})$. The filled disk is the event
horizon and the dashed circle the photon sphere.}
\label{figray3}
\end{figure*}

To examine the null trajectories, we numerically integrated the second-order
equation (\ref{orbit}) for rays initialized at $r=25$, with impact parameters
between $0.3\,b_{c}$ and $1.9\,b_{c}$. This is a finite-radius initial-value
setup, consistent with the AdS geometry; it is not an assumption of an
asymptotically flat incoming state. The results are shown in
Fig.~\ref{figray3}. For the plotted initial conditions, rays with $b<b_{c}$ (blue) cross the
photon-sphere region and fall through the event horizon, whereas rays with
$b>b_{c}$ (black) reach a radial turning point and return toward larger radius. The two red rays, with $b=0.999\,b_{c}$ and
$1.001\,b_{c}$, wind several times around the dashed circle $r=r_{\rm ph}$
before they either plunge or escape. This is the unstable orbit that forms
the shadow boundary. Comparing the two panels, raising $\sigma$ from $0.02$ to
$0.1$ enlarges both the horizon and the photon sphere and bends the scattered
rays more strongly. Even rays with $b\simeq1.9\,b_{c}$ are deflected by a
large angle, since the quintessence term keeps $f(r)$ well below its
$\sigma=0$ value far from the hole.

For the reference family and parameter ranges shown, the string-cloud and
quintessence sectors enlarge the shadow, whereas increasing $|q|$ or positive
$\beta$ reduces it. These are numerical trends, not universal monotonicity
theorems. The cosmological constant cancels from the photon-sphere equation
for this metric ansatz but remains in $f(r_{\rm ph})$, $b_c$, and the
finite-observer angle $\theta_{\rm sh}$. Thus a shadow measurement alone
cannot determine all model parameters; it constrains a degeneracy surface.
Combining that constraint with the horizon, temperature, and local-stability
conditions derived above can reduce the parameter region, but does not by
itself establish global thermodynamic stability.

\section{Wave optics and effective refractive index}
\label{sec:waveoptics}

The ray analysis identifies the unstable null orbit and capture threshold; the wave problem tests how that picture is modified at finite frequency. In addition to the angular barrier, scalar waves respond to radial derivatives of $f(r)$, producing turning points, evanescent intervals, tunnelling, and possible partial confinement. We characterize them by reducing the massless
Klein--Gordon equation to a one-dimensional Helmholtz equation and interpreting
its coefficient as a frequency-dependent effective refractive index
\cite{21,22,23,24,Konoplya2011}. This representation complements, rather than
duplicates, the null-geodesic analysis in Sec.~\ref{sec:rayoptics}: its
high-frequency limit recovers the optical geometry, while its finite-frequency
terms resolve wave effects that rays cannot describe. Related optical-medium
formulations have also been used for wormholes, curved optical surfaces,
topological defects, rotating backgrounds, dark-matter environments, and
modified-gravity black holes
\cite{12,13,14,15,19,Spiral2026,Godel2026,CJP2026,EMSchw2026,EslamPanah2026bumblebee}.

\subsection{Helmholtz form of the Klein--Gordon equation}

Let a minimally coupled, massless test scalar field satisfy
\begin{equation}
\Box\Phi\equiv
\frac{1}{\sqrt{-g}}\partial_\mu
\left(\sqrt{-g}\,g^{\mu\nu}\partial_\nu\Phi\right)=0
\end{equation}
on the spacetime (\ref{metric}). We separate variables as
\begin{equation}
\Phi(t,r,\theta,\varphi)
=e^{-i\varpi t}Y_{\ell m}(\theta,\varphi)R_\ell(r),
\end{equation}
where \(\varpi>0\) denotes the wave frequency and is distinct from the
quintessence state parameter \(\omega\). The radial equation is
\begin{equation}
\frac{1}{r^2}\frac{d}{dr}
\left(r^2 f(r)\frac{dR_\ell}{dr}\right)
+\left[\frac{\varpi^2}{f(r)}
-\frac{\ell(\ell+1)}{r^2}\right]R_\ell=0.
\label{KGradial}
\end{equation}
In the static exterior, where \(f(r)>0\), the substitution
\(R_\ell(r)=\psi_\ell(r)/[r\sqrt{f(r)}]\) removes the first derivative and
puts the equation in the radial-coordinate Helmholtz form
\begin{equation}
\frac{d^2\psi_\ell}{dr^2}
+\varpi^2 n_\ell^2(r,\varpi)\psi_\ell=0,
\label{Helmholtz}
\end{equation}
with
\begin{align}
n_\ell^2(r,\varpi)
&=\frac{\varpi^2-U_\ell(r)}{\varpi^2 f^2(r)}, \label{neff}\\
U_\ell(r)
&=\frac{\ell(\ell+1)}{r^2}f(r)
+\frac{f(r)f''(r)}{2}
+\frac{f(r)f'(r)}{r}
-\frac{[f'(r)]^2}{4}. \label{Uwave}
\end{align}
Here \(U_\ell\) is the effective wave potential in this particular
radial-coordinate representation. It is not the same function as the
standard Schrödinger potential written in the tortoise coordinate
\(dr_*/dr=f^{-1}\); the two descriptions are related by a coordinate change
and a field redefinition. This distinction matters when interpreting the
potential or comparing numerical values across formulations.

The derivatives needed in $U_\ell$ are already given in Eqs.~(\ref{fprime}) and (\ref{fsecond}); we use those expressions throughout, avoiding a second notation for the same derivatives. Whenever \(\beta=0\), terms containing
\(\beta\ln(r/|\beta|)\) are interpreted through their continuous limit and
vanish.

The sign of \(n_\ell^2\) distinguishes oscillatory and evanescent radial
regions: \(n_\ell^2>0\) permits local oscillatory solutions, whereas
\(n_\ell^2<0\) gives exponential behavior. Turning points occur where
\[
n_\ell^2(r,\varpi)=0
\quad\Longleftrightarrow\quad
U_\ell(r)=\varpi^2,
\]
provided \(f(r)\ne0\). This local classification does not by itself determine
a global mode: the boundary conditions at the horizon and at the outer
boundary must also be imposed.

Three limits clarify the physical interpretation.

First, at fixed \(r\) in the static exterior and for
\(\varpi^2\gg |U_\ell(r)|\),
\begin{equation}
n_\ell^2(r,\varpi)\longrightarrow \frac{1}{f^2(r)},
\qquad n_\ell(r,\varpi)\longrightarrow\frac{1}{f(r)}.
\label{ngeo}
\end{equation}
This is the radial refractive index associated with the optical metric
\(dl_{\rm opt}^2=dr^2/f^2+r^2d\Omega^2/f\), and recovers the
geometric-optics description \cite{8}. The limit is local: it need not be
uniform near a horizon or in a region whose radius grows with frequency.

Second, for a non-extremal outer horizon \(r=r_+\), let
\(f(r_+)=0\) and \(f'(r_+)=2\kappa>0\), where
\(\kappa=2\pi T\) in the normalization of Sec.~\ref{sec:thermo}. Writing
\(x=r-r_+\), one has \(f=2\kappa x+O(x^2)\) and
\(U_\ell(r_+)=-\kappa^2\). Consequently,
\begin{equation}
n_\ell^2(r,\varpi)
\simeq
\frac{\varpi^2+\kappa^2}
{4\varpi^2\kappa^2(r-r_+)^2},
\qquad r\to r_+.
\label{nhorizon}
\end{equation}
The coordinate refractive index therefore diverges at the horizon. This is
consistent with the infinite radial optical distance in the static
coordinates; absorption is fixed by imposing an ingoing condition at the
future horizon, rather than inferred from the divergence alone. The surface
gravity sets the leading near-horizon coefficient and thereby links this
wave equation to the black-hole temperature.

Third, for \(\Lambda<0\) and \(-1<\omega<-1/3\), the AdS term dominates the
large-\(r\) metric function:
\[
f(r)\simeq ar^2,\qquad a=-\frac{\Lambda}{3}>0.
\]
For fixed finite \(\varpi\) and \(\ell\), the leading terms in the effective
potential give
\begin{equation}
U_\ell(r)\simeq 2a^2r^2+O(1),
\qquad
n_\ell^2(r,\varpi)\simeq-\frac{2}{\varpi^2r^2}<0,
\qquad r\to\infty.
\label{nAdS}
\end{equation}
Thus the asymptotic radial solution is evanescent in this representation.
For the usual normalizable massless-scalar boundary condition, the field
vanishes at the AdS boundary. The reflecting character of the AdS boundary,
combined with the horizon boundary condition, permits confined or long-lived
wave modes; their existence and decay rates require a boundary-value
calculation, not merely the sign of \(n_\ell^2\) \cite{Horowitz2000}.

\subsection{Barrier, cavity, and the eikonal limit}

\begin{figure*}[tbp]
\centering
\IfFileExists{fig_wave_pot.eps}{\includegraphics[width=0.94\textwidth]{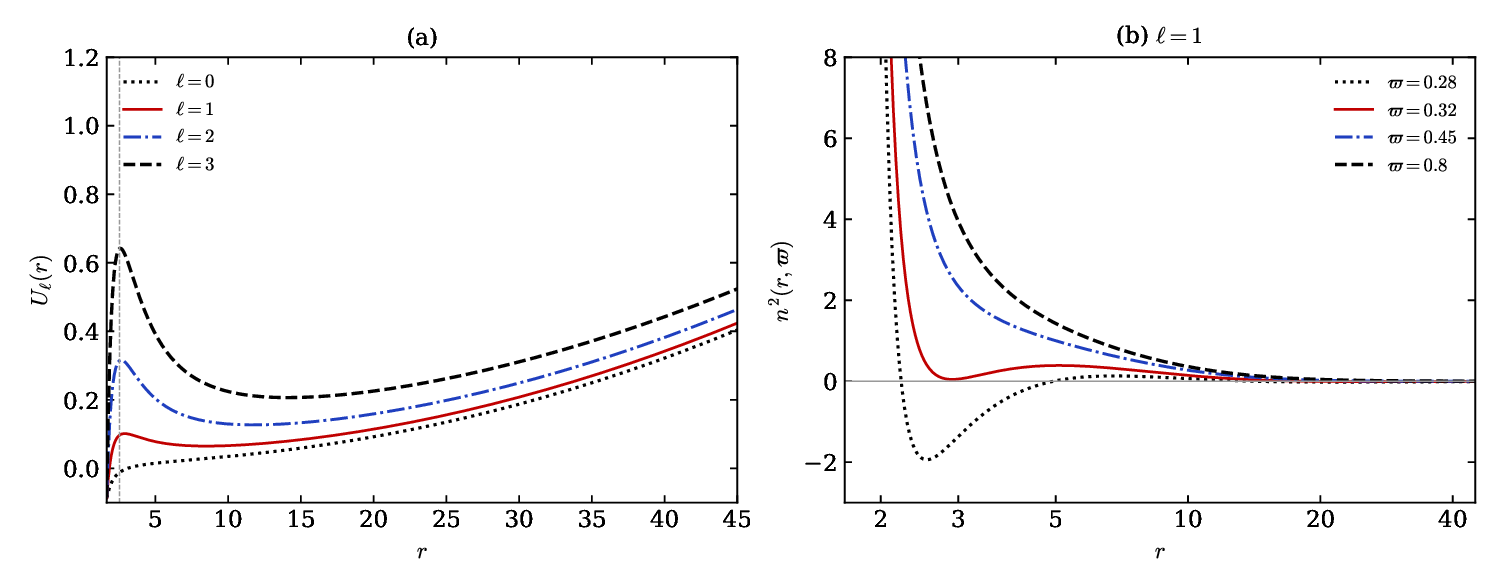}}{\MissingFigure{fig_wave_pot.eps}{0.92\textwidth}{0.20\textheight}}
\caption{(a) Effective radial potential \(U_\ell(r)\) defined in
Eq.~(\ref{Uwave}) for \(\ell=0,1,2,3\); the vertical dashed line marks the
photon-sphere radius \(r_{\rm ph}\). (b) The corresponding squared effective
refractive index \(n_\ell^2(r,\varpi)\) for \(\ell=1\) at selected
frequencies. The reference parameters are those of Table~\ref{tabray}.}
\label{figwave1}
\end{figure*}

For the reference parameter set, the numerical profiles in
Fig.~\ref{figwave1}(a) show a centrifugal barrier for \(\ell\geq1\), followed
by a shallow minimum and the large-radius rise of \(U_\ell\) associated with
the AdS asymptotics. The \(\ell=0\) profile has no corresponding centrifugal
barrier for this parameter set. For \(\ell=1\), the reported barrier maximum
is at \(r\simeq2.89\), with
\(\varpi_{\rm top}=\sqrt{U_{\rm max}}\simeq0.319\); the minimum is near
\(r\simeq8.52\), with \(\varpi_{\rm min}\simeq0.255\).

The turning-point structure is frequency dependent. At
\(\varpi=0.28\), between these two characteristic frequencies, the equation
\(U_\ell=\varpi^2\) has three reported roots,
\(r\simeq2.23,\ 4.95,\ 13.86\). The interval
\(4.95\lesssim r\lesssim13.86\) is oscillatory and is bounded on its inner
side by an evanescent barrier and on its outer side by an evanescent region
toward the AdS boundary. A wave localized in this cavity can lose amplitude
through tunnelling toward the horizon. This is a qualitative trapping
picture, not a determination of a mode frequency: actual resonances require
solving the radial equation with ingoing horizon behavior and the appropriate
normalizable AdS boundary condition. At \(\varpi=0.32\), just above the
reported barrier-top frequency, the inner pair of turning points merges or
disappears within numerical resolution, leaving an outer turning point near
\(r\simeq18.3\). For \(\varpi=0.45\) and \(0.8\), the reported outer turning
points move to approximately \(r\simeq29.5\) and \(r\simeq56\), respectively.

\begin{figure*}[tbp]
\centering
\IfFileExists{fig_wave_n2.eps}{\includegraphics[width=0.94\textwidth]{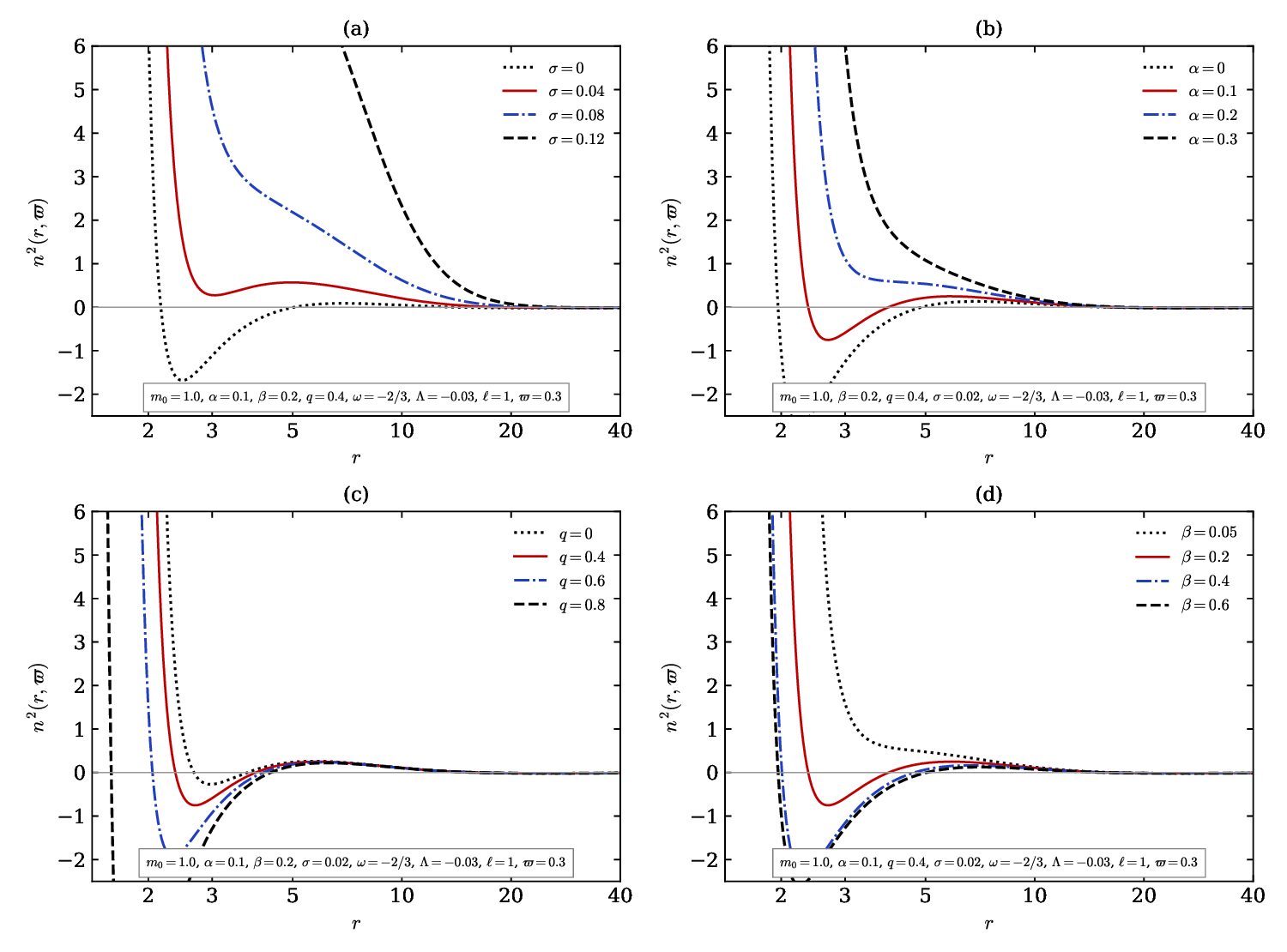}}{\MissingFigure{fig_wave_n2.eps}{0.92\textwidth}{0.20\textheight}}
\caption{Squared effective refractive index \(n_\ell^2(r,\varpi)\) for
\(\ell=1\) and \(\varpi=0.3\), for variations of (a) \(\sigma\), (b)
\(\alpha\), (c) \(q\), and (d) \(\beta\). The remaining parameters are fixed
to the values stated in the corresponding panels.}
\label{figwave2}
\end{figure*}

Figure~\ref{figwave2} compares the radial propagation regions at fixed
frequency. For the parameter ranges displayed, increasing \(\sigma\) or
\(\alpha\) lowers the barrier and can remove the evanescent interval around
the photon sphere; the plotted scans report this interval absent for
\(\sigma\geq0.04\) or \(\alpha\geq0.2\). Increasing \(q\) or \(\beta\) raises
the barrier over the displayed ranges and enlarges the forbidden region.
These are conditional numerical trends for the stated background and
frequency, not universal monotonicity statements for arbitrary parameters.
At large radius the curves approach the common asymptotic behavior in
Eq.~(\ref{nAdS}), because the leading term is set by \(\Lambda\) and
\(\varpi\), rather than by the finite-radius matter-sector details.

\begin{figure*}[tbp]
\centering
\IfFileExists{fig_wave_eik.eps}{\includegraphics[width=0.94\textwidth]{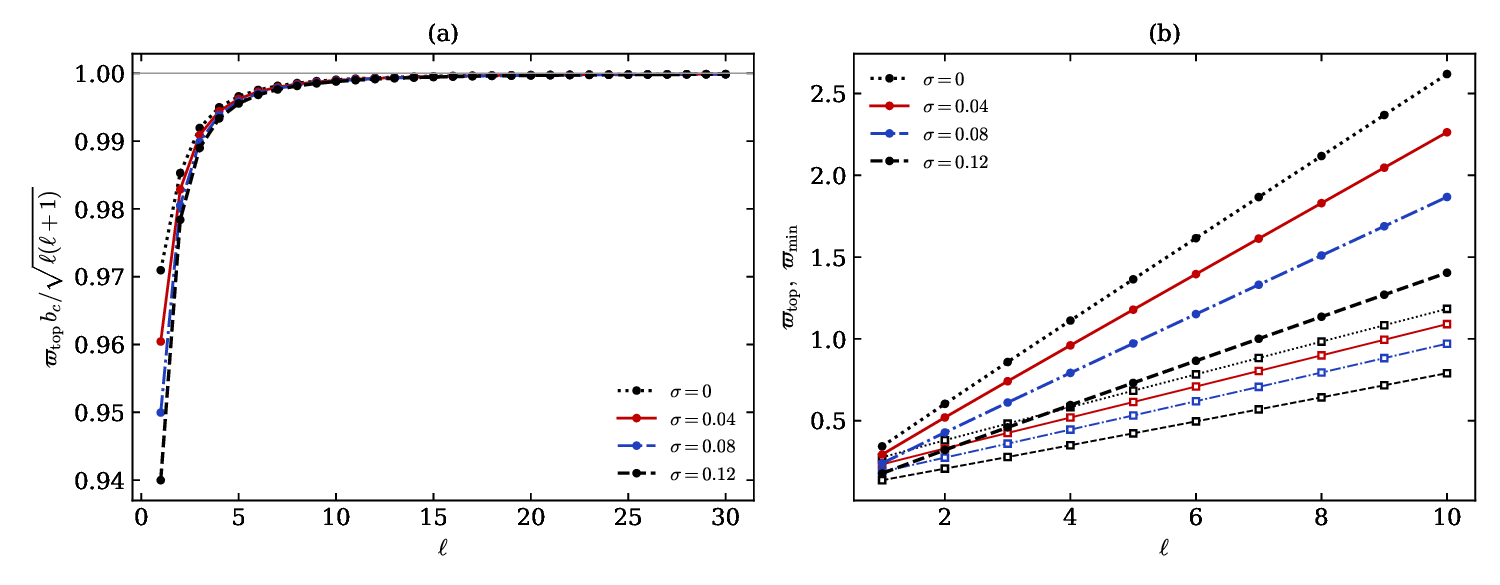}}{\MissingFigure{fig_wave_eik.eps}{0.92\textwidth}{0.20\textheight}}
\caption{(a) Barrier-top frequency normalized by its eikonal estimate,
\(\varpi_{\rm top}b_c/\sqrt{\ell(\ell+1)}\), versus \(\ell\) for selected
values of \(\sigma\). (b) Barrier-top frequency \(\varpi_{\rm top}\)
(filled symbols) and cavity-minimum frequency \(\varpi_{\rm min}\) (open
symbols) versus \(\ell\). Other parameters are fixed to their reference
values.}
\label{figwave3}
\end{figure*}

The eikonal limit is obtained by taking \(\ell\gg1\) at fixed background
parameters. In that limit the angular term dominates the derivative terms in
\(U_\ell\):
\begin{equation}
U_\ell(r)=\ell(\ell+1)\frac{f(r)}{r^2}+O(\ell^0).
\label{Ueikonal}
\end{equation}
The leading term is proportional to the null-geodesic effective potential
\(f(r)/r^2\), whose exterior maximum occurs at the photon sphere when the
relevant unstable circular null orbit exists. Since
\(b_c=r_{\rm ph}/\sqrt{f(r_{\rm ph})}\), the barrier-top frequency obeys
\begin{equation}
\varpi_{\rm top}
\simeq\frac{\sqrt{\ell(\ell+1)}}{b_c},
\qquad \ell\gg1.
\label{eikonal}
\end{equation}
Equivalently, the dimensionless ratio
\(\varpi_{\rm top}b_c/\sqrt{\ell(\ell+1)}\) tends to unity. This is the wave-optical limit of the capture threshold obtained in Sec.~\ref{sec:rayoptics}, and is the standard photon-sphere/eikonal correspondence \cite{Cardoso2009}. It fixes the leading real barrier-top scale, not the complex quasinormal frequencies or their damping rates.

For the numerical data shown in Fig.~\ref{figwave3}(a), the normalized ratio
is reported to be within about \(1\%\) of unity by \(\ell=4\) and within
about \(0.1\%\) by \(\ell=10\) for the displayed \(\sigma\) values. Such
finite-\(\ell\) accuracy is background- and definition-dependent; it should
be read as a result of the plotted scan rather than as a general error bound.
The remaining low-multipole deviations arise from the \(O(\ell^0)\) terms in
Eq.~(\ref{Ueikonal}), including
\(ff''/2+ff'/r-(f')^2/4\), which are absent from the null-geodesic potential.

Figure~\ref{figwave3}(b) also shows how the interval between the
cavity-minimum and barrier-top frequencies varies with \(\ell\) and
\(\sigma\). For \(\ell=1\), the reported interval shifts from
\(0.276<\varpi<0.343\) at \(\sigma=0\) to
\(0.136<\varpi<0.178\) at \(\sigma=0.12\), for the other parameters held at
their reference values. These intervals identify candidate frequencies for
partial trapping in the plotted effective potential. Establishing long-lived
modes, however, requires a complex-frequency calculation with ingoing
behavior at the future horizon and a normalizable condition at the AdS
boundary. We do not claim such a spectrum here; determining it is a separate
boundary-value problem.

\section{Conclusions}
\label{sec:conclusions}

We have analyzed a four-dimensional static, spherically symmetric
Maxwell-charged black-hole geometry with a Letelier string cloud, a
phenomenological PFDM source, and a Kiselev-type quintessence term. The
metric function follows from the specified effective stress tensors, and
the Ricci and Kretschmann scalars provide complementary diagnostics of its
curvature. Horizon locations and extremality must be determined from
$f(r_+)=0$ and the double-root conditions $f(r_e)=f'(r_e)=0$; parameter
trends reported here apply only to the stated scans.

The thermodynamic quantities satisfy the first-law identity for the
horizon-derived mass function at fixed external parameters. In extended
phase space, the corresponding conjugates and weighted Smarr identity
follow from the same function. These are internal thermodynamic
consistency results. Because the PFDM logarithm and quintessence term affect
the asymptotic behavior, identifying this function with a conserved AdS
mass requires a separate boundary-charge and normalization analysis.
Similarly, heat-capacity signs diagnose local thermal response, whereas
global phase preference requires a free-energy comparison between
admissible branches in a specified ensemble.

For the explicitly stated rectangular cycle, the calculation gives
$W=0.095000$ and $\eta\simeq15.39\%$, with the first-law identity
$Q_{\rm H}-Q_{\rm C}=W$ satisfied to the precision reported. These values
belong to the selected cycle and horizon-thermodynamic prescription; they
are not a universal efficiency enhancement. The benchmark value
$\beta=0.2$ also corresponds to negative PFDM density under the convention
used here. A positive-density PFDM interpretation instead requires
$\beta<0$, and the source and parameter domain must be specified whenever
the results are assigned a physical interpretation.

The optical analysis yields the photon-sphere condition, critical impact
parameter, finite-observer shadow angle, and representative ray trajectories.
The cosmological-constant term cancels from the circular-null-orbit equation
for this metric ansatz, but it remains in the critical impact parameter and
in the finite-radius angle. Consequently, the predicted angle depends on
the observer's location, and a single shadow measurement cannot determine
all model parameters independently.

For scalar waves, the radial Helmholtz representation identifies
frequency-dependent barriers and turning points. Its large-$\ell$ barrier
scale obeys
$\varpi_{\rm top}\simeq\sqrt{\ell(\ell+1)}/b_c$, connecting the leading
wave barrier to the null-geodesic capture scale. The effective potential
and turning-point structure do not determine complex quasinormal
frequencies or damping rates; those require solving the radial
boundary-value problem with ingoing horizon behavior and a specified
normalizable AdS-boundary condition.

The results are therefore a parameter-specific, cross-sector analysis of
the adopted phenomenological metric, not an observational constraint or a
complete spectral study. The principal qualifications are the PFDM
density sign, the unresolved asymptotic-charge interpretation of the
horizon-derived mass, and the absence of a computed complex-frequency
spectrum. These distinctions delimit what follows from the calculations
presented here.


\begin{thebibliography}{0}%
\makeatletter
\providecommand \@ifxundefined [1]{%
 \@ifx{#1\undefined}
}%
\providecommand \@ifnum [1]{%
 \ifnum #1\expandafter \@firstoftwo
 \else \expandafter \@secondoftwo
 \fi
}%
\providecommand \@ifx [1]{%
 \ifx #1\expandafter \@firstoftwo
 \else \expandafter \@secondoftwo
 \fi
}%
\providecommand \natexlab [1]{#1}%
\providecommand \enquote  [1]{``#1''}%
\providecommand \bibnamefont  [1]{#1}%
\providecommand \bibfnamefont [1]{#1}%
\providecommand \citenamefont [1]{#1}%
\providecommand \href@noop [0]{\@secondoftwo}%
\providecommand \href [0]{\begingroup \@sanitize@url \@href}%
\providecommand \@href[1]{\@@startlink{#1}\@@href}%
\providecommand \@@href[1]{\endgroup#1\@@endlink}%
\providecommand \@sanitize@url [0]{\catcode `\\12\catcode `\$12\catcode
  `\&12\catcode `\#12\catcode `\^12\catcode `\_12\catcode `\%12\relax}%
\providecommand \@@startlink[1]{}%
\providecommand \@@endlink[0]{}%
\providecommand \url  [0]{\begingroup\@sanitize@url \@url }%
\providecommand \@url [1]{\endgroup\@href {#1}{\urlprefix }}%
\providecommand \urlprefix  [0]{URL }%
\providecommand \Eprint [0]{\href }%
\providecommand \doibase [0]{https://doi.org/}%
\providecommand \selectlanguage [0]{\@gobble}%
\providecommand \bibinfo  [0]{\@secondoftwo}%
\providecommand \bibfield  [0]{\@secondoftwo}%
\providecommand \translation [1]{[#1]}%
\providecommand \BibitemOpen [0]{}%
\providecommand \bibitemStop [0]{}%
\providecommand \bibitemNoStop [0]{.\EOS\space}%
\providecommand \EOS [0]{\spacefactor3000\relax}%
\providecommand \BibitemShut  [1]{\csname bibitem#1\endcsname}%
\let\auto@bib@innerbib\@empty
\end{thebibliography}%


\begin{thebibliography}{999}

\bibitem{Penrose1965}
R.~Penrose, ``Gravitational collapse and space-time singularities,''
Phys.\ Rev.\ Lett.\ \textbf{14}, 57 (1965).
\url{https://doi.org/10.1103/PhysRevLett.14.57}

\bibitem{Hawking1970}
S.~W.~Hawking and R.~Penrose, ``The singularities of gravitational collapse and cosmology,''
Proc.\ Roy.\ Soc.\ Lond.\ A \textbf{314}, 529--548 (1970).
\url{https://doi.org/10.1098/rspa.1970.0021}

\bibitem{Berti2015}
E.~Berti et al., ``Testing general relativity with present and future astrophysical observations,''
Class.\ Quantum Grav.\ \textbf{32}, 243001 (2015).
\url{https://doi.org/10.1088/0264-9381/32/24/243001}

\bibitem{LIGO2016}
B.~P.~Abbott et al. (LIGO Scientific Collaboration and Virgo Collaboration),
``Observation of gravitational waves from a binary black hole merger,''
Phys.\ Rev.\ Lett.\ \textbf{116}, 061102 (2016).
\url{https://doi.org/10.1103/PhysRevLett.116.061102}

\bibitem{LIGO2017}
B.~P.~Abbott et al. (LIGO Scientific Collaboration and Virgo Collaboration),
``GW170814: A three-detector observation of gravitational waves from a binary black hole coalescence,''
Phys.\ Rev.\ Lett.\ \textbf{119}, 141101 (2017).
\url{https://doi.org/10.1103/PhysRevLett.119.141101}

\bibitem{LIGO2020}
R.~Abbott et al. (LIGO Scientific Collaboration and Virgo Collaboration),
``Population properties of compact objects from the second LIGO--Virgo Gravitational-Wave Transient Catalog,''
Astrophys.\ J.\ Lett.\ \textbf{896}, L44 (2020).
\url{https://doi.org/10.3847/2041-8213/ab960f}

\bibitem{EHT2019}
K.~Akiyama et al. (Event Horizon Telescope Collaboration),
``First M87 Event Horizon Telescope results. I. The shadow of the supermassive black hole,''
Astrophys.\ J.\ Lett.\ \textbf{875}, L1 (2019).
\url{https://doi.org/10.3847/2041-8213/ab0ec7}

\bibitem{EHT2022}
K.~Akiyama et al. (Event Horizon Telescope Collaboration),
``First Sagittarius A* Event Horizon Telescope results. I. The shadow of the supermassive black hole in the center of the Milky Way,''
Astrophys.\ J.\ Lett.\ \textbf{930}, L12 (2022).
\url{https://doi.org/10.3847/2041-8213/ac6674}

\bibitem{Medved2004}
A.~J.~M.~Medved, D.~Martin, and M.~Visser,
``Dirty black holes: Quasinormal modes,''
Class.\ Quantum Grav.\ \textbf{21}, 1393--1405 (2004).
\url{https://doi.org/10.1088/0264-9381/21/6/008}

\bibitem{Narayan2005}
R.~Narayan, ``Black holes in astrophysics,''
New J.\ Phys.\ \textbf{7}, 199 (2005).
\url{https://doi.org/10.1088/1367-2630/7/1/199}

\bibitem{Barausse2014}
E.~Barausse, V.~Cardoso, and P.~Pani,
``Can environmental effects spoil precision gravitational-wave astrophysics?''
Phys.\ Rev.\ D \textbf{89}, 104059 (2014).
\url{https://doi.org/10.1103/PhysRevD.89.104059}

\bibitem{Macedo2016}
C.~F.~B.~Macedo, L.~C.~S.~Leite, and L.~C.~B.~Crispino,
``Absorption by dirty black holes: Null geodesics and scalar waves,''
Phys.\ Rev.\ D \textbf{93}, 024027 (2016).
\url{https://doi.org/10.1103/PhysRevD.93.024027}

\bibitem{Macedo2024}
C.~F.~B.~Macedo, J.~L.~Rosa, and D.~Rubiera-Garcia,
``Optical appearance of black holes surrounded by a dark matter halo,''
JCAP \textbf{07}, 046 (2024).
\url{https://doi.org/10.1088/1475-7516/2024/07/046}

\bibitem{Cardoso2022}
V.~Cardoso, K.~Destounis, F.~Duque, R.~P.~Macedo, and A.~Maselli,
``Black holes in galaxies: Environmental impact on gravitational-wave generation and propagation,''
Phys.\ Rev.\ D \textbf{105}, L061501 (2022).
\url{https://doi.org/10.1103/PhysRevD.105.L061501}

\bibitem{Cunha2015}
P.~V.~P.~Cunha, C.~A.~R.~Herdeiro, E.~Radu, and H.~F.~Rúnarsson,
``Shadows of Kerr black holes with scalar hair,''
Phys.\ Rev.\ Lett.\ \textbf{115}, 211102 (2015).
\url{https://doi.org/10.1103/PhysRevLett.115.211102}

\bibitem{Konoplya2021}
R.~A.~Konoplya, ``Black holes in galactic centers: Quasinormal ringing, grey-body factors and Unruh temperature,''
Phys.\ Lett.\ B \textbf{823}, 136734 (2021).
\url{https://doi.org/10.1016/j.physletb.2021.136734}

\bibitem{Rahmatov2025}
B.~Rahmatov, I.~Egamberdiev, S.~Murodov, J.~Rayimbaev, I.~Ibragimov,
E.~Davletov, and S.~Djumanov,
``Gravitational lensing by black holes surrounded by PFDM in Kalb--Ramond gravity in plasma medium,''
Phys.\ Dark Univ.\ \textbf{50}, 102152 (2025).
\url{https://doi.org/10.1016/j.dark.2025.102152}

\bibitem{Hoshimov2025}
H.~Hoshimov, S.~Orzuev, F.~Atamurotov, and A.~Abdujabbarov,
``Gravitational weak lensing and shadow radius of black holes in Horndeski gravity: Effects of plasma and scalar modifications,''
Ann.\ Phys.\ \textbf{482}, 170209 (2025).
\url{https://doi.org/10.1016/j.aop.2025.170209}

\bibitem{Bakhodirov2025}
A.~Bakhodirov, B.~Shodikulov, F.~Atamurotov, A.~Abdujabbarov,
C.~Zhou, and C.~Yuan,
``Optical signatures of hairy black holes surrounded by plasma: Shadows and weak lensing,''
Phys.\ Dark Univ.\ \textbf{49}, 102008 (2025).
\url{https://doi.org/10.1016/j.dark.2025.102008}

\bibitem{Letelier1979}
P.~S.~Letelier,
``Clouds of strings in general relativity,''
Phys.\ Rev.\ D \textbf{20}, 1294 (1979).
\url{https://doi.org/10.1103/PhysRevD.20.1294}

\bibitem{Ghosh2014}
S.~G.~Ghosh, U.~Papnoi, and S.~D.~Maharaj,
``Cloud of strings in third order Lovelock gravity,''
Phys.\ Rev.\ D \textbf{90}, 044068 (2014).
\url{https://doi.org/10.1103/PhysRevD.90.044068}

\bibitem{Toledo2018}
J.~de M.~Toledo and V.~B.~Bezerra,
``Black holes with cloud of strings and quintessence in Lovelock gravity,''
Eur.\ Phys.\ J.\ C \textbf{78}, 534 (2018).
\url{https://doi.org/10.1140/epjc/s10052-018-6001-z}

\bibitem{Waseem2023}
A.~Waseem, F.~Javed, M.~Z.~Gul, G.~Mustafa, and A.~Errehymy,
``Impact of quintessence and cloud of strings on self-consistent $d$-dimensional charged thin-shell wormholes,''
Eur.\ Phys.\ J.\ C \textbf{83}, 1088 (2023).
\url{https://doi.org/10.1140/epjc/s10052-023-12239-7}

\bibitem{Rodrigues2022}
M.~E.~Rodrigues and H.~A.~Vieira,
``Bardeen solution with a cloud of strings,''
Phys.\ Rev.\ D \textbf{106}, 084015 (2022).
\url{https://doi.org/10.1103/PhysRevD.106.084015}

\bibitem{Muniz2025}
C.~R.~Muniz, J.~A.~Rebouças, L.~T.~De Oliveira, F.~T.~B.~Sampaio, and F.~B.~Lustosa,
``Regularized black hole solution from a new string cloud source,''
Phys.\ Dark Univ.\ \textbf{52}, 102272 (2026).
\url{https://doi.org/10.1016/j.dark.2026.102272}

\bibitem{Mustafa2022}
G.~Mustafa, F.~Atamurotov, I.~Hussain, S.~Shaymatov, and A.~Övgün,
``Shadows and gravitational weak lensing by the Schwarzschild black hole in the string cloud background with quintessential field,''
Chin.\ Phys.\ C \textbf{46}, 125107 (2022).
\url{https://doi.org/10.1088/1674-1137/ac917f}

\bibitem{Atamurotov2022}
F.~Atamurotov, I.~Hussain, G.~Mustafa, and K.~Jusufi,
``Shadow and quasinormal modes of the Kerr--Newman--Kiselev--Letelier black hole,''
Eur.\ Phys.\ J.\ C \textbf{82}, 831 (2022).
\url{https://doi.org/10.1140/epjc/s10052-022-10782-3}

\bibitem{Rincon2018}
A.~Rincón and G.~Panotopoulos,
``Greybody factors and quasinormal modes for a nonminimally coupled scalar field in a cloud of strings in $(2+1)$-dimensional background,''
Eur.\ Phys.\ J.\ C \textbf{78}, 858 (2018).
\url{https://doi.org/10.1140/epjc/s10052-018-6352-5}

\bibitem{Li2021}
Z.~Li and T.~Zhou,
``Kerr black hole surrounded by a cloud of strings and its weak gravitational lensing in Rastall gravity,''
Phys.\ Rev.\ D \textbf{104}, 104044 (2021).
\url{https://doi.org/10.1103/PhysRevD.104.104044}

\bibitem{Belhaj2022}
A.~Belhaj and Y.~Sekhmani,
``Shadows of rotating quintessential black holes in Einstein--Gauss--Bonnet gravity with a cloud of strings,''
Gen.\ Relativ.\ Gravit.\ \textbf{54}, 17 (2022).
\url{https://doi.org/10.1007/s10714-022-02902-x}

\bibitem{He2022}
A.~He, J.~Tao, Y.~Xue, and L.~Zhang,
``Shadow and photon sphere of black hole in clouds of strings and quintessence,''
Chin.\ Phys.\ C \textbf{46}, 065102 (2022).
\url{https://doi.org/10.1088/1674-1137/ac56cf}

\bibitem{Toledo2019}
J.~M.~Toledo and V.~B.~Bezerra,
``Some remarks on the thermodynamics of charged AdS black holes with cloud of strings and quintessence,''
Eur.\ Phys.\ J.\ C \textbf{79}, 110 (2019).
\url{https://doi.org/10.1140/epjc/s10052-019-6616-8}


\bibitem{Toledo2019b}
J.~M.~Toledo and V.~B.~Bezerra,
``Some remarks on the thermodynamics of charged AdS black holes with cloud of strings and quintessence,''
Eur.\ Phys.\ J.\ C \textbf{79}, 110 (2019).
\url{https://doi.org/10.1140/epjc/s10052-019-6616-8}

\bibitem{Singh2020}
D.~V.~Singh, S.~G.~Ghosh, and S.~D.~Maharaj,
``Clouds of strings in 4D Einstein--Gauss--Bonnet black holes,''
Phys.\ Dark Univ.\ \textbf{30}, 100730 (2020).
\url{https://doi.org/10.1016/j.dark.2020.100730}

\bibitem{Rodrigues2022b}
M.~E.~Rodrigues and M.~V.~de S.~Silva,
``Embedding regular black holes and black bounces in a cloud of strings,''
Phys.\ Rev.\ D \textbf{106}, 084016 (2022).
\url{https://doi.org/10.1103/PhysRevD.106.084016}

\bibitem{Sadeghi2024}
J.~Sadeghi, S.~N.~Gashti, I.~Sakallı, and B.~Pourhassan,
``Weak gravity conjecture of charged-rotating-AdS black hole surrounded by quintessence and string cloud,''
Nucl.\ Phys.\ B \textbf{1004}, 116581 (2024).
\url{https://doi.org/10.1016/j.nuclphysb.2024.116581}

\bibitem{Dias2019} M. M. Dias e Costa, J. M. Toledo, and V. B. Bezerra,
The Letelier spacetime with quintessence: solution, thermodynamics and Hawking radiation,
Int. J. Mod. Phys. D \textbf{28}, 1950074 (2019).
\url{https://doi.org/10.1142/S0218271819500743}

\bibitem{Zafar2025} U. Zafar, K. Bamba, T. Rasheed, and K. Bhattacharya,
Thermodynamic analysis of black holes with cloud of strings and quintessence via Barrow entropy,
Phys. Lett. B \textbf{864}, 139446 (2025).
\url{https://doi.org/10.1016/j.physletb.2025.139446}

\bibitem{AhmedArX2025} F. Ahmed and E. O. Silva,
Letelier-AdS black hole surrounded by a perfect fluid dark matter in the presence of quintessence,
arXiv:2510.21948 [gr-qc] (2025).
\url{https://arxiv.org/abs/2510.21948}

\bibitem{PFDM1}
F.~Rahaman, K.~K.~Nandi, A.~Bhadra, M.~Kalam, and K.~Chakraborty,
``Perfect fluid dark matter,''
Phys.\ Lett.\ B \textbf{694}, 10--15 (2010).
\url{https://doi.org/10.1016/j.physletb.2010.09.038}

\bibitem{PFDM2}
J.~Barranco, A.~Bernal, and D.~Núñez,
``Dark matter equation of state from rotational curves of galaxies,''
Mon.\ Not.\ R.\ Astron.\ Soc.\ \textbf{449}, 403--413 (2015).
\url{https://doi.org/10.1093/mnras/stv302}

\bibitem{PFDM3}
A.~A.~Potapov, G.~M.~Garipova, and K.~K.~Nandi,
``Revisiting perfect fluid dark matter: Observational constraints from our galaxy,''
Phys.\ Lett.\ B \textbf{753}, 140--146 (2016).
\url{https://doi.org/10.1016/j.physletb.2015.11.057}

\bibitem{BHPFDM1}
V.~V.~Kiselev,
``Quintessence and black holes,''
Class.\ Quantum Grav.\ \textbf{20}, 1187--1198 (2003).
\url{https://doi.org/10.1088/0264-9381/20/6/310}

\bibitem{BHPFDM2}
M.-H.~Li and K.-C.~Yang,
``Galactic dark matter in the phantom field,''
Phys.\ Rev.\ D \textbf{86}, 123015 (2012).
\url{https://doi.org/10.1103/PhysRevD.86.123015}

\bibitem{BHPFDM3}
Y.~Heydarzade and F.~Darabi,
``Black hole solutions surrounded by perfect fluid in Rastall theory,''
Phys.\ Lett.\ B \textbf{771}, 365--373 (2017).
\url{https://doi.org/10.1016/j.physletb.2017.05.064}

\bibitem{BHPFDM4}
A.~Das, A.~Saha, and S.~Gangopadhyay,
``Investigation of circular geodesics in a rotating charged black hole in the presence of perfect fluid dark matter,''
Class.\ Quantum Grav.\ \textbf{38}, 065015 (2021).
\url{https://doi.org/10.1088/1361-6382/abd95b}

\bibitem{ProPFDM1}
Z.~Xu, X.~Hou, and J.~Wang,
``Kerr--anti-de Sitter/de Sitter black hole in perfect fluid dark matter background,''
Class.\ Quantum Grav.\ \textbf{35}, 115003 (2018).
\url{https://doi.org/10.1088/1361-6382/aabcb6}

\bibitem{ProPFDM2}
Z.~Xu, X.~Hou, X.~Gong, and J.~Wang,
``Kerr--Newman-AdS black hole surrounded by perfect fluid matter in Rastall gravity,''
Eur.\ Phys.\ J.\ C \textbf{78}, 513 (2018).
\url{https://doi.org/10.1140/epjc/s10052-018-5991-x}

\bibitem{ProPFDM3}
M.~Rizwan, M.~Jamil, and K.~Jusufi,
``Distinguishing a Kerr-like black hole and a naked singularity in perfect fluid dark matter via precession frequencies,''
Phys.\ Rev.\ D \textbf{99}, 024050 (2019).
\url{https://doi.org/10.1103/PhysRevD.99.024050}

\bibitem{ProPFDM4}
S.~Haroon, M.~Jamil, K.~Jusufi, K.~Lin, and R.~B.~Mann,
``Shadow and deflection angle of rotating black holes in perfect fluid dark matter with a cosmological constant,''
Phys.\ Rev.\ D \textbf{99}, 044015 (2019).
\url{https://doi.org/10.1103/PhysRevD.99.044015}

\bibitem{ProPFDM5}
B.~Narzilloev, J.~Rayimbaev, S.~Shaymatov, A.~Abdujabbarov, B.~Ahmedov, and C.~Bambi,
``Dynamics of test particles around a Bardeen black hole surrounded by perfect fluid dark matter,''
Phys.\ Rev.\ D \textbf{102}, 104062 (2020).
\url{https://doi.org/10.1103/PhysRevD.102.104062}

\bibitem{ProPFDM6}
K.~Jusufi,
``Quasinormal modes of black holes surrounded by dark matter and their connection with the shadow radius,''
Phys.\ Rev.\ D \textbf{101}, 084055 (2020).
\url{https://doi.org/10.1103/PhysRevD.101.084055}

\bibitem{ProPFDM7}
B.~Narzilloev, J.~Rayimbaev, S.~Shaymatov, A.~Abdujabbarov, B.~Ahmedov, and C.~Bambi,
``Dynamics of test particles around a Bardeen black hole surrounded by perfect fluid dark matter,''
Phys.\ Rev.\ D \textbf{102}, 104062 (2020).
\url{https://doi.org/10.1103/PhysRevD.102.104062}

\bibitem{ProPFDM8}
F.~Atamurotov, A.~Abdujabbarov, and W.-B.~Han,
``Effect of plasma on gravitational lensing by a Schwarzschild black hole immersed in perfect fluid dark matter,''
Phys.\ Rev.\ D \textbf{104}, 084015 (2021).
\url{https://doi.org/10.1103/PhysRevD.104.084015}

\bibitem{ProPFDM9}
T.-C.~Ma, H.-X.~Zhang, P.-Z.~He, H.-R.~Zhang, Y.~Chen, and J.-B.~Deng,
``Shadow cast by a rotating and nonlinear magnetic-charged black hole in perfect fluid dark matter,''
Mod.\ Phys.\ Lett.\ A \textbf{36}, 2150112 (2021).
\url{https://doi.org/10.1142/S0217732321501121}

\bibitem{ProPFDM10}
A.~Das, A.~Saha, and S.~Gangopadhyay,
``Study of circular geodesics and shadow of rotating charged black hole surrounded by perfect fluid dark matter immersed in plasma,''
Class.\ Quantum Grav.\ \textbf{39}, 075005 (2022).
\url{https://doi.org/10.1088/1361-6382/ac50ed}

\bibitem{ProPFDM11}
X.~Zhou, Y.~Xue, B.~Mu, and J.~Tao,
``Temporal and spatial chaos of RN-AdS black holes immersed in perfect fluid dark matter,''
Phys.\ Dark Univ.\ \textbf{39}, 101168 (2023).
\url{https://doi.org/10.1016/j.dark.2023.101168}

\bibitem{ProPFDM12}
X.-J.~Gao, X.-K.~Yan, Y.~Yin, and Y.-P.~Hu,
``Gravitational lensing by a charged spherically symmetric black hole immersed in thin dark matter,''
Eur.\ Phys.\ J.\ C \textbf{83}, 281 (2023).
\url{https://doi.org/10.1140/epjc/s10052-023-11414-0}

\bibitem{ProPFDM13}
M.~Heydari-Fard, S.~G.~Honarvar, and M.~Heydari-Fard,
``Thin accretion disc luminosity and its image around rotating black holes in perfect fluid dark matter,''
Mon.\ Not.\ R.\ Astron.\ Soc.\ \textbf{521}, 708--721 (2023).
\url{https://doi.org/10.1093/mnras/stad145}

\bibitem{ProPFDM14}
G.~Abbas and R.~H.~Ali,
``Thermal fluctuations, quasi-normal modes and phase transition of the charged AdS black hole with perfect fluid dark matter,''
Eur.\ Phys.\ J.\ C \textbf{83}, 407 (2023).
\url{https://doi.org/10.1140/epjc/s10052-023-11580-1}

\bibitem{ProPFDM15}
X.~Liang, Y.-P.~Hu, C.-H.~Wu, and Y.-S.~An,
``Thermodynamics and evaporation of perfect fluid dark matter black hole in phantom background,''
Eur.\ Phys.\ J.\ C \textbf{83}, 1009 (2023).
\url{https://doi.org/10.1140/epjc/s10052-023-12200-8}

\bibitem{ProPFDM16}
X.~Yang,
``Observational appearance of the spherically symmetric black hole in PFDM,''
Phys.\ Dark Univ.\ \textbf{44}, 101467 (2024).
\url{https://doi.org/10.1016/j.dark.2024.101467}

\bibitem{ProPFDM17}
Y.~Feng, A.~Ditta, G.~Mustafa, S.~K.~Maurya, A.~Mahmood, and F.~Atamurotov,
``Non-commutative Schwarzschild black hole surrounded by perfect fluid dark matter: Plasma lensing and thermodynamics analysis,''
Nucl.\ Phys.\ B \textbf{1008}, 116713 (2024).
\url{https://doi.org/10.1016/j.nuclphysb.2024.116713}

\bibitem{ProPFDM18}
C.-K.~Qiao and M.~Zhou,
``Gravitational lensing of Schwarzschild and charged black holes immersed in perfect fluid dark matter halo,''
JCAP \textbf{12}, 005 (2023).
\url{https://doi.org/10.1088/1475-7516/2023/12/005}

\bibitem{ProPFDM19}
F.~Javed, S.~Mumtaz, A.~Raza, B.~Almutairi, G.~Mustafa, and G.~Fatima,
``Greybody factor of uncharged black hole in symmetric teleparallel gravity,''
Phys.\ Dark Univ.\ \textbf{46}, 101656 (2024).
\url{https://doi.org/10.1016/j.dark.2024.101656}

\bibitem{ProPFDM20}
Y.~Y.~Sekhmani, S.~K.~Maurya, M.~K.~Jasim, and \.{I}.~Sakall{\i},
``Thermodynamics and phase transition of anti de Sitter black holes with ModMax nonlinear electrodynamics and perfect fluid dark matter,''
Eur.\ Phys.\ J.\ C \textbf{85}, 229 (2025).
\url{https://doi.org/10.1140/epjc/s10052-025-13932-5}

\bibitem{ProPFDM21}
M.~R.~Shahzad, G.~Abbas, T.~Zhu, R.~H.~Ali, A.~Ashraf, and B.~S.~Al-Kahtani,
``Optical features and thermal properties of charged AdS Euler--Heisenberg black hole surrounded by perfect fluid dark matter,''
Eur.\ Phys.\ J.\ C \textbf{85}, 164 (2025).
\url{https://doi.org/10.1140/epjc/s10052-025-13876-w}

\bibitem{ProPFDM22}
Q.~Tan, D.~Liu, J.~Liang, and Z.-W.~Long,
``Testing black holes in a perfect fluid dark matter environment using quasinormal modes,''
Eur.\ Phys.\ J.\ C \textbf{85}, 687 (2025).
\url{https://doi.org/10.1140/epjc/s10052-025-14407-3}

\bibitem{ProPFDM23}
S.~Kumar Jha,
``Black hole surrounded by perfect fluid dark matter with a background Kalb--Ramond field,''
J. Cosmol. Astropart. Phys. \textbf{09}, 069 (2025).
\url{https://doi.org/10.1088/1475-7516/2025/09/069}

\bibitem{ProPFDM24}
B.~Eslam Panah, B.~Hamil, and M.~E.~Rodrigues,
``Born--Infeld AdS black holes surrounded by perfect fluid dark matter,''
Eur.\ Phys.\ J.\ C \textbf{86}, 632 (2026).
\url{https://doi.org/10.1140/epjc/s10052-026-15823-9}

\bibitem{ProPFDM25}
B.~Hamil and B.~C.~L{\"u}tf{\"u}o\u{g}lu,
``Schwarzschild black hole surrounded by a cloud of strings in the background of perfect fluid dark matter,''
Chin.\ Phys.\ C \textbf{49}, 025107 (2025).
\url{https://doi.org/10.1088/1674-1137/ad9894}

\bibitem{ProPFDM26}
H.~Feng, Z.~Cai, H.-P.~Yan, R.-J.~Yang, and J.~Zhang,
Dark Universe \textbf{54}, 102388 (2026).
\url{https://doi.org/10.1016/j.dark.2026.102388}

\bibitem{ProPFDM27}
A.~Al-Badawi, F.~Ahmed, and \.{I}.~Sakall{\i},
``$P$--$V$ criticality, Joule--Thomson expansion, and holographic heat engine of charged Hayward--AdS black holes with a cloud of strings and perfect fluid dark matter,''
arXiv:2603.02276 (2026).
\url{https://arxiv.org/abs/2603.02276}

\bibitem{ProPFDM28}
F.~Ahmed, C.~E.~Romero-Figueroa, and H.~Quevedo,
Dark Universe \textbf{54}, 102377 (2026).
\url{https://doi.org/10.1016/j.dark.2026.102377}

\bibitem{ProPFDM29}
D.~J.~Gogoi, J.~Bora, and A.~{\"O}vg{\"u}n,
``Equatorial periodic orbits and gravitational wave signatures in Euler--Heisenberg black holes surrounded by perfect fluid dark matter,''
JCAP \textbf{07}, 087 (2026).
\url{https://doi.org/10.1088/1475-7516/2026/07/087}

\bibitem{ProPFDM30}
E.~O.~Silva and F.~Ahmed,
``Shadow, quasinormal modes, sparsity, and energy emission rate of Euler--Heisenberg black hole surrounded by perfect fluid dark matter,''
Phys.\ Lett.\ B \textbf{880}, 140755 (2026).
\url{https://doi.org/10.1016/j.physletb.2026.140755}

\bibitem{Bekenstein1973}
J.~D.~Bekenstein,
``Black holes and entropy,''
Phys.\ Rev.\ D \textbf{7}, 2333--2346 (1973).
\url{https://doi.org/10.1103/PhysRevD.7.2333}

\bibitem{Hawking1974}
S.~W.~Hawking,
``Black hole explosions?,''
Nature \textbf{248}, 30--31 (1974).
\url{https://doi.org/10.1038/248030a0}

\bibitem{Callen1985}
H.~B.~Callen,
\textit{Thermodynamics and an Introduction to Thermostatistics}, 2nd ed.
(John Wiley \& Sons, New York, 1985).
\url{https://books.google.com/books?q=Thermodynamics+and+an+Introduction+to+Thermostatistics+Callen+1985}

\bibitem{Davies1977}
P.~C.~W.~Davies,
``The thermodynamic theory of black holes,''
Proc.\ R.\ Soc.\ Lond.\ A \textbf{353}, 499--521 (1977).
\url{https://doi.org/10.1098/rspa.1977.0047}

\bibitem{Hawking1983}
S.~W.~Hawking and D.~N.~Page,
``Thermodynamics of black holes in anti-de Sitter space,''
Commun.\ Math.\ Phys.\ \textbf{87}, 577--588 (1983).
\url{https://doi.org/10.1007/BF01208266}

\bibitem{Chamblin1999}
A.~Chamblin, R.~Emparan, C.~V.~Johnson, and R.~C.~Myers,
``Charged AdS black holes and catastrophic holography,''
Phys.\ Rev.\ D \textbf{60}, 064018 (1999).
\url{https://doi.org/10.1103/PhysRevD.60.064018}

\bibitem{Witten1998}
E.~Witten,
``Anti-de Sitter space, thermal phase transition, and confinement in gauge theories,''
Adv.\ Theor.\ Math.\ Phys.\ \textbf{2}, 505--532 (1998).
\url{https://doi.org/10.4310/ATMP.1998.v2.n3.a3}

\bibitem{Kastor2009}
D.~Kastor, S.~Ray, and J.~Traschen,
``Enthalpy and the mechanics of AdS black holes,''
Class.\ Quantum Grav.\ \textbf{26}, 195011 (2009).
\url{https://doi.org/10.1088/0264-9381/26/19/195011}

\bibitem{PLambda2}
B.~P.~Dolan,
``The cosmological constant and the black-hole equation of state,''
Class.\ Quantum Grav.\ \textbf{28}, 125020 (2011).
\url{https://doi.org/10.1088/0264-9381/28/12/020}

\bibitem{PLambda3}
B.~P.~Dolan,
``Pressure and volume in the first law of black hole thermodynamics,''
Class.\ Quantum Grav.\ \textbf{28}, 235017 (2011).
\url{https://doi.org/10.1088/0264-9381/28/23/017}

\bibitem{PLambda4}
D.~Kubiz{\v n}{\'a}k and R.~B.~Mann,
``$P$--$V$ criticality of charged AdS black holes,''
JHEP \textbf{07}, 033 (2012).
\url{https://doi.org/10.1007/JHEP07(2012)033}

\bibitem{PV1}
D.~D.~Doneva, S.~S.~Yazadjiev, K.~D.~Kokkotas, I.~Z.~Stefanov, and M.~D.~Todorov,
``Charged anti-de Sitter scalar-tensor black holes and their thermodynamic phase structure,''
Phys.\ Rev.\ D \textbf{81}, 104030 (2010).
\url{https://doi.org/10.1103/PhysRevD.81.104030}

\bibitem{PV2}
S.~Gunasekaran, D.~Kubiz{\v n}{\'a}k, and R.~B.~Mann,
``Extended phase space thermodynamics for charged and rotating black holes and Born--Infeld vacuum polarization,''
JHEP \textbf{11}, 110 (2012).
\url{https://doi.org/10.1007/JHEP11(2012)110}

\bibitem{PV3}
S.~H.~Hendi and M.~H.~Vahidinia,
``Extended phase space thermodynamics and criticality of black holes with a nonlinear source,''
Phys.\ Rev.\ D \textbf{88}, 084045 (2013).
\url{https://doi.org/10.1103/PhysRevD.88.084045}

\bibitem{PV4}
R.-G.~Cai, L.-M.~Cao, L.~Li, and R.-Q.~Yang,
``$P$--$V$ criticality in the extended phase space of Gauss--Bonnet black holes in AdS space,''
JHEP \textbf{09}, 005 (2013).
\url{https://doi.org/10.1007/JHEP09(2013)005}

\bibitem{PV5}
S.~Chen, X.~Liu, C.~Liu, and J.~Jing,
``$P$--$V$ criticality of an AdS black hole in $f(R)$ gravity,''
Chin.\ Phys.\ Lett.\ \textbf{30}, 060401 (2013).
\url{https://doi.org/10.1088/0256-307X/30/6/060401}

\bibitem{PV6}
S.-W.~Wei and Y.-X.~Liu,
``Triple points and phase diagrams in the extended phase space of charged Gauss--Bonnet black holes in AdS space,''
Phys.\ Rev.\ D \textbf{90}, 044057 (2014).
\url{https://doi.org/10.1103/PhysRevD.90.044057}

\bibitem{PV7}
B.~Mirza and Z.~Sherkatghanad,
``Phase transition, critical behavior, and critical exponents of Myers--Perry black holes,''
Phys.\ Rev.\ D \textbf{90}, 084006 (2014).
\url{https://doi.org/10.1103/PhysRevD.90.084006}

\bibitem{PV8} J. Xu, L. M. Cao, and Y. P. Hu,
P--V criticality in the extended phase space of black holes in massive gravity,
Phys. Rev. D \textbf{91}, 124033 (2015).
\url{https://doi.org/10.1103/PhysRevD.91.124033}

\bibitem{PV9} M. Zhang, Z.-Y. Yang, D.-C. Zou, W. Xu, and R.-H. Yue,
$P$-$V$ criticality of AdS black hole in the Einstein--Maxwell-power-Yang--Mills gravity,
Gen. Relativ. Gravit. \textbf{47}, 14 (2015).
\href{https://doi.org/10.1007/s10714-015-1851-2}{https://doi.org/10.1007/s10714-015-1851-2}

\bibitem{PV10} S. H. Hendi, S. Panahiyan, and B. Eslam Panah,
$P$-$V$ criticality and geometrical thermodynamics of black holes with Born--Infeld type nonlinear electrodynamics,
Int. J. Mod. Phys. D \textbf{25}, 1650010 (2016).
\url{https://doi.org/10.1142/S0218271816500103}

\bibitem{PV11} S. H. Hendi, B. Eslam Panah, and S. Panahiyan,
Massive charged BTZ black holes in asymptotically (a)dS spacetimes,
J. High Energy Phys. \textbf{05}, 029 (2016).
\url{https://doi.org/10.1007/JHEP05(2016)029}

\bibitem{PV12} D. Kubiznak, R. B. Mann, and M. Teo,
Black hole chemistry: thermodynamics with $\Lambda$,
Class. Quantum Grav. \textbf{34}, 063001 (2017).
\url{https://doi.org/10.1088/1361-6382/aa5c69}

\bibitem{PV13} S. H. Hendi, R. B. Mann, S. Panahiyan, and B. Eslam Panah,
Van der Waals like behavior of topological AdS black holes in massive gravity,
Phys. Rev. D \textbf{95}, 021501(R) (2017).
\url{https://doi.org/10.1103/PhysRevD.95.021501}

\bibitem{PV14} S. H. Hendi, B. Eslam Panah, and S. Panahiyan,
Topological charged black holes in massive gravity's rainbow and their thermodynamical analysis through various approaches,
Phys. Lett. B \textbf{769}, 191 (2017).
\url{https://doi.org/10.1016/j.physletb.2017.03.051}

\bibitem{PV15} H.-L. Li, Z.-W. Feng, and X.-T. Zu,
Holographic Van der Waals phase transition of the higher-dimensional electrically charged hairy black hole,
Eur. Phys. J. C \textbf{78}, 49 (2018).
\url{https://doi.org/10.1140/epjc/s10052-018-5515-8}

\bibitem{PV16} D. V. Singh and S. Siwach,
Thermodynamics and $P$-$v$ criticality of Bardeen-AdS black hole in 4D Einstein-Gauss-Bonnet gravity,
Phys. Lett. B \textbf{808}, 135658 (2020).
\url{https://doi.org/10.1016/j.physletb.2020.135658}

\bibitem{PV17} B. Eslam Panah and Kh. Jafarzade,
Thermal stability, $P$-$V$ criticality and heat engine of charged rotating accelerating black holes,
Gen. Relativ. Gravit. \textbf{54}, 19 (2022).
\url{https://doi.org/10.1007/s10714-022-02904-9}

\bibitem{PV18} Kh. Jafarzade, B. Eslam Panah, and M. E. Rodrigues,
Thermodynamics and optical properties of phantom AdS black holes in massive gravity,
Class. Quantum Grav. \textbf{41}, 065007 (2024).
\url{https://doi.org/10.1088/1361-6382/ad242e}

\bibitem{PV19} B. Eslam Panah, B. Hamil, and M. E. Rodrigues,
Thermodynamics, phase transitions, and geodesic structure of $F(R)$-phantom Banados--Teitelboim--Zanelli (BTZ) black holes,
Prog. Theor. Exp. Phys. \textbf{2026}(5), 053E03 (2026).
\url{https://doi.org/10.1093/ptep/ptag067}

\bibitem{Li2014} G.-Q. Li,
Effects of dark energy on $P$-$V$ criticality of charged AdS black holes,
Phys. Lett. B \textbf{735}, 256--260 (2014).
\url{https://doi.org/10.1016/j.physletb.2014.06.047}

\bibitem{liu2017} H. Liu and X.-H. Meng,
Effects of dark energy on the efficiency of charged AdS black holes as heat engines,
Eur. Phys. J. C \textbf{77}, 556 (2017).
\url{https://doi.org/10.1140/epjc/s10052-017-5134-9}

\bibitem{Xu2018} W. Xu and Y. Wu,
``Exact phase structure of AdS black holes surrounded by quintessence dark energy,''
\textit{EPL} \textbf{121}, 40001 (2018).
\url{https://doi.org/10.1209/0295-5075/121/40001}

\bibitem{1} S. Chandrasekhar,
\textit{The Mathematical Theory of Black Holes}
(Clarendon Press, Oxford; Oxford University Press, New York, 1983).
\url{https://books.google.com/books?id=oLN-AAAAIAAJ}

\bibitem{2} V. Perlick,
\textit{Ray Optics, Fermat's Principle, and Applications to General Relativity}
(Springer, Berlin, 2000).
\url{https://doi.org/10.1007/3-540-46662-2}

\bibitem{3} K. S. Virbhadra and G. F. R. Ellis,
``Gravitational lensing by naked singularities,''
\textit{Phys. Rev. D} \textbf{65}, 103004 (2002).
\url{https://doi.org/10.1103/PhysRevD.65.103004}

\bibitem{4} V. Bozza,
``Gravitational lensing in the strong field limit,''
\textit{Phys. Rev. D} \textbf{66}, 103001 (2002).
\url{https://doi.org/10.1103/PhysRevD.66.103001}

\bibitem{Claudel2001} C. M. Claudel, K. S. Virbhadra, and G. F. R. Ellis,
``The geometry of photon surfaces,''
\textit{J. Math. Phys.} \textbf{42}, 818--838 (2001).
\url{https://doi.org/10.1063/1.1308507}

\bibitem{8} G. W. Gibbons and M. C. Werner,
``Applications of the Gauss--Bonnet theorem to gravitational lensing,''
\textit{Class. Quantum Grav.} \textbf{25}, 235009 (2008).
\url{https://doi.org/10.1088/0264-9381/25/23/235009}

\bibitem{9} M. C. Werner,
``Gravitational lensing in the Kerr--Randers optical geometry,''
\textit{Gen. Relativ. Gravit.} \textbf{44}, 3047--3057 (2012).
\url{https://doi.org/10.1007/s10714-012-1458-9}

\bibitem{10} G. W. Gibbons, C. A. R. Herdeiro, C. M. Warnick, and M. C. Werner,
``Stationary metrics and optical Zermelo--Randers--Finsler geometry,''
\textit{Phys. Rev. D} \textbf{79}, 044022 (2009).
\url{https://doi.org/10.1103/PhysRevD.79.044022}

\bibitem{11} K. Jusufi, M. C. Werner, A. Banerjee, and A. \\"Ovgu\\"n,
``Light deflection by a rotating global monopole spacetime,''
\textit{Phys. Rev. D} \textbf{95}, 104012 (2017).
\url{https://doi.org/10.1103/PhysRevD.95.104012}

\bibitem{25} P. V. P. Cunha and C. A. R. Herdeiro,
``Shadows and strong gravitational lensing: a brief review,''
\textit{Gen. Relativ. Gravit.} \textbf{50}, 42 (2018).
\url{https://doi.org/10.1007/s10714-018-2361-9}

\bibitem{Perlick2022} V. Perlick and O. Yu. Tsupko,
``Calculating black hole shadows: Review of analytical studies,''
\textit{Phys. Rep.} \textbf{947}, 1--39 (2022).
\url{https://doi.org/10.1016/j.physrep.2021.10.004}

\bibitem{Synge1966} J. L. Synge,
``The escape of photons from gravitationally intense stars,''
\textit{Mon. Not. R. Astron. Soc.} \textbf{131}, 463--466 (1966).
\url{https://doi.org/10.1093/mnras/131.3.463}

\bibitem{12} S. Gurtas Dogan, A. Guvendi, and O. Mustafa,
``Ray geodesics and wave propagation on the Beltrami surface: optics of an optical wormhole,''
\textit{Eur. Phys. J. C} \textbf{85}, 896 (2025).
\url{https://doi.org/10.1140/epjc/s10052-025-14644-6}

\bibitem{13} S. Gurtas Dogan, A. Guvendi, and O. Mustafa,
``Ray and wave optics in an optical wormhole,''
\textit{Phys. Lett. B} \textbf{868}, 139626 (2025).
\url{https://doi.org/10.1016/j.physletb.2025.139626}

\bibitem{14} S. Gurtas Dogan, A. Guvendi, and O. Mustafa,
``Geometric and wave optics in a BTZ optical metric-based wormhole,''
\textit{Phys. Lett. B} \textbf{868}, 139824 (2025).
\url{https://doi.org/10.1016/j.physletb.2025.139824}

\bibitem{15} S. Gurtas Dogan, A. Guvendi, and O. Mustafa,
``Ray trajectories and wave optics in a traversable wormhole: Optics of a minimal surface,''
\textit{Nucl. Phys. B} \textbf{1018}, 117000 (2025).
\url{https://doi.org/10.1016/j.nuclphysb.2025.117000}

\bibitem{19} S. Gurtas Dogan, K. M. Kaya, O. T. Kaya, U. E. Kaya, and O. Mustafa,
``Wave optics on helicoidal graphene nanoribbons,''
\textit{Phys. Scr.} \textbf{100}, 095102 (2025).
\url{https://doi.org/10.1088/1402-4896/adfe2d}

\bibitem{26} A. Errehymy, A. Guvendi, S. Gurtas Dogan, and O. Mustafa,
``Null geodesics and shadows of slowly rotating wormholes immersed in dark matter halos,''
\textit{Eur. Phys. J. C} \textbf{86}, 124 (2026).
\url{https://doi.org/10.1140/epjc/s10052-026-15356-1}

\bibitem{17} F. Ahmed and A. Bouzenada,
``Geometric and wave optics in a topologically charged Perry--Mann type wormhole with disclinations,''
\textit{Phys. Lett. B} \textbf{868}, 139704 (2025).
\url{https://doi.org/10.1016/j.physletb.2025.139704}

\bibitem{18} F. Ahmed and A. Bouzenada,
``Lorentz-violating and topological effects on gravitational lensing phenomena and wave optics in wormhole backgrounds,''
\textit{Phys. Dark Universe} \textbf{50}, 102111 (2025).
\url{https://doi.org/10.1016/j.dark.2025.102111}

\bibitem{20} S. Gurtas Dogan, A. Guvendi, and O. Mustafa,
``Wave and ray optics in EiBI global monopole geometry,''
\textit{Phys. Lett. B} \textbf{878}, 140560 (2026).
\url{https://doi.org/10.1016/j.physletb.2026.140560}

\bibitem{Spiral2026} S. Gurtas Dogan, O. Mustafa, A. Guvendi, and H. Hassanabadi,
``Optics in spiral dislocation spacetime: Torsion as a geometric waveguide and frequency-filtering mechanism,''
\textit{Eur. Phys. J. C} \textbf{86}, 31 (2026).
\url{https://doi.org/10.1140/epjc/s10052-025-15239-x}

\bibitem{Godel2026} O. Mustafa, A. Guvendi, and S. Gurtas Dogan,
``Optics in the Godel universe: Effects of global rotation and Planck-scale corrections,''
\textit{Nucl. Phys. B} \textbf{1024}, 117347 (2026).
\url{https://doi.org/10.1016/j.nuclphysb.2026.117347}

\bibitem{CJP2026} A. Guvendi, O. Mustafa, S. Gurtas Dogan, A. Errehymy, and H. Hassanabadi,
``Scalar-wave signatures of wormholes in dark matter halos,''
\textit{Chin. J. Phys.} \textbf{102}, 633--647 (2026).
\url{https://doi.org/10.1016/j.cjph.2026.04.031}

\bibitem{EMSchw2026} A. Guvendi, O. Mustafa, S. Gurtas Dogan, and H. Hassanabadi,
``Electromagnetic wave propagation in static black hole spacetimes: An effective refractive index description in Schwarzschild geometry,''
\textit{Phys. Scr.} \textbf{101}, 285302 (2026).
\url{https://doi.org/10.1088/1402-4896/ae8726}

\bibitem{EslamPanah2026bumblebee} B. Eslam Panah, A. Guvendi, S. Gurtas Dogan, and O. Mustafa,
``Optical and thermodynamic signatures of Lorentz symmetry breaking in bumblebee AdS black holes,''
\textit{Chin. J. Phys.} \textbf{103}, 1698--1713 (2026).
\url{https://doi.org/10.1016/j.cjph.2026.07.024}

\bibitem{21} N. D. Birrell and P. C. W. Davies,
\textit{Quantum Fields in Curved Space}
(Cambridge University Press, Cambridge, 1982).
\url{https://doi.org/10.1017/CBO9780511622632}

\bibitem{22} S. Chandrasekhar and S. Detweiler,
``The quasi-normal modes of the Schwarzschild black hole,''
\textit{Proc. R. Soc. Lond. A} \textbf{344}, 441--452 (1975).
\url{https://doi.org/10.1098/rspa.1975.0112}

\bibitem{23} B. F. Schutz and C. M. Will,
``Black hole normal modes: A semianalytic approach,''
\textit{Astrophys. J. Lett.} \textbf{291}, L33--L36 (1985).
\url{https://doi.org/10.1086/184453}

\bibitem{24} S. Iyer and C. M. Will,
``Black-hole normal modes: A WKB approach. I. Foundations and application of a higher-order WKB analysis of potential-barrier scattering,''
\textit{Phys. Rev. D} \textbf{35}, 3621--3631 (1987).
\url{https://doi.org/10.1103/PhysRevD.35.3621}

\bibitem{Konoplya2011} R. A. Konoplya and A. Zhidenko,
``Quasinormal modes of black holes: From astrophysics to string theory,''
\textit{Rev. Mod. Phys.} \textbf{83}, 793--836 (2011).
\url{https://doi.org/10.1103/RevModPhys.83.793}

\bibitem{Horowitz2000} G. T. Horowitz and V. E. Hubeny,
``Quasinormal modes of AdS black holes and the approach to thermal equilibrium,''
\textit{Phys. Rev. D} \textbf{62}, 024027 (2000).
\url{https://doi.org/10.1103/PhysRevD.62.024027}

\bibitem{Cardoso2009} V. Cardoso, A. S. Miranda, E. Berti, H. Witek, and V. T. Zanchin,
``Geodesic stability, Lyapunov exponents, and quasinormal modes,''
\textit{Phys. Rev. D} \textbf{79}, 064016 (2009).
\url{https://doi.org/10.1103/PhysRevD.79.064016}

\bibitem{Zou2017} D.-C. Zou, Y. Liu, and R.-H. Yue,
Behavior of quasinormal modes and Van der Waals-like phase transition of charged AdS black holes in massive gravity,
Eur.\ Phys.\ J.\ C \textbf{77}, 365 (2017).
\url{https://doi.org/10.1140/epjc/s10052-017-4937-z}

\bibitem{Wang2026} Y. Wang, H. Chen, J. Chen, and Y. Wang,
Thermodynamical phase structures and particle emission rate of charged AdS
black hole surrounded by string cloud and quintessence via shadow formalism,
arXiv:2606.03263 [gr-qc] (2026).
\url{https://arxiv.org/abs/2606.03263}

\bibitem{AhmedSilva2026} F. Ahmed and E. O. Silva,
Thermodynamic and radiative properties of Euler--Heisenberg AdS black holes
surrounded by quintessence and dark matter with a cloud of strings,
arXiv:2604.25629 [gr-qc] (2026).
\url{https://arxiv.org/abs/2604.25629}

\end{thebibliography}
\end{document}